\documentclass[runningheads]{llncs}
\usepackage{style}
\begin{document}
\title{Counting Cycles on Planar Graphs in Subexponential Time}
%
%
\author{Jin-Yi Cai\inst{1} \and
Ashwin Maran\inst{1}}
\authorrunning{J. Cai, A. Maran}
%
\institute{University of Wisconsin-Madison, USA}
\maketitle              
\begin{abstract}
    We study the problem of counting all cycles or self-avoiding walks (SAWs) on triangulated planar graphs.
    We present a subexponential $2^{O(\sqrt{n})}$ time algorithm for this counting problem.
    Among the technical ingredients used in this algorithm are the planar separator theorem and a delicate analysis using pairs of Motzkin paths and Motzkin numbers.
    We can then adapt this algorithm to uniformly sample SAWs, 
    in subexponential time. 
    Our work is motivated by the problem of gerrymandered districting maps.

\keywords{counting cycles \and planar graphs \and planar separator \and Motzkin paths}
\end{abstract}
\section{Introduction}\label{sec:intro}

We consider the  problem of counting
all (simple) cycles, or   self-avoiding  walks  (SAWs), in a planar undirected graph.
well known algorithms by Tiernan~\cite{tiernan1970efficient}, Tarjan~\cite{tarjan1973enumeration}, and Johnson~\cite{johnson1975finding} enumerate cycles in a graph, such that it
outputs one cycle per polynomial number of steps.
However, many planar graphs (e.g.,
 $n \times  O(1)$ grid graphs) have $2^{\Omega(n)}$ cycles.
 Bousquet-M\'{e}lou et al.~\cite{bousquet2005self}
 solved the problem of counting SAWs for
 $n \times n$ grid graphs exactly\footnote{The algorithm in~\cite{bousquet2005self} is applicable for grid graphs only,
 and it was explicitly calculated that the 
number of self avoiding walks connecting two diagonal corners in a
$19 \times 19$ grid graph is 
 $1523344971704879993080742810319229690899454255323294555776029866737355060592877569255844> 10^{88}$. 
 Our algorithm
 for  planar graphs is 
based on a recursive, thus different, approach.}.
Dorn et al.~\cite{dorn2010efficient} solved the decision version of the weighted Hamiltonian cycle problem
on planar graphs in $2^{O(\sqrt{n})}$ time using planar branch decomposition, with $n= |V(G)|$, as well as the decision problem of
cycles of length $\ge k$  in time
 $O(2^{13.6 \sqrt{k}} n + n^3)$ on planar graphs (see also~\cite{bodlaender2015deterministic}).
We give a $2^{O(\sqrt{n})}$ time algorithm for the problem of counting all cycles in a planar undirected 3-regular graph.
Compared to the approaches in~\cite{dorn2010efficient,bodlaender2015deterministic},
our algorithm is more elementary in the use of
structural graph properties.
It is well known~\cite{jerrum1986random}
that
counting leads to
uniform sampling.
We also extend our algorithm to count the number of ways a planar
graph can be partitioned into two simply connected components, in $2^{O(\sqrt{n})}$ time.
It is known~\cite{najt2019complexity} that this problem cannot be solved in polynomial time unless $\NP = \RP$.


\subsection{Overview}\label{sec: overview}

A basic ingredient to achieving this time complexity
is the planar separator theorem~\cite{lipton1979separator,miller1986finding,alon1994planar,spielman1996disk,har2011simple}.
We consider our plane graph $H$ is embedded on a sphere, and let $G = H^*$ be its dual graph.
The overall idea is to divide the sphere by
a cycle separator of $G$ and solve a more general problem  in two smaller 
topological semi-spheres.
We use (a version of) the planar separator theorem to find a 
planar separator which is a cycle on $G$.
We then use the fact that cycles on  $G$ correspond to cuts on the primal $H$,
to derive  the number of cycles on the primal graph based on 
structural properties of certain quantities we recursively compute
for the topological semi-spheres.

A Motzkin path has many equivalent
definitions. We can think of
a (cyclic) Motzkin path as encoding a (cyclic) string on the alphabet set
$\{\texttt{0}, \texttt{(}, \texttt{)}\}$ of matching
left and right parentheses.
We make the crucial observation that each cycle 
 on the original primal graph, being  self-avoiding, 
 intersects with the cycle separator in a way that corresponds
 to a Motzkin path on the cycle separator (with the symbol ``$\texttt{0}$''
 indicating non-crossing at that location). 
 Therefore the number of cycles is intimately related to
  the Motzkin number which counts Motzkin paths on the cut 
  generated by the cycle separator.
It is  well known~\cite{aigner1998motzkin,motzkin1948relations} that the Motzkin number of a cycle of size $k$ is $2^{O(k)}$, 
and that the Motzkin paths can be enumerated. Note that our separator
has size $k = O(\sqrt{n})$.
We use this fact to efficiently keep track of all possible ways that cycles on the primal graph
can intersect with the cut-set generated by the cyclic planar separator.

We then need to somehow count the number of 
cycles on the primal graph that intersects with the cut-set in a  way specified by a Motzkin path.
We look at  each of the two subgraphs that the planar separator divides the dual graph into.
We would like to count the number of traces that are
the intersections of cycles
on each of these subgraphs
that traverse  the cut-set specified by a Motzkin path, and then combine them to count the number of cycles on the original primal graph.
However, this problem turns out to be a little tricky.
It turns out that just keeping track of the intersection of the cycles with the cut-set is not enough.
Given a trace which is an intersection of a cycle with
one of the two subgraphs,
it is possible to have extensions into the subgraph on the other semi-sphere, that form
cycles,
but that are topologically distinct (see \cref{fig:ell1notUnique}).
This complication makes it harder to disentangle what happens in the two subgraphs, and in turn, 
this makes it harder to synthesize the counts from the two subgraphs and find the number of cycles on the original graph.

We solve this problem by keeping track of not just the intersections of the
traces with the cut-set, 
but also the \emph{order} in which they are crossed.
At face value, this seems to introduce an additional factor  $k!$ for $k$ crossings. 
This would ruin our $2^{O(\sqrt{n})}$ time bound.  However,
we argue that by using a pair of Motzkin paths this difficulty can be
overcome.
Abstractly,
suppose $K, K' \subset S^2$ are  homeomorphic to $S^1$ and  they intersect at $k$ points transversely, then there are only  $2^{O(k)}$
many topologically distinct ways this can happen
(see \cref{remark:pairMotzkinPaths}).

In \cref{sec:preliminary},
we introduce necessary
definitions and notations, and we 
show in \cref{sec:reductionSection} 
that the problem of counting cycles on a 3-regular planar graph can be reduced to 
the problem of counting
self-avoiding walks (SAWs)
on subgraphs that pass through some fixed edges of the subgraph in a given (cyclic) order. 
In \cref{sec:stitchingSection}, we show that this problem can be further reduced to 
counting cycles on a smaller planar graph, that pass through some fixed edges in a given order. 
In \cref{sec:inductionSection}, we inductively solve this problem, 
and show that the number of cycles on a planar graph can be counted in subexponential time.
There is a 1-1 correspondence between cycles and partitions of a planar
graph (on $S^2$) into two simply connected regions (viewed in the dual graph). 
In \cref{sec:extensions}, we extend the algorithm to count the number of partitions of a planar graph  into two simply connected regions. It is well known that
counting leads to uniform sampling~\cite{jerrum1986random}. 
Our work is motivated by the problem of detecting and
certifying gerrymandered districting maps.
Sampling can provide a provably unbiased argument that a given gerrymandered map is an extreme outlier as measured by various fairness criteria; see~\cite{ellenberg2021geometry}.

\section{Self-Avoiding Walks and Partial Self-Avoiding Walks}\label{sec:preliminary}

Let $H$ be a connected 3-regular planar graph with an embedding
on the sphere $S^2$.  Our technical problem
is to count the number of cycles on the graph $H$.
(Our graph $H$ can have parallel edges but no loops.)
We denote by $G = (V, E, F)$ the plane dual graph 
of $H$, and denote $H$ by  $H = (\tilde{F}, \tilde{E}, \tilde{V})$ (see \cref{remark:dualRemark} in the \cref{sec:appendix_a}).
We will use induction on this graph $G$ to find the number of cycles on $H$.
So, we will define the set of cycles on the graph $H$ using its dual graph $G$ as  index.
 
\begin{definition}\label{def:SAW}
	A simple path, or \textit{self-avoiding walk} (SAW), in $H$ is a vertex (and edge) disjoint path (also a sequence of distinct adjacent triangular faces in
	$G$). 	A cycle is a special case of SAW with the same beginning and ending vertex.
	Denote by $\mathcal{W}_{G}$  the set of all cycles on the graph $H$.
\end{definition}

While our goal is to count cycles on $H$, it turns out that cycles on  $G$ are extremely helpful.
Let $A = (V_{A}, E_{A})$ be any cycle in $G$. This cycle divides $S^{2}$ into two parts
$D_{1}$ and $D_{2}$,
each homeomorphic to a disc with $A$ as its boundary. Thus, this cycle $A$ gives rise to two subgraphs $G_{D_{i}} = (V_{D_{i}}, E_{D_{i}}, F_{D_{i}})$ ($i = 1, 2$)
on these two discs.

Let $D$ be one of  $D_1$ or  $D_2$, and let
$D'$ be the other one. 
Consider the graph $G_{D}$, and its dual $H_{D}$. Note that $H_D$ has a special vertex $v_{\infty}$ (of degree $|V_A|=|E_A|$) which corresponds to  the disc $D'$ on the sphere.
Suppose  $\{u, v\} \in \tilde{E}$ is an edge from $H$,
such that
$u \in \tilde{F}_{D}$ but $v \in \tilde{F}_{D'}$. 
This is precisely when
$\{u, v\} \in \tilde{E}_{A}$. Now we modify
$H_{D}$ as follows: 
We remove $v_{\infty}$  and its incident edges, and for each $\{u, v\} \in \tilde{E}_{A}$ with  $v \in \tilde{F}_{D'}$ we add a \emph{dangling edge}
$\{u, v^*\}$ to $\tilde{E}_{D}$ with $v^*$ being a new \emph{dangling vertex} (see \cref{remark:danglingEdgeRemark} in the \cref{sec:appendix_a}).

\begin{figure}
 \centering
 \begin{subfigure}[b]{0.3\textwidth}
     \centering
     \includegraphics[width=4cm]{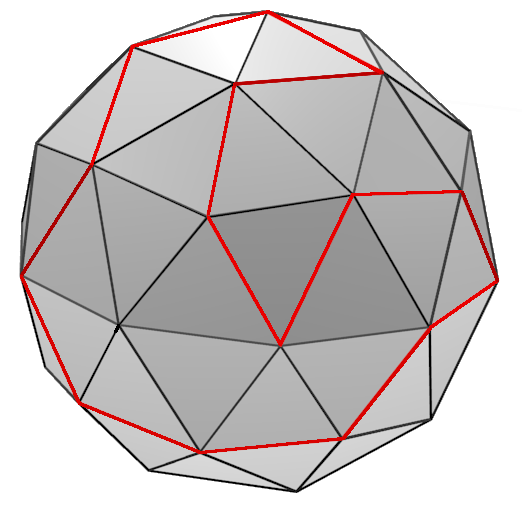}
 \end{subfigure}
 \hfill
 \begin{subfigure}[b]{0.3\textwidth}
     \centering
     \includegraphics[width=4cm]{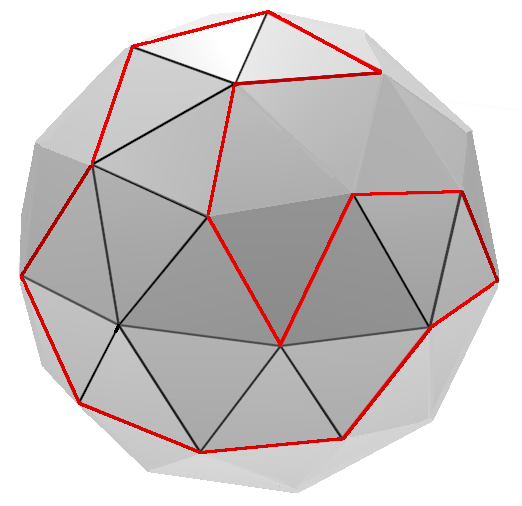}
 \end{subfigure}
 \hfill
 \begin{subfigure}[b]{0.3\textwidth}
     \centering
     \includegraphics[width=4cm]{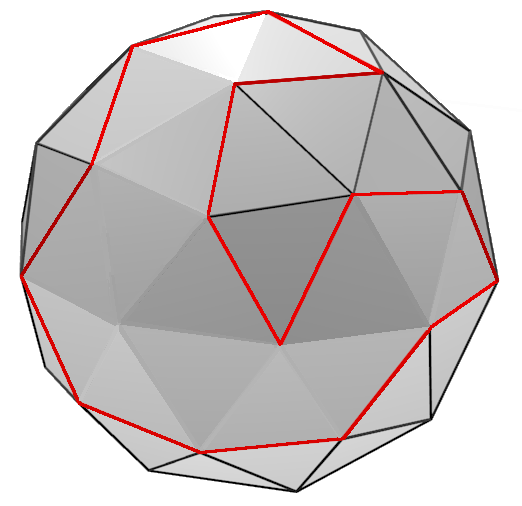}
 \end{subfigure}
    \caption{The red cycle $A$ dividing the graph $G$ embedded on $S^{2}$, and generating the two subgraphs $G_{D_{1}}$ and $G_{D_{2}}$.}
    \label{fig:sphereExample}
\end{figure}

Since $A$ is a cycle on the graph $G$, the edges $\tilde{E}_{A}$ in the graph $H$ act as a cut between the vertices in $\tilde{F}_{D_{1}}$ and $\tilde{F}_{D_{2}}$
(see \cref{fig:sphereExample}).
Therefore, if any $W \in \mathcal{W}_{G}$ has to go from $\tilde{F}_{D_{1}}$ to $\tilde{F}_{D_{2}}$, it has to go across $\tilde{E}_{A}$
(a consequence of the Jordan Curve Theorem). Also the fact that $W$ is non-self-intersecting places strong constraints on how it will interact with $\tilde{E}_{A}$.

Let $W \in \mathcal{W}_{G}$, and let 
$P = W \cap \tilde{E}_{D}$,
for  $D \in \{D_1, D_2\}$. Then for this $D$ we 
will use this set of edges $P$ in $H$ to define a function $\ell_{P}$ on $E_{A} \subseteq E$.	
Consider any edge $e \in E_{A}$. Let its corresponding edge in the graph $H$ be $\tilde{e} \in \tilde{E}_{A} \subseteq \tilde{E}$. If $\tilde{e} \notin P$, then we define $\ell_{P}(e) = 0$. Suppose $\tilde{e} \in P$.

\begin{restatable}{lemma}{labellingExists}\label{lemma:labellingExists}
	Let $\tilde{e} = \{u_{1}, u_{2}\} \in P \cap \tilde{E}_{A}$. Without loss of generality  $u_{1} \notin \tilde{F}_{D}$, and $u_{2} \in \tilde{F}_{D}$. Then there exist unique vertices $u_{3}, \dots, u_{k - 1} \in \tilde{F}_{D}$, and $u_{k} \notin \tilde{F}_{D}$ ($k \ge 3$), such that $\{u_{i}, u_{i + 1}\} \in P$ for $1 \leq i < k$.
\end{restatable}

The proof can be found in the \cref{sec:appendix_a}.

	

Let $\tilde{e}' = \{u_{k - 1}, u_{k}\}$  be the edge such that $u_{k - 1} \in \tilde{F}_{D}$ and $u_{k} \notin \tilde{F}_{D}$
as guaranteed by \cref{lemma:labellingExists}. We note that $\tilde{e}' \not =
\tilde{e}$ (see \cref{remark:labellingRemark} in the \cref{sec:appendix_a}).
Then 
$\tilde{e}' \in P \cap \tilde{E}_A$, and let
$e' \in E_{A}$ be the corresponding edge in $E_{A}$. Now we  define $\ell_{P}(e) = e'$, which completes the definition of  the function $\ell_{P} : E_{A} \rightarrow E_{A} \cup \{0\}$.



For any $P = W \cap \tilde{E}_{D}$ where $W \in \mathcal{W}_{G}$, the function $\ell_{P}$ represents a set of non-intersecting chords on the cycle $A$. Now, let us characterize the set of all functions that represent non-intersecting chords on the cycle $A$. Consider the following three  properties for a function $\ell: E_{A} \rightarrow E_{A} \cup \{0\}$:
\begin{enumerate}
	\item For any $e \in E_A$,
	$\ell(e) \ne e$.
	\item For any $e, e' \in E_A$,
	$\ell(e) = e' \iff \ell(e') = e$.
	\item  For any $e, e' \in E_A$,
	if $\ell(e) = e'$,
	then edges $e, e'$ partition the cycle $E_{A}$ into two segments  $\sigma_{1} \sqcup \sigma_{2} = E_{A} \setminus \{e, e'\}$ such that for no two edges $e_{1} \in \sigma_{1}, e_{2} \in \sigma_{2}$ is $\ell(e_{1}) = e_{2}$.
\end{enumerate}

Let $\mathcal{L}_{A}$ be the set of all functions $\ell: E_{A} \rightarrow E_{A} \cup \{0\}$ that satisfy the above properties. 	Every $\ell_{P}$ defined by $P = W \cap \tilde{E}_{D}$ for any $W \in \mathcal{W}_{G}$ belongs to
$\mathcal{L}_{A}$. Thus,
$$\{\ell_{W \cap \tilde{E}_{D}}: W \in \mathcal{W}_{G} \} \subseteq \mathcal{L}_{A},\ \forall\ D \in \{D_{1}, D_{2}\}.$$

Each $\ell \in \mathcal{L}_{A}$ represents a Motzkin path on $E_{A}$ (see \cref{lemma:symmetryOfEll} in the \cref{sec:appendix_a}), and $\left\lvert \mathcal{L}_{A} \right\rvert$ is a Motzkin number~\cite{motzkin1948relations,aigner1998motzkin}. The following is a well known fact about Motzkin numbers.

\begin{fact}\label{fact:MotzkinBound}
	The number of Motzkin paths of length $n$ is $2^{O(n)}$. Thus, there exists a constant $\kappa$  such that~\footnote{It is known that asymptotically
	we have $ \left\lvert \mathcal{L}_{A} \right\rvert \leq O(3^{|E_{A}|})$.} 
	for any planar triangulated graph $G$ and a cycle $A$ on $G$, $ \left\lvert \mathcal{L}_{A} \right\rvert \leq \kappa^{|E_{A}|}$.
\end{fact}

Now, let
$$\mathcal{C}(\ell_{1}, \ell_{2}) = \{W \in \mathcal{W}_{G}: \ell_{W \cap \tilde{E}_{D_{1}}} = \ell_{1} ~~\mbox{and}~~ \ell_{W \cap \tilde{E}_{D_{2}}} = \ell_{2} \}.$$

The following lemma shows that $\left\lvert \mathcal{W}_{G} \right\rvert$
is determined by $\lvert \mathcal{C}(\ell_{1}, \ell_{2}) \rvert$ for all $\ell_{1}, \ell_{2} \in \mathcal{L}_{A}$. 

\begin{restatable}{lemma}{compatibilityPartitionSimple}\label{lemma:compatibilityPartitionSimple}
	$$\left\lvert \mathcal{W}_{G} \right\rvert = \sum_{\ell_{1} \in \mathcal{L}_{A}} \sum_{\ell_{2} \in \mathcal{L}_{A}}\left\lvert \mathcal{C}(\ell_{1}, \ell_{2}) \right\rvert.$$
\end{restatable}

The proof can be found in the \cref{sec:appendix_a}.

	

Now, let us see how we can compute $\lvert \mathcal{C}(\ell_{1}, \ell_{2}) \rvert$. Essentially,
for $\ell_1, \ell_2 \in  \mathcal{L}_{A}$, we will   
 count $\lvert \{W \cap \tilde{E}_{D_1}: W \in \mathcal{W}_{G}, \ell_{W \cap \tilde{E}_{D_1}} = \ell_1\} \rvert$
 and $\lvert \{W \cap \tilde{E}_{D_2}: W \in \mathcal{W}_{G}, \ell_{W \cap \tilde{E}_{D_2}} = \ell_2\} \rvert$.
With some care, we will be 
able to combine them to find the value of $\lvert \mathcal{C}(\ell_{1}, \ell_{2}) \rvert$.

In order to formalize this notion, we will now define \textit{partial self-avoiding walks} (PSAW) on the disc $D$ (see \cref{remark:PSAWRemark} in
the \cref{sec:appendix_a}).
Let $\ell \in \mathcal{L}_{A}$ be a labelling.
We saw that $\ell$ represents a Motzkin path, which corresponds to a set of non-intersecting chords on $E_{A}$.
Let $D$ and $D'$ be as before.
We will view this labelling $\ell$ as being embedded in the
disc $D'$.
For any $e \in E_{A}$,
we have a dangling edge $\tilde{e}$
incident to a newly created dangling vertex $v^*$ not in 
$\tilde{F}_{D}$, which we will
denote by  $\pi_{D}(e)$.
If $\ell(e) = e'$ we also have
a dangling vertex $\pi_{D}(e')$.

Now, let
$$\mu_{D}(\ell) = \big\{ \{\pi_{D}(e), \pi_{D}(e') \}: e, e' \in E_{A} \text{ such that } \ell(e) = e' \text{ and } \ell(e') = e \big\}.$$
We can think of the 
dangling vertices as belonging to the other disc $D'$ (see \cref{fig:partialWalkExample} in \cref{sec:appendix_a} for an example of the dangling edges). Then, $\mu_{D}(\ell)$ can be viewed as a set of edges between these dangling vertices that are embedded on $D'$.

\begin{definition}\label{def:PSAW}
	Let  $D$ be one of $D_1$ or $D_2$. 
	A set of edges $P
	\subseteq \tilde{E}_{D}$ is called a \textit{partial self-avoiding walk} (PSAW) on the graph $H_{D}$, if there exists an $\ell \in \mathcal{L}_{A}$, such that $P \cup \mu_{D}(\ell)$ is a cycle. 
	Denote by $\mathcal{P}_{D}$  the set of all PSAW on $H_{D}$. 
\end{definition}

Note that $P$ is a set of edges in $\tilde{E}_{D}$, some of which may be incident on the dangling vertices
which are thought to be in $D'$, and $\mu_{D}(\ell)$ is a set of edges between these vertices as described above. (Despite the notation $\mu_{D}(\ell)$,
intuitively one should think of $\ell \in \mathcal{L}_{A}$
as referring to pairings that take place in  $D'$. See the proof of \cref{lemma:partialSAWwellDefined} in the \cref{sec:appendix_a}.)
But we observe that the definition of $P$ being a PSAW does not explicitly refer to the existence of
some cycle $W \in \mathcal{W}_{G}$.

\begin{restatable}{lemma}{partialSAWwellDefined}\label{lemma:partialSAWwellDefined}
	For any planar triangulated graph $G$ and any $D \in \{D_{1}, D_{2}\}$ 
	determined by a cycle $A$ on $G$,
	$$\{W \cap \tilde{E}_{D}: W \in \mathcal{W}_{G} \} \subseteq \mathcal{P}_{D}.$$
\end{restatable}

The proof can be found in the \cref{sec:appendix_a}. An example of $W \cap \tilde{E}_{D} \in \mathcal{P}_{D}$ can also be seen in \cref{fig:partialWalkExample} in \cref{sec:appendix_a}.


More importantly, as a corollary of \cref{lemma:labellingExists},  given $P \in \mathcal{P}_{D}$ for any $D \in \{D_{1}, D_{2}\}$, since there exists an $\ell \in \mathcal{L}_{A}$ such that $P \cup \mu_{D}(\ell)$ is a cycle
with $(P \cup \mu_{D}(\ell)) \cap \tilde{E}_{D} = P$, we see that $\ell_{P}$ is well-defined and $\ell_{P} \in \mathcal{L}_{A}$. We can then define, for any $\ell \in \mathcal{L}_{A}$, 
$$\mathcal{C}_{D}(\ell) = \{P \in \mathcal{P}_{D}: \ell_{P} = \ell \}.$$

\begin{figure}
	\centering
	\scalebox{0.75}{\begin{tikzpicture}[line join=miter, fill opacity=0.2, draw opacity=1]

\draw[line width=0.5mm, black] (-3, 4.5) -- (-3, -1.5);
\draw[line width=0.5mm, black] (-2, 4.5) -- (-2, -1.5);
\draw[line width=0.5mm, black] (-1, 4.5) -- (-1, -1.5);
\draw[line width=0.5mm, black] (0, 4.5) -- (0, 2);
\draw[line width=0.5mm, black] (0, 0) -- (0, -1.5);
\draw[line width=0.5mm, black] (1, 4.5) -- (1, 2);
\draw[line width=0.5mm, black] (1, 0) -- (1, -1.5);
\draw[line width=0.5mm, black] (2, 4.5) -- (2, -1.5);
\draw[line width=0.5mm, black] (3, 4.5) -- (3, -1.5);
\draw[line width=0.5mm, black] (4, 4.5) -- (4, -1.5);

\draw[line width=0.5mm, black] (-3.5, 4) -- (4.5, 4);
\draw[line width=0.5mm, black] (-3.5, 3) -- (4.5, 3);
\draw[line width=0.5mm, black] (-3.5, 2) -- (4.5, 2);
\draw[line width=0.5mm, black] (-3.5, 1) -- (-1, 1);
\draw[line width=0.5mm, black] (2, 1) -- (4.5, 1);
\draw[line width=0.5mm, black] (-3.5, 0) -- (4.5, 0);
\draw[line width=0.5mm, black] (-3.5, -1) -- (4.5, -1);

\draw[line width=0.5mm, black] (-1, 1) -- (1, 3);
\draw[line width=0.5mm, black] (-2, 1) -- (1, 4);
\draw[line width=0.5mm, black] (-3, 1) -- (0.5, 4.5);
\draw[line width=0.5mm, black] (-3.5, 1.5) -- (-0.5, 4.5);
\draw[line width=0.5mm, black] (-3.5, 2.5) -- (-1.5, 4.5);
\draw[line width=0.5mm, black] (-3.5, 3.5) -- (-2.5, 4.5);

\draw[line width=0.5mm, black] (-1, 1) -- (0, 0);
\draw[line width=0.5mm, black] (-2, 1) -- (0, -1);
\draw[line width=0.5mm, black] (-3, 1) -- (-0.5, -1.5);
\draw[line width=0.5mm, black] (-3.5, 0.5) -- (-1.5, -1.5);
\draw[line width=0.5mm, black] (-3.5, -0.5) -- (-2.5, -1.5);

\draw[line width=0.5mm, black] (2, 1) -- (1, 2);
\draw[line width=0.5mm, black] (3, 1) -- (1, 3);
\draw[line width=0.5mm, black] (4, 1) -- (1, 4);
\draw[line width=0.5mm, black] (4.5, 1.5) -- (1.5, 4.5);
\draw[line width=0.5mm, black] (4.5, 2.5) -- (2.5, 4.5);
\draw[line width=0.5mm, black] (4.5, 3.5) -- (3.5, 4.5);

\draw[line width=0.5mm, black] (2, 1) -- (0, -1);
\draw[line width=0.5mm, black] (3, 1) -- (0.5, -1.5);
\draw[line width=0.5mm, black] (4, 1) -- (1.5, -1.5);
\draw[line width=0.5mm, black] (4.5, 0.5) -- (2.5, -1.5);
\draw[line width=0.5mm, black] (4.5, -0.5) -- (3.5, -1.5);

\fill[yellow, even odd rule] (-3.5, 1) -- (-1, 1) -- (0, 0) -- (1, 0) -- (2, 1) -- (4.5, 1) -- (4.5, -1.5) -- (-3.5, -1.5) -- cycle;
\fill[yellow, even odd rule] (-3.5, 1) -- (-1, 1) -- (0, 2) -- (1, 2) -- (2, 1) -- (4.5, 1) -- (4.5, 4.5) -- (-3.5, 4.5) -- cycle;

\draw[line width=0.8mm, red, dash pattern={on 7pt off 7pt}] (-1, 1) -- (0, 0);
\draw[line width=0.8mm, red, dash pattern={on 7pt off 7pt}] (1, 0) -- (0, 0);
\draw[line width=0.8mm, red, dash pattern={on 7pt off 7pt}] (1, 0) -- (2, 1);
\draw[line width=0.8mm, red, dash pattern={on 7pt off 7pt}] (1, 2) -- (2, 1);
\draw[line width=0.8mm, red, dash pattern={on 7pt off 7pt}] (0, 2) -- (1, 2);
\draw[line width=0.8mm, red, dash pattern={on 7pt off 7pt}] (0, 2) -- (-1, 1);

\draw[line width=0.8mm, brown] (-0.67, 0.33) -- (-0.33, 0.67);
\draw[line width=0.8mm, cyan] (-0.33, 1.33) -- (-0.33, 0.67);
\draw[line width=0.8mm, brown] (-0.33, 1.33) -- (-0.67, 1.67);
\draw[line width=0.8mm, blue] (-0.33, 2.33) -- (-0.67, 1.67);
\draw[line width=0.8mm, blue]  (-0.33, 2.33) -- (0.33, 2.67);
\draw[line width=0.8mm, blue] (0.67, 2.33) -- (0.33, 2.67);
\draw[line width=0.8mm, brown] (0.67, 2.33) -- (0.67, 1.67);
\draw[line width=0.8mm, cyan] (0.67, 0.33) -- (0.67, 1.67);
\draw[line width=0.8mm, brown] (0.67, 0.33) -- (0.33, -0.33);
\draw[line width=0.8mm, blue] (0.67, -0.67) -- (0.33, -0.33);
\draw[line width=0.8mm, blue] (0.67, -0.67) -- (1.33, -0.33);
\draw[line width=0.8mm, blue] (1.67, 0.33) -- (1.33, -0.33);
\draw[line width=0.8mm, brown] (1.67, 0.33) -- (1.33, 0.67);
\draw[line width=0.8mm, cyan] (1.33, 1.33) -- (1.33, 0.67);
\draw[line width=0.8mm, brown] (1.33, 1.33) -- (1.67, 1.67);
\draw[line width=0.8mm, blue] (1.33, 2.33) -- (1.67, 1.67);
\draw[line width=0.8mm, blue] (1.33, 2.33) -- (1.67, 2.67);
\draw[line width=0.8mm, blue] (1.33, 3.33) -- (1.67, 2.67);
\draw[line width=0.8mm, blue] (1.33, 3.33) -- (0.67, 3.33);
\draw[line width=0.8mm, blue] (0.33, 3.67) -- (0.67, 3.33);
\draw[line width=0.8mm, blue] (0.33, 3.67) -- (-0.33, 3.33);
\draw[line width=0.8mm, blue] (-0.67, 2.67) -- (-0.33, 3.33);
\draw[line width=0.8mm, blue] (-0.67, 2.67) -- (-1.33, 2.33);
\draw[line width=0.8mm, blue] (-1.67, 1.67) -- (-1.33, 2.33);
\draw[line width=0.8mm, blue] (-1.67, 1.67) -- (-1.33, 1.33);
\draw[line width=0.8mm, blue] (-1.33, 0.67) -- (-1.33, 1.33);
\draw[line width=0.8mm, blue] (-1.33, 0.67) -- (-0.67, 0.33);

\draw[line width=0.5mm, gray, ->] (-0.3, 1.5) to [in=90, out=0] (0, 1);
\draw[line width=0.5mm, gray, ->] (-0.3, 0.5) to [in=-90, out=0] (0, 1);
\draw[line width=0.5mm, gray, ->] (0.45, 0.1) to (0.45, 1);
\draw[line width=0.5mm, gray, ->] (0.45, 1.9) to (0.45, 1);
\draw[line width=0.5mm, gray, ->] (1.3, 1.5) to [in=90, out=180] (1, 1);
\draw[line width=0.5mm, gray, ->] (1.3, 0.5) to [in=-90, out=180] (1, 1);


\begin{scope}[xshift = 10cm]

\draw[line width=0.5mm, black] (-3, 4.5) -- (-3, -1.5);
\draw[line width=0.5mm, black] (-2, 4.5) -- (-2, -1.5);
\draw[line width=0.5mm, black] (-1, 4.5) -- (-1, -1.5);
\draw[line width=0.5mm, black] (0, 4.5) -- (0, -1.5);
\draw[line width=0.5mm, black] (1, 4.5) -- (1, -1.5);
\draw[line width=0.5mm, black] (2, 4.5) -- (2, -1.5);
\draw[line width=0.5mm, black] (3, 4.5) -- (3, -1.5);
\draw[line width=0.5mm, black] (4, 4.5) -- (4, -1.5);

\draw[line width=0.5mm, black] (-3.5, 4) -- (4.5, 4);
\draw[line width=0.5mm, black] (-3.5, 3) -- (4.5, 3);
\draw[line width=0.5mm, black] (-3.5, 2) -- (-1, 2);
\draw[line width=0.5mm, black] (2, 2) -- (4.5, 2);
\draw[line width=0.5mm, black] (-3.5, 1) -- (4.5, 1);
\draw[line width=0.5mm, black] (-3.5, 0) -- (-1, 0);
\draw[line width=0.5mm, black] (2, 0) -- (4.5, 0);
\draw[line width=0.5mm, black] (-3.5, -1) -- (4.5, -1);

\draw[line width=0.5mm, black] (-2, 1) -- (1, 4);
\draw[line width=0.5mm, black] (-3, 1) -- (0.5, 4.5);
\draw[line width=0.5mm, black] (-3.5, 1.5) -- (-0.5, 4.5);
\draw[line width=0.5mm, black] (-3.5, 2.5) -- (-1.5, 4.5);
\draw[line width=0.5mm, black] (-3.5, 3.5) -- (-2.5, 4.5);

\draw[line width=0.5mm, black] (-2, 1) -- (0, -1);
\draw[line width=0.5mm, black] (-3, 1) -- (-0.5, -1.5);
\draw[line width=0.5mm, black] (-3.5, 0.5) -- (-1.5, -1.5);
\draw[line width=0.5mm, black] (-3.5, -0.5) -- (-2.5, -1.5);

\draw[line width=0.5mm, black] (3, 1) -- (1, 3);
\draw[line width=0.5mm, black] (4, 1) -- (1, 4);
\draw[line width=0.5mm, black] (4.5, 1.5) -- (1.5, 4.5);
\draw[line width=0.5mm, black] (4.5, 2.5) -- (2.5, 4.5);
\draw[line width=0.5mm, black] (4.5, 3.5) -- (3.5, 4.5);

\draw[line width=0.5mm, black] (3, 1) -- (0.5, -1.5);
\draw[line width=0.5mm, black] (4, 1) -- (1.5, -1.5);
\draw[line width=0.5mm, black] (4.5, 0.5) -- (2.5, -1.5);
\draw[line width=0.5mm, black] (4.5, -0.5) -- (3.5, -1.5);

\draw[line width=0.5mm, black] (-1, 0) -- (0, 1);
\draw[line width=0.5mm, black] (-1, 2) -- (0, 1);
\draw[line width=0.5mm, black] (2, 0) -- (1, 1);
\draw[line width=0.5mm, black] (2, 2) -- (1, 1);
\draw[line width=0.5mm, black] (1, 3) -- (0, 1);
\draw[line width=0.5mm, black] (0, -1) -- (1, 1);

\fill[yellow, even odd rule] (-3.5, 4.5) -- (-3.5, -1.5) -- (4.5, -1.5) -- (4.5, 4.5) -- cycle;

\draw[line width=0.8mm, red, dash pattern={on 7pt off 7pt}] (-1, 1) -- (2, 1);

\draw[line width=0.8mm, brown] (-0.67, 0.67) -- (-0.67, 1.33);
\draw[line width=0.8mm, blue] (-0.33, 2) -- (-0.67, 1.33);
\draw[line width=0.8mm, blue]  (-0.33, 2) -- (0.33, 2.33);
\draw[line width=0.8mm, blue]  (0.67, 1.67) -- (0.33, 2.33);
\draw[line width=0.8mm, brown]  (0.67, 1.67) -- (0.33, 0.33);
\draw[line width=0.8mm, blue] (0.67, -0.67) -- (0.33, 0.33);
\draw[line width=0.8mm, blue] (0.67, -0.67) -- (1.33, 0);
\draw[line width=0.8mm, blue] (1.67, 0.67) -- (1.33, 0);
\draw[line width=0.8mm, brown] (1.67, 0.67) -- (1.67, 1.33);
\draw[line width=0.8mm, blue] (1.33, 2) -- (1.67, 1.33);
\draw[line width=0.8mm, blue] (1.33, 2) -- (1.67, 2.67);
\draw[line width=0.8mm, blue] (1.33, 3.33) -- (1.67, 2.67);
\draw[line width=0.8mm, blue] (1.33, 3.33) -- (0.67, 3.33);
\draw[line width=0.8mm, blue] (0.33, 3.67) -- (0.67, 3.33);
\draw[line width=0.8mm, blue] (0.33, 3.67) -- (-0.33, 3.33);
\draw[line width=0.8mm, blue] (-0.67, 2.67) -- (-0.33, 3.33);
\draw[line width=0.8mm, blue] (-0.67, 2.67) -- (-1.33, 2.33);
\draw[line width=0.8mm, blue] (-1.67, 1.67) -- (-1.33, 2.33);
\draw[line width=0.8mm, blue] (-1.67, 1.67) -- (-1.33, 1.33);
\draw[line width=0.8mm, blue] (-1.33, 0.67) -- (-1.33, 1.33);
\draw[line width=0.8mm, blue] (-1.33, 0.67) -- (-0.67, 0.67);

\end{scope}

\end{tikzpicture}}
	\caption{During the stitching process, edges $e, e'$ such that $\ell_{2}(e) = e'$ and $\ell_{2}(e') = e$ are identified with each other. SAWs in the original graph that induce the labelling $\ell_{2}$ on $D_{2}$ correspond to SAWs in the stitched graph that pass through the edges that were identified.}
	\label{fig:contractionExample}
\end{figure}
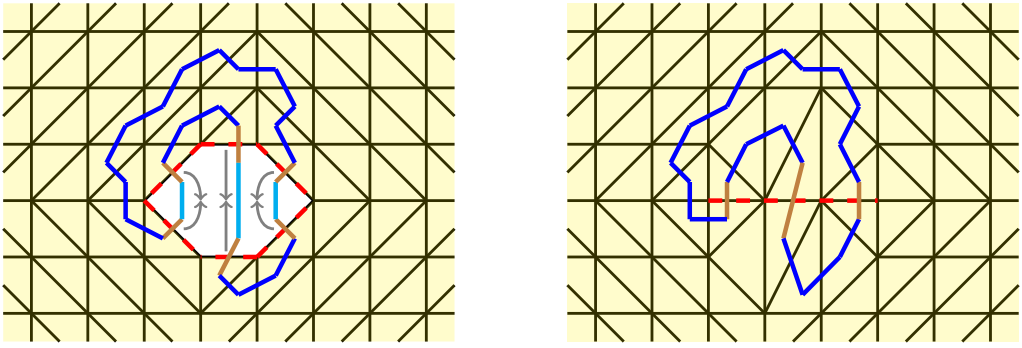

Note that, if $ \mathcal{C}(\ell_{1}, \ell_{2}) \ne \emptyset$, then  $\ell_1$ and $\ell_2$ must assign
the same set of edges $e \in E_A$ to nonzero, even though
the pairings by  $\ell_1$ and $\ell_2$  are typically different. Thus the nonzero places of $\ell_1$ and $\ell_2$ mutually determine each other. 
Our  idea is to try to compute $\lvert \mathcal{C}_{D_{1}}(\ell_{1}) \rvert$ and $\lvert \mathcal{C}_{D_{2}}(\ell_{2}) \rvert$, and then to use it to compute $\lvert \mathcal{C}(\ell_{1}, \ell_{2}) \rvert$. 
Intuitively, we will use $\ell_{2}$ to ``stitch'' up the disc $D_{2}$ in $H_{D_{1}}$ to construct a new planar triangulated graph $G_{1}$ (see \cref{fig:contractionExample} for an example, a more detailed discussion on the exact procedure to be followed can be found in \cref{sec:stitchingSection}). If we can somehow count all the cycles in the graph $H_{1}$ that passes through certain special edges, we will be able to compute $\lvert \{P \in \mathcal{P}_{D_{1}}: P \cup \mu_{D_{1}}(\ell_{2}) \text{ is a cycle} \}\rvert$.

\begin{figure}
	\centering
	\scalebox{0.7}{\input{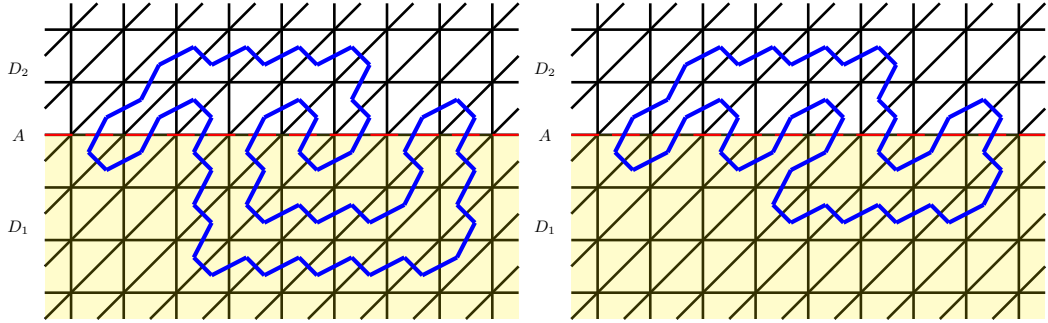}}
	\caption{In this graph $G$, the blue cycles are $W_{1}, W_{2} \in \mathcal{W}_{G}$. $\ell_{W_{1} \cap \tilde{E}_{D_{1}}} \neq \ell_{W_{2} \cap \tilde{E}_{D_{1}}}$, but $\ell_{W_{1} \cap \tilde{E}_{D_{2}}} = \ell_{W_{2} \cap \tilde{E}_{D_{2}}}$. Therefore, given a labelling $\ell_{2} \in \mathcal{L}_{A}$, there can be multiple  $\ell_{1}$ such that $P \cup \mu_{D_{1}}(\ell_{2})$ is a cycle for all $P \in \mathcal{C}_{D_{1}}(\ell_{1})$.}
	\label{fig:ell1notUnique}
\end{figure}

However, as the example in \cref{fig:ell1notUnique} shows,
given  $\ell_{2} \in \mathcal{L}_{A}$, there can be multiple  $\ell_{1}$ such that $P \cup \mu_{D_{1}}(\ell_{2})$ is a cycle for all $P \in \mathcal{C}_{D_{1}}(\ell_{1})$.
Therefore, the stitching  alone unfortunately does not allow us to compute $\lvert \mathcal{C}_{D_{1}}(\ell_{1}) \rvert$ directly. However, the next key insight is that given $\ell_1, \ell_2 \in \mathcal{L}_{A}$ with $\mathcal{C}(\ell_1, \ell_2) \neq \emptyset$, any cycle $W \in \mathcal{C}(\ell_1, \ell_2)$ crosses the same set of edges from $\tilde{E}_{A}$, and more importantly, these edges from $\tilde{E}_{A}$ are crossed in the same cyclic order (upto reversal) by all cycles from $\mathcal{C}(\ell_1, \ell_2)$ (see \cref{remark:conformRemark} in the \cref{sec:appendix_a}).

For any   two adjacent vertices  $u, v$ in $H$ connected by an edge $\{u, v\}$
(or two triangular faces of $G$ that share an edge),
we  use $(u \rightarrow v)$ to refer to an \textit{arc} in $H$ from  $u$ to $v$ in that  direction. Given an arc $\tilde{a} = (u \rightarrow v)$ in $H$, let $\varepsilon(\tilde{a}) \in \tilde{E}$ denote the underlying edge $\{u, v\}$. Let  $S = (\tilde{a}_{1}, \dots, \tilde{a}_{k})$ be an ordered set of arcs in the graph $H$. We  
denote by $\varepsilon(S) = \{\varepsilon(\tilde{a}_1), \dots, \varepsilon(\tilde{a}_k) \}$  the 
unordered set of edges corresponding to the arcs in $S$.

\begin{definition}\label{def:conforms}
	Let $S = ((u_{1}\rightarrow v_{1}), \dots, (u_{k}\rightarrow v_{k}))$
	be an ordered set of arcs in $H$. We say that a cycle $W$ {conforms} to $S$ if $\varepsilon(S) \subseteq W$, and if in one cyclic traversal of the cycle $W$ we encounter the vertices $u_{1}, v_{1}, u_{2}, v_{2}, \dots, u_{k}$, and $v_{k}$ in this order.
\end{definition}

Note that \cref{def:conforms} is defined in terms of some cyclic traversal of $W$.
If $W$ \textit{conforms} to $S$, then $W$ also {conforms} to 
$((u_{2} \rightarrow v_{2}), \dots, (u_{k} \rightarrow v_{k}), (u_{1} \rightarrow v_{1}))$,
as well as its reversal.

We can define a relation $\sim$ on ordered sets of arcs as follows:
$$((u_{1} \rightarrow v_{1}), (u_{2} \rightarrow v_{2}), \dots, (u_{k} \rightarrow v_{k})) \sim ((u_{2} \rightarrow v_{2}), \dots, (u_{k} \rightarrow v_{k}), (u_{1} \rightarrow v_{1})).$$
$$((u_{1} \rightarrow v_{1}), \dots, (u_{k} \rightarrow v_{k})) \sim ((v_{k} \rightarrow u_{k}), \dots, (v_{1} \rightarrow u_{1})).$$
For brevity we will also  denote by $\sim$ its reflexive, symmetric, and transitive closure. Thus,
for $S \sim T$, we have $W$ conforms to $S$ iff $W$ conforms to $T$.

We will now define a few terms to refer to cycles that conform to a given ordered set of arcs.	
\begin{definition}\label{def:WGS}
	Let $S$ be an ordered set of arcs in $H$. We define
	$$\mathcal{W}_{G, S} = \big\{ W \in \mathcal{W}_{G}: W ~\mbox{\rm  conforms to } S \big\}.$$
	For any $\ell_{1}, \ell_{2} \in \mathcal{L}_{A}$,
	$$\mathcal{C}_{S}(\ell_{1}, \ell_{2}) = \big\{ W \in \mathcal{W}_{G, S}: \ell_{W \cap \tilde{E}_{D_{1}}} = \ell_{1} ~\mbox{\rm and}~ \ell_{W \cap \tilde{E}_{D_{2}}} = \ell_{2} \big\}.$$
\end{definition}

\begin{definition}\label{def:PDS-CDS}
	Let $D \in \{D_{1}, D_{2}\}$. Let $S$ be an ordered subset of arcs in $H_{D}$ such that $\varepsilon(S) \subseteq \tilde{E}_{D}$. We define
	$$\mathcal{P}_{D, S} = \big\{P \in \mathcal{P}_{D}: P \cup \mu_{D}(\ell) ~\mbox{\rm conforms to $S$, for some }~\ell \in \mathcal{L}_{A} \big\}.$$
	Finally, for any $\ell \in \mathcal{L}_{A}$, 
	$$\mathcal{C}_{D, S}(\ell) = \{P \in \mathcal{P}_{D, S}: \ell_{P} = \ell \}.$$
\end{definition}

Our main technical result in  this paper is
an  algorithm  to compute $\left\lvert \mathcal{W}_{G,S} \right\rvert$ for a general ordered set $S$ of arcs in $H$.
This will give us a subexponential time
algorithm  to compute
$\left\lvert \mathcal{W}_{G} \right\rvert$.
\section{The Algorithm}\label{sec:algorithm}

Continuing from the previous section, consider an ordered set of arcs $S$ from the graph $H$. Our overall design of the algorithm to compute $\left\lvert \mathcal{W}_{G, S} \right\rvert$ is a divide-and-conquer strategy, and the following planar separator theorem of Lipton and Tarjan~\cite{lipton1979separator} plays a pivotal role. Here we use a version of the  theorem due to Alon et al. using a simple cycle separator~\cite{miller1986finding,alon1994planar,spielman1996disk,har2011simple}
\footnote{For simplicity in stating our results,
	we are not using  optimal constants, see~\cite{alon1994planar,spielman1996disk,djidjev1997reduced} for more details.}.

\begin{theorem}\label{theorem:AlonCurveExists}
	Given a planar triangulated graph $G = (V, E ,F)$, there exists a cycle $A = (V_{A}, E_{A})$ in $G$ with vertices $V_{A} \subseteq V$ and edges $E_{A} \subseteq E$ of size $|V_{A}| = |E_{A}| \leq \sqrt{8|V|}$, which partitions the vertices $V \setminus V_{A}$ into two parts $V_{B}, V_{C}$ such that $|V_{B}|, |V_{C}| \leq \frac{2|V|}{3}$.
	Moreover, this cycle $A$ can be found in polynomial time, as a function of $|V|$.
\end{theorem}

We use \cref{theorem:AlonCurveExists} to find the cycle $A$ on the graph $G$. We may assume that the vertices $V_{A} \cup V_{B}$ belong to the disc $D_{1}$, and the vertices $V_{A} \cup V_{C}$ belong to the disc $D_{2}$.
The following is a generalization  of \cref{lemma:compatibilityPartitionSimple}; the proof is a simple adaptation.
It implies that if we can compute $\left\lvert \mathcal{C}_{S}(\ell_{1}, \ell_{2}) \right\rvert$ for all $\ell_{1}, \ell_{2} \in \mathcal{L}_{A}$, we will be able to compute $\left\lvert \mathcal{W}_{G, S} \right\rvert$.

\begin{lemma}\label{lemma:compatibilityPartition}
	$$\left\lvert \mathcal{W}_{G, S} \right\rvert = \sum_{\ell_{1} \in \mathcal{L}_{A}} \sum_{\ell_{2} \in \mathcal{L}_{A}}\left\lvert \mathcal{C}_{S}(\ell_{1}, \ell_{2}) \right\rvert.$$
\end{lemma}

\subsection{Computing $\left\lvert \mathcal{C}_{S}(\ell_{1}, \ell_{2}) \right\rvert$}\label{sec:reductionSection}
	
We now need to compute $\left\lvert \mathcal{C}_{S}(\ell_{1}, \ell_{2}) \right\rvert$. Given labellings $\ell_{1}, \ell_{2} \in \mathcal{L}_{A}$, we will say they are \textit{compatible} if they pass the test defined by
the algorithm \texttt{CompatibilityCheck}.

\begin{figure}
\centering
\scalebox{0.67}{\begin{tikzpicture}[line join=miter, draw opacity=1]

\draw[line width=0.5mm, black] (-2.5, 0) -- (3.5, 0);
\draw[line width=0.8mm, red, dash pattern={on 7pt off 7pt}] (-2.5, 0) -- (3.5, 0);
\fill[opacity=0.2, yellow, even odd rule] (-2.5, 0) -- (3.5, 0) -- (3.5, -2) -- (-2.5, -2) -- cycle;

\draw[line width=0.8mm, purple] (-2, 0.2) -- (-2, -0.2);
\draw[line width=0.8mm, purple] (-1, 0.2) -- (-1, -0.2);
\draw[line width=0.8mm, purple] (0, 0.2) -- (0, -0.2);
\draw[line width=0.8mm, purple] (1, 0.2) -- (1, -0.2);
\draw[line width=0.8mm, purple] (2, 0.2) -- (2, -0.2);
\draw[line width=0.8mm, purple] (3, 0.2) -- (3, -0.2);

\draw[line width=0.88mm, cyan] (-2, 0.2) .. controls(-1, 1.5) and (0, 1.5) .. (1, 0.2);
\draw[line width=0.88mm, cyan] (-1, 0.2) .. controls(-0.5, 0.75) .. (0, 0.2);
\draw[line width=0.88mm, cyan] (2, 0.2) .. controls(2.5, 0.75) .. (3, 0.2);

\draw[line width=0.88mm, cyan] (-2, -0.2) .. controls(-1, -2.25) and (2, -2.25) .. (3, -0.2);
\draw[line width=0.88mm, cyan] (-1, -0.2) .. controls(0, -1.5) and (1, -1.5) .. (2, -0.2);
\draw[line width=0.88mm, cyan] (0, -0.2) .. controls(0.5, -0.75) .. (1, -0.2);

\node[draw=none] at (-3,-0.5) {$D_{1}$};
\node[draw=none] at (-3,0.5) {$D_{2}$};
\node[draw=none] at (1.2,-0.5) {$\ell_{1}$};
\node[draw=none] at (1.2,0.5) {$\ell_{2}$};

\begin{scope}[xshift=7cm]

\draw[line width=0.5mm, black] (-2.5, 0) -- (3.5, 0);
\draw[line width=0.8mm, red, dash pattern={on 7pt off 7pt}] (-2.5, 0) -- (3.5, 0);
\fill[opacity=0.2, yellow, even odd rule] (-2.5, 0) -- (3.5, 0) -- (3.5, -2) -- (-2.5, -2) -- cycle;

\draw[line width=0.8mm, purple] (-2, 0.2) -- (-2, -0.2);
\draw[line width=0.8mm, purple] (-1, 0.2) -- (-1, -0.2);
\draw[line width=0.8mm, purple] (0, 0.2) -- (0, -0.2);
\draw[line width=0.8mm, purple] (1, 0.2) -- (1, -0.2);
\draw[line width=0.8mm, purple] (2, 0.2) -- (2, -0.2);
\draw[line width=0.8mm, purple] (3, 0.2) -- (3, -0.2);

\draw[line width=0.88mm, cyan] (-2, 0.2) .. controls(-1, 1.5) and (0, 1.5) .. (1, 0.2);
\draw[line width=0.88mm, cyan] (-1, 0.2) .. controls(-0.5, 0.75) .. (0, 0.2);
\draw[line width=0.88mm, cyan] (2, 0.2) .. controls(2.5, 0.75) .. (3, 0.2);

\draw[line width=0.88mm, cyan] (-2, -0.2) .. controls(-1.5, -0.75) .. (-1, -0.2);
\draw[line width=0.88mm, cyan] (0, -0.2) .. controls(1, -1.5) and (2, -1.5) .. (3, -0.2);
\draw[line width=0.88mm, cyan] (1, -0.2) .. controls(1.5, -0.75) .. (2, -0.2);

\node[draw=none] at (-3,-0.5) {$D_{1}$};
\node[draw=none] at (-3,0.5) {$D_{2}$};
\node[draw=none] at (0.75,-0.5) {$\ell'_{1}$};
\node[draw=none] at (1.2,0.5) {$\ell_{2}$};

\end{scope}

\begin{scope}[xshift=14cm]

\draw[line width=0.5mm, black] (-2.5, 0) -- (3.5, 0);
\draw[line width=0.8mm, red, dash pattern={on 7pt off 7pt}] (-2.5, 0) -- (3.5, 0);
\fill[opacity=0.2, yellow, even odd rule] (-2.5, 0) -- (3.5, 0) -- (3.5, -2) -- (-2.5, -2) -- cycle;

\draw[line width=0.8mm, purple] (-2, 0.2) -- (-2, -0.2);
\draw[line width=0.8mm, purple] (-1, 0.2) -- (-1, -0.2);
\draw[line width=0.8mm, purple] (0, 0.2) -- (0, -0.2);
\draw[line width=0.8mm, purple] (1, 0.2) -- (1, -0.2);
\draw[line width=0.8mm, purple] (2, 0.2) -- (2, -0.2);
\draw[line width=0.8mm, purple] (3, 0.2) -- (3, -0.2);

\draw[line width=0.88mm, cyan] (-2, 0.2) .. controls(-1, 1.5) and (0, 1.5) .. (1, 0.2);
\draw[line width=0.88mm, cyan] (-1, 0.2) .. controls(-0.5, 0.75) .. (0, 0.2);
\draw[line width=0.88mm, cyan] (2, 0.2) .. controls(2.5, 0.75) .. (3, 0.2);

\draw[line width=0.88mm, cyan] (-2, -0.2) .. controls(-1, -2.25) and (2, -2.25) .. (3, -0.2);
\draw[line width=0.88mm, cyan] (-1, -0.2) .. controls(-0.5, -0.75) .. (0, -0.2);
\draw[line width=0.88mm, cyan] (1, -0.2) .. controls(1.5, -0.75) .. (2, -0.2);

\node[draw=none] at (-3,-0.5) {$D_{1}$};
\node[draw=none] at (-3,0.5) {$D_{2}$};
\node[draw=none] at (0.75,-0.5) {$\ell''_{1}$};
\node[draw=none] at (1.2,0.5) {$\ell_{2}$};

\end{scope}
\end{tikzpicture}}
\caption{Given $\ell_{2} \in \mathcal{L}_{A}$, we see two examples $\ell_{1}, \ell_{1}' \in \mathcal{L}_{A}$ that are compatible with $\ell_{2}$, 
but one example $\ell_{1}''$ 
is not.}
\label{fig:compatibilityExample}
\end{figure}
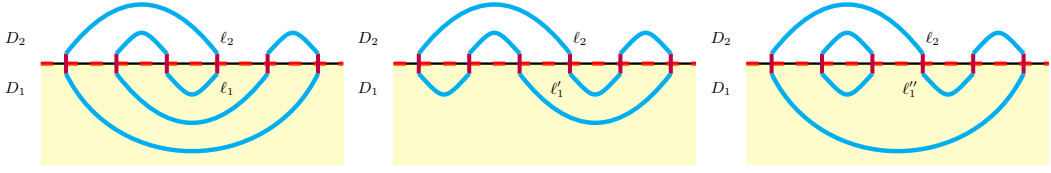

We note that if the labellings $\ell_{1}, \ell_{2}$ are defined by some $W \in  \mathcal{W}_{G, S}$ then they are compatible.
The test is trying to reflect how  $W \in \mathcal{C}_{S}(\ell_{1}, \ell_{2})$ crosses  $\tilde{E}_{A}$ and forms a
cycle (see \cref{fig:compatibilityExample}). We use a pair of Motzkin paths
to get around the difficulty described in
\cref{fig:ell1notUnique}.
While \texttt{CompatibilityCheck}
is not sufficient to conclude that $\left\lvert \mathcal{C}_{S}(\ell_{1}, \ell_{2}) \right\rvert > 0$, it is a necessary condition. 
By dealing only with
 pairs of Motzkin paths of length $k$, which have order $2^{O(k)}$,
we avoid the complexity of order $k!$.

\renewcommand{\thealgorithm}{}
\begin{algorithm}
	\caption{\texttt{CompatibilityCheck}\Big($G = (V, E, F), A = (V_{A}, E_{A}), \ell_{1} \in \mathcal{L}_{A}, \ell_{2} \in \mathcal{L}_{A}$\Big)} 
	\begin{algorithmic}[1]
		\If {$\{e \in E_{A}: \ell_{1}(e) = 0\} \neq \{e \in E_{A}: \ell_{2}(e) = 0\}$}
			\State\Return \textbf{false}
		\EndIf
		\If {$\ell_{1}(e) = 0\ \forall\ e \in E_{A}$}
			\State\Return \textbf{true}
		\EndIf
		\State Pick $e \in E_{A}$ such that $\ell_{1}(e) \neq 0$
		\State $M \leftarrow \{e\}$
		\While {$\ell_{1}(e) \notin M$}
			\State $e' \leftarrow \ell_{1}(e)$
			\State $e \leftarrow \ell_{2}(e')$
			\State $M \leftarrow M \cup \{e', e \}$
		\EndWhile
		\If {$M \neq \{e \in E_{A}: \ell_{1}(e) \neq 0 \}$}
			\State\Return \textbf{false}
		\EndIf
		\State\Return \textbf{true}
	\end{algorithmic} 
\end{algorithm}

\begin{restatable}{lemma}{incompatibilityBad}\label{lemma:incompatibilityBad}
	If $\mathcal{C}_{S}(\ell_{1}, \ell_{2}) \not = \emptyset$,
    then $\ell_{1}$ and $\ell_{2}$ are  compatibile as defined by 
    \emph{\texttt{CompatibilityCheck}}.
\end{restatable}

The proof can be found in the \cref{sec:appendix_b}.

Combining 	\cref{lemma:incompatibilityBad} and \cref{lemma:compatibilityPartition}, we have
$$\left\lvert \mathcal{W}_{G, S} \right\rvert = \sum_{\text{compatible } \ell_{1}, \ell_{2} \in \mathcal{L}_{A}}\left\lvert \mathcal{C}_{S}(\ell_{1}, \ell_{2}) \right\rvert.$$
Now consider 
two
compatible labellings $\ell_{1}$ and $\ell_{2}$.
We deal with the special case, where $\ell_{1} = \ell_{2} = 0$ in \cref{lemma:zeroSpecialCase} in \cref{sec:appendix_b}.
We will instead consider here, the more interesting case where $(\ell_{1}, \ell_{2}) \neq (0, 0)$. Before we see how to compute $\left\lvert \mathcal{C}_{S}(\ell_{1}, \ell_{2}) \right\rvert$, let us first tackle the simpler problem of computing $\left\lvert \mathcal{C}_{\emptyset}(\ell_{1}, \ell_{2}) \right\rvert$.
We know that any cycle $W \in \mathcal{C}_{\emptyset}(\ell_{1}, \ell_{2})$   satisfies the following condition:
$$W \cap \tilde{E}_{A} = \{\tilde{e} \in \tilde{E}_{A}: \ell_{1}(e) \neq 0 \} = \{\tilde{e} \in \tilde{E}_{A}: \ell_{2}(e) \neq 0 \}.$$
So, let us consider $\{\tilde{e} \in \tilde{E}_{A}: \ell_{1}(e) \neq 0 \}$,
and assume it is nonempty.
We can now create an ordered set $T= T(\ell_{1}, \ell_{2})$ of arcs on $H$, representing the order in which any $W \in \mathcal{C}_{\emptyset}(\ell_{1}, \ell_{2})$ would go across these edges.
We construct $T$ as follows: First, we arbitrarily pick some $e_{1} \in E_{A}$ such that $\ell_{1}(e_{1}) \neq 0$. Let $\tilde{e}_{1} = \{u_{1}, v_{1}\}$, such that
$u_{1} \in \tilde{F}_{D_{2}}, v_{1} \in \tilde{F}_{D_{1}}$.
Then, we add the arc $(u_{1}\rightarrow v_{1})$ to $T$.
Next, let $e_{2} = \ell_{1}(e_1)$ and let  $\tilde{e}_{2} = \{u_{2}, v_{2}\}$,  such that $u_{2} \in \tilde{F}_{D_{1}}, v_{2} \in \tilde{F}_{D_{2}}$. Then we add the arc $(u_{2} \rightarrow v_{2})$ as the next element to $T$.
We then move on to $e_{3} = \ell_{2}(e_{2})$. If $e_3 = e_1$ then $T$ is completed; otherwise,
let  $\tilde{e}_{3} = \{u_{3}, v_{3}\}$, such that  $u_{3} \in \tilde{F}_{D_{2}}, v_{3} \in \tilde{F}_{D_{1}}$. Then we  add $(u_{3} \rightarrow v_{3})$ to $T$.
We repeat this process just as in the algorithm \texttt{CompatibilityCheck}, until an edge is visited again, at which point we stop and $T$ is completed. 

As a cyclically ordered set of arcs
under the equivalence of
$\sim$, the definition of $T$ does not depend on the choice of $e_1 \in E_A$.
 It is important to note that the ordered set $T$ is a function of the labellings $\ell_{1}$ and $\ell_{2}$;
 a crucial point is that a compatible pair $(\ell_1, \ell_2)$ together determines not only the subset of $E_A$ where  $\ell_1$ and $\ell_2$ are nonzero, but also the order they are to be traversed by any $W \in  \mathcal{C}_{\emptyset}(\ell_{1}, \ell_{2})$.
This $T$ represents the order and orientation in which these edges on $\tilde{E}_{A}$ will be visited in a cyclic traversal by any $W \in \mathcal{C}_{\emptyset}(\ell_{1}, \ell_{2})$.
When there is a need to indicate this dependence, we will use $T(\ell_{1}, \ell_{2})$ to denote this ordered set of arcs (see \cref{remark:TWellDefinedRemark} in the \cref{sec:appendix_b}).

\begin{restatable}{lemma}{PhiTDecomposition}\label{lemma: PhiTDecomposition}
    Given a compatible $(\ell_{1}, \ell_{2}) \neq (0, 0)$,
    $$ \left\lvert \mathcal{C}_{\emptyset}(\ell_{1}, \ell_{2}) \right\rvert = \left\lvert \mathcal{C}_{D_{1}, T}(\ell_{1}) \right\rvert \times \left\lvert \mathcal{C}_{D_{2}, T}(\ell_{2}) \right\rvert.$$
\end{restatable}

The proof can be found in \cref{sec:appendix_b}.

\begin{remark}\label{remark:pairMotzkinPaths}
Going back to the statement at the end of \cref{sec:intro} that
two cycles 
homeomorphic to $S^1$ intersecting at $k$ points transversely
can have only $2^{O(k)}$
many topologically distinct ways, we have shown how to
handle this problem using
two compatible  Motzkin paths. 
We showed that to compute $\lvert\mathcal{C}_{D_{1}}(\ell_{1})\rvert$,
we do not need to find the union of $\mathcal{C}_{D_{1}, T'}(\ell_{1})$ over all ordered sets $T'$ such that $\varepsilon(T') = \{\tilde{e} \in \tilde{E}_{A}: \ell_{1}(e) \neq 0 \}$.
If $|\varepsilon(T')| = k$
then there are up to $k!$ many ordered sets $T'$ to consider.
Instead,  we can restrict our attention to just $\{ T(\ell_{1}, \ell_{2}):  \ell_{2} \in \mathcal{L}_{A} ~\mbox{compatible with $\ell_1$}\}$,
which has size only $2^{O(k)}$.
On the other hand, to achieve this saving, the definition of
these sets and the underlying concept
become more delicate and we have had to use
some more involved notations.
\end{remark}



 We have successfully reduced the problem of computing $\left\lvert \mathcal{C}_{\emptyset}(\ell_{1}, \ell_{2}) \right\rvert$
to the problem of computing $\left\lvert \mathcal{C}_{D, T}(\ell) \right\rvert$, as promised at the beginning of this section.
Now, we need to extend this to the problem of computing $\left\lvert \mathcal{C}_{S}(\ell_{1}, \ell_{2}) \right\rvert$, for a general $S \neq \emptyset$.

 Given an ordered set $S$ of arcs in $H$, and the ordered set $T$ described above, we now need to somehow combine these two ordered sets into one.
Informally, we say that $I$ is an interleaving of $S$ and $T$, if $I$ contains the arcs from $S$ and $T$ in the same cyclic order upto reversals.
The main idea here is that if a cycle  $W$ conforms to both $S$ and $T$, it will have to conform to some interleaving $I$ of $S$ and $T$ as well.
In order to formally define it, we will need the equivalence relation $\sim$ from earlier.
We say an ordered set $I$ of arcs in $H$ is an \textit{interleaving} of $S$ and $T$ if the following conditions hold:
\begin{enumerate}
	\item $\varepsilon(I) = \varepsilon(S) \cup \varepsilon(T)$.
	\item $I \cap \varepsilon(S) \sim S$, and $I \cap \varepsilon(T) \sim T$.
\end{enumerate}

 Here, $I \cap \varepsilon(S)$ denotes  the ordered set of arcs on the underlying (unordered) edge set  $\varepsilon(S)$ with
the induced orientation and order from $I$, and $I \cap \varepsilon(T)$ is similarly defined.
Note that in item 2 the equivalence relation $\sim$ allows taking reversals independently for $S$ and $T$.
We then use $\mathcal{I}_{S, T}$ to denote the set of all the equivalence classes of interleavings of $S$ and $T$ with each equivalence class represented by a single interleaving.

Now that we have defined interleavings, we can show that 
$$\left\lvert \mathcal{C}_{S}(\ell_{1}, \ell_{2}) \right\rvert = \sum_{I \in \mathcal{I}_{S, T}}\left\lvert \mathcal{C}_{I}(\ell_{1}, \ell_{2}) \right\rvert.$$

The formal proof of this claim can be found as \cref{lemma:interleavingsPartition} in \cref{sec:appendix_b}.


We showed in \cref{lemma:incompatibilityBad} that incompatible labelling pairs $\ell_{1}, \ell_{2} \in \mathcal{L}_{A}$ can be omitted, 
since 	in that case $\mathcal{C}_{S}(\ell_{1}, \ell_{2}) = \emptyset$.
Going one step further,
even if $\ell_{1}, \ell_{2} \in \mathcal{L}_{A}$ are compatible we may still have 
$\mathcal{C}_{I}(\ell_{1}, \ell_{2}) = \emptyset$ if $I$ is not an appropriate interleaving of $S$ and $T = T(\ell_{1}, \ell_{2})$.
Now, we will define what a \emph{valid} interleaving is, and then show that $\mathcal{C}_{I}(\ell_{1}, \ell_{2}) = \emptyset$ for the invalid ones.

\begin{definition}
    Let $\ell_{1}, \ell_{2} \in \mathcal{L}_{A}$  be compatible labellings, and $T = T(\ell_{1}, \ell_{2})$.
	We say $I \in \mathcal{I}_{S, T}$  is a \emph{valid interleaving} of $S$ and $T$ if the following properties hold:
	\begin{enumerate}
		\item 
			There exists some ordered subset $T'$ of $T$ such that
			$$S \cap \tilde{E}_{A} \sim T'.$$
		\item
			If $\Big((u^{T}_{1} \rightarrow v^{T}_{1}), (u^{S}_{1} \rightarrow v^{S}_{1}), \dots, (u^{S}_{k} \rightarrow v^{S}_{k}), (u^{T}_{2} \rightarrow v^{T}_{2})\Big)$ are consecutive arcs in $I' \sim I$ such that $\Big\{\{u^{T}_{1}, v^{T}_{1}\}, \{u^{T}_{2}, v^{T}_{2}\}\Big\} \subseteq I' \cap \varepsilon(T)$, and $\Big\{\{u^{S}_{1}, v^{S}_{1}\}, \dots, \{u^{S}_{k}, v^{S}_{k}\}\Big\} \subseteq I' \cap \big(\varepsilon(S) \setminus \varepsilon(T)\big)$, then
			$$u^{T}_{1}, v^{T}_{2} \in \tilde{F}_{D}, \text{ and } v^{T}_{1}, u^{S}_{1}, v^{S}_{1}, \dots, u^{S}_{k}, v^{S}_{k}, u^{T}_{1} \notin \tilde{F}_{D} \text{ for some } D \in \{D_{1}, D_{2}\}.$$
	\end{enumerate}			
\end{definition}

 Informally, condition (1) says that the subset of $S$ that are on   $\tilde{E}_{A}$, 
must  appear in $T$ in the right order. 
Condition (2) constrains that walks in $\mathcal{C}_{I}(\ell_{1}, \ell_{2})$ can only jump between discs $D_{1}$ and $D_{2}$ using the arcs from $T$. Formally, we have the following lemma, whose proof can be found in \cref{sec:appendix_b}.

\begin{restatable}{lemma}{invalidBad}\label{lemma:invalidBad}
	Given compatible labellings $\ell_{1}, \ell_{2} \in \mathcal{L}_{A}$ such that $(\ell_{1}, \ell_{2}) \neq (0, 0)$, $$\left\lvert \mathcal{C}_{S}(\ell_{1}, \ell_{2}) \right\rvert = \sum_{\text{valid } I \in \mathcal{I}_{S, T}}\left\lvert \mathcal{C}_{I}(\ell_{1}, \ell_{2}) \right\rvert,$$
    where $T = T(\ell_{1}, \ell_{2})$.
\end{restatable}

Finally, we have the main lemma of this section, which proves that the interleavings allows us to circumvent the issue we observed in \cref{fig:ell1notUnique}.

\begin{restatable}{lemma}{validIDecomposition}\label{lemma:validIDecomposition}
    Given compatible $\ell_{1}, \ell_{2} \in \mathcal{L}_{A}$ such that $(\ell_{1}, \ell_{2}) \neq (0, 0)$, and $T = T(\ell_{1}, \ell_{2})$,

    $$\left\lvert \mathcal{C}_{S}(\ell_{1}, \ell_{2}) \right\rvert = \sum_{\text{valid } I \in \mathcal{I}_{S, T}}\left\lvert \mathcal{C}_{D_{1}, I \cap \tilde{E}_{D_{1}}}(\ell_{1}) \right\rvert \times \left\lvert \mathcal{C}_{D_{2}, I \cap \tilde{E}_{D_{2}}}(\ell_{2}) \right\rvert.$$
\end{restatable}

The proof can be found in \cref{sec:appendix_b}.

 We have now succeeded in reducing the problem of computing $\left\lvert \mathcal{W}_{G, S} \right\rvert$ 
to the problem of computing $\left\lvert \mathcal{C}_{D, I \cap \tilde{E}_{D}}(\ell) \right\rvert$.
\subsection{Computing $\left\lvert \mathcal{C}_{D, S}(\ell) \right\rvert$}\label{sec:stitchingSection}

In this section, given compatible $(\ell_{1}, \ell_{2})$ (including the case $(\ell_{1}, \ell_{2}) = (0, 0)$), and valid $I \in \mathcal{I}_{S, T}$,
we will construct two graphs $G_{1}, G_{2}$, and two ordered sets $S_{1}, S_{2}$ of arcs on the duals $H_{1}, H_{2}$, and use $\left\lvert \mathcal{W}_{G_{1}, S_{1}} \right\rvert$ and $\left\lvert \mathcal{W}_{G_{2}, S_{2}} \right\rvert$ to compute 
$\left\lvert \mathcal{C}_{D_{1}, I \cap \tilde{E}_{D_{1}}}(\ell_{1}) \right\rvert$, and $\left\lvert \mathcal{C}_{D_{2}, I \cap \tilde{E}_{D_{2}}}(\ell_{2}) \right\rvert$ respectively.

The reason why we want this reduction is  that $G_{1}$ and $G_{2}$ will have substantially fewer vertices than $G$, 
and $S_{1}$ and $S_{2}$ will not be much larger than $S$. 
Therefore, this reduction will allow us to reduce the problem of computing the number of cycles on a graph $G$ to that of computing the number of cycles  on smaller graphs.

We will briefly describe the construction of $G_{1}$ here. 
Intuitively, we will use the labelling $\ell_{2}$ to \emph{stitch up} the disc $D_{2}$.
The construction of $G_{2}$ will be analogous, using $\ell_1$ to stitch up the disc $D_{1}$. 

\begin{figure}
 \centering
 \begin{subfigure}[b]{0.3\textwidth}
     \centering
     \scalebox{0.65}{\begin{tikzpicture}[line join=miter, draw opacity=1]

\draw[line width=0.5mm, black] (-1, 1) -- (-1, -1.5);
\draw[line width=0.5mm, black] (0, 0) -- (0, -1.5);
\draw[line width=0.5mm, black] (1, 1) -- (1, -1.5);

\draw[line width=0.5mm, black] (-1.5, 1) -- (1.5, 1);
\draw[line width=0.5mm, black] (-1.5, 0) -- (1.5, 0);
\draw[line width=0.5mm, black] (-1.5, -1) -- (1.5, -1);

\draw[line width=0.5mm, black] (-1, 1) -- (0, 0);
\draw[line width=0.5mm, black] (-1.5, 0.5) -- (0, -1);
\draw[line width=0.5mm, black] (-1.5, -0.5) -- (-0.5, -1.5);

\draw[line width=0.5mm, black] (1, 1) -- (0, 0);
\draw[line width=0.5mm, black] (1.5, 0.5) -- (0, -1);
\draw[line width=0.5mm, black] (1.5, -0.5) -- (0.5, -1.5);

\fill[opacity=0.2, yellow, even odd rule] (-1.5,1) -- (1.5,1) -- (1.5, -1.5) -- (-1.5, -1.5) -- cycle;

\draw[line width=0.8mm, red, dash pattern={on 7pt off 7pt}] (-1.5, 1) -- (1.5, 1);

\node[draw=none] at (0,1.2) {$e$};
\node[draw=none] at (-0.25,0.6) {$e'$};
\node[draw=none] at (0.30,0.6) {$e''$};

\node[draw=none] at (-1,1.2) {$a$};
\node[draw=none] at (1,1.2) {$b$};
\node[draw=none] at (-0.2,-0.2) {$c$};

\begin{scope}[xshift=3.5cm]

\draw[line width=0.5mm, black] (-1, 0) -- (-1, -1.5);
\draw[line width=0.5mm, black] (0, 1) -- (0, -1.5);
\draw[line width=0.5mm, black] (1, 0) -- (1, -1.5);

\draw[line width=0.5mm, black] (-1.5, 1) -- (1.5, 1);
\draw[line width=0.5mm, black] (-1.5, 0) -- (1.5, 0);
\draw[line width=0.5mm, black] (-1.5, -1) -- (1.5, -1);

\draw[line width=0.5mm, black] (-1, 0) -- (0, 1);
\draw[line width=0.5mm, black] (-1.5, 0.5) -- (0, -1);
\draw[line width=0.5mm, black] (-1.5, -0.5) -- (-0.5, -1.5);

\draw[line width=0.5mm, black] (1, 0) -- (0, 1);
\draw[line width=0.5mm, black] (1.5, 0.5) -- (0, -1);
\draw[line width=0.5mm, black] (1.5, -0.5) -- (0.5, -1.5);

\fill[opacity=0.2, yellow, even odd rule] (-1.5,1) -- (1.5,1) -- (1.5,-1.5) -- (-1.5,-1.5) -- cycle;

\draw[line width=0.8mm, red, dash pattern={on 7pt off 7pt}] (-1.5, 1) -- (1.5, 1);

\node[draw=none] at (0,1.2) {$a = b$};
\node[draw=none] at (-0.2,0.3) {$e'$};
\node[draw=none] at (0.25,0.3) {$e''$};

\node[draw=none] at (-0.2,-0.2) {$c$};

\end{scope}

\end{tikzpicture}}
     \caption{single-edge contraction}
     \label{fig:singleEdgeExample}
 \end{subfigure}
 \hfill
 \begin{subfigure}[b]{0.3\textwidth}
     \centering
     \scalebox{0.65}{\begin{tikzpicture}[line join=miter, draw opacity=1]

\draw[line width=0.5mm, black] (-1, 1) -- (-1, -1.5);
\draw[line width=0.5mm, black] (0, 0) -- (0, -1.5);
\draw[line width=0.5mm, black] (1, 1) -- (1, -1.5);

\draw[line width=0.5mm, black] (-1.5, 1) -- (-1, 1);
\draw[line width=0.5mm, black] (1, 1) -- (1.5, 1);
\draw[line width=0.5mm, black] (-1.5, 0) -- (1.5, 0);
\draw[line width=0.5mm, black] (-1.5, -1) -- (1.5, -1);

\draw[line width=0.5mm, black] (-1, 1) -- (0, 0);
\draw[line width=0.5mm, black] (-1.5, 0.5) -- (0, -1);
\draw[line width=0.5mm, black] (-1.5, -0.5) -- (-0.5, -1.5);

\draw[line width=0.5mm, black] (1, 1) -- (0, 0);
\draw[line width=0.5mm, black] (1.5, 0.5) -- (0, -1);
\draw[line width=0.5mm, black] (1.5, -0.5) -- (0.5, -1.5);

\fill[opacity=0.2, yellow, even odd rule] (-1.5,1) -- (-1,1) -- (0,0) -- (1,1) -- (1.5,1) -- (1.5, -1.5) -- (-1.5, -1.5) -- cycle;

\draw[line width=0.8mm, red, dash pattern={on 7pt off 7pt}] (-1.5, 1) -- (-1, 1);
\draw[line width=0.8mm, red, dash pattern={on 7pt off 7pt}] (-1, 1) -- (0, 0);
\draw[line width=0.8mm, red, dash pattern={on 7pt off 7pt}] (0, 0) -- (1, 1);
\draw[line width=0.8mm, red, dash pattern={on 7pt off 7pt}] (1, 1) -- (1.5, 1);

\node[draw=none] at (-0.25,0.55) {$e$};
\node[draw=none] at (0.30,0.6) {$e'$};

\node[draw=none] at (-1,1.2) {$a$};
\node[draw=none] at (1,1.2) {$c$};
\node[draw=none] at (-0.2,-0.2) {$b$};

\begin{scope}[xshift=3.5cm]

\draw[line width=0.5mm, black] (-1, 0) -- (-1, -1.5);
\draw[line width=0.5mm, black] (0, 1) -- (0, -1.5);
\draw[line width=0.5mm, black] (1, 0) -- (1, -1.5);

\draw[line width=0.5mm, black] (-1.5, 1) -- (1.5, 1);
\draw[line width=0.5mm, black] (-1.5, 0) -- (1.5, 0);
\draw[line width=0.5mm, black] (-1.5, -1) -- (1.5, -1);

\draw[line width=0.5mm, black] (-1, 0) -- (0, 1);
\draw[line width=0.5mm, black] (-1.5, 0.5) -- (0, -1);
\draw[line width=0.5mm, black] (-1.5, -0.5) -- (-0.5, -1.5);

\draw[line width=0.5mm, black] (1, 0) -- (0, 1);
\draw[line width=0.5mm, black] (1.5, 0.5) -- (0, -1);
\draw[line width=0.5mm, black] (1.5, -0.5) -- (0.5, -1.5);

\fill[opacity=0.2, yellow, even odd rule] (-1.5,1) -- (1.5,1) -- (1.5,-1.5) -- (-1.5,-1.5) -- cycle;

\draw[line width=0.8mm, red, dash pattern={on 7pt off 7pt}] (-1.5, 1) -- (1.5, 1);

\node[draw=none] at (-0.2,0.25) {$e$};
\node[draw=none] at (0.25,0.3) {$e'$};

\node[draw=none] at (0,1.2) {$a = c$};
\node[draw=none] at (-0.2,-0.2) {$b$};

\end{scope}

\end{tikzpicture}}
     \caption{edge-pair contraction}
     \label{fig:edgePairExample}
 \end{subfigure}
 \hfill
 \begin{subfigure}[b]{0.3\textwidth}
     \centering
     \scalebox{0.65}{\begin{tikzpicture}[line join=miter, draw opacity=1]

\draw[line width=0.5mm, black] (-1, 1) -- (-1, -1.5);
\draw[line width=0.5mm, black] (0, -1) -- (0, -1.5);
\draw[line width=0.5mm, black] (1, 1) -- (1, -1.5);

\draw[line width=0.5mm, black] (-1.5, 1) -- (1.5, 1);
\draw[line width=0.5mm, black] (-1.5, 0) -- (-1, 0);
\draw[line width=0.5mm, black] (1, 0) -- (1.5, 0);
\draw[line width=0.5mm, black] (-1.5, -1) -- (1.5, -1);

\draw[line width=0.5mm, black] (-1.5, 0.5) -- (0, -1);
\draw[line width=0.5mm, black] (-1.5, -0.5) -- (-0.5, -1.5);

\draw[line width=0.5mm, black] (1.5, 0.5) -- (0, -1);
\draw[line width=0.5mm, black] (1.5, -0.5) -- (0.5, -1.5);

\draw[line width=0.5mm, black] (-1, 1) -- (0, -1);
\draw[line width=0.5mm, black] (1, 1) -- (0, -1);
\draw[line width=0.5mm, black] (-1, 1) -- (0, 0.25);
\draw[line width=0.5mm, black] (1, 1) -- (0, 0.25);
\draw[line width=0.5mm, black] (-1, 1) .. controls(0, -0.5).. (1, 1);

\fill[opacity=0.2, yellow, even odd rule] (-1.5,1) -- (1.5,1) -- (1.5, -1.5) -- (-1.5, -1.5) -- cycle;

\draw[line width=0.8mm, red, dash pattern={on 7pt off 7pt}] (-1.5, 1) -- (1.5, 1);

\node[draw=none] at (0,1.2) {$e$};
\node[draw=none] at (-0.25,0.75) {$e''$};
\node[draw=none] at (0.30,0.75) {$e'''$};
\node[draw=none] at (0,-0.3) {$e'$};

\node[draw=none] at (0,0.1) {$c$};
\node[draw=none] at (-1,1.2) {$a$};
\node[draw=none] at (1,1.2) {$b$};

\begin{scope}[xshift=3.5cm]

\draw[line width=0.5mm, black] (-1, 1) -- (-1, -1.5);
\draw[line width=0.5mm, black] (0, -1) -- (0, -1.5);
\draw[line width=0.5mm, black] (1, 1) -- (1, -1.5);

\draw[line width=0.5mm, black] (-1.5, 1) -- (1.5, 1);
\draw[line width=0.5mm, black] (-1.5, 0) -- (-1, 0);
\draw[line width=0.5mm, black] (1, 0) -- (1.5, 0);
\draw[line width=0.5mm, black] (-1.5, -1) -- (1.5, -1);

\draw[line width=0.5mm, black] (-1.5, 0.5) -- (0, -1);
\draw[line width=0.5mm, black] (-1.5, -0.5) -- (-0.5, -1.5);

\draw[line width=0.5mm, black] (1.5, 0.5) -- (0, -1);
\draw[line width=0.5mm, black] (1.5, -0.5) -- (0.5, -1.5);

\draw[line width=0.5mm, black] (-1, 1) -- (0, -1);
\draw[line width=0.5mm, black] (1, 1) -- (0, -1);

\fill[opacity=0.2, yellow, even odd rule] (-1.5,1) -- (1.5,1) -- (1.5,-1.5) -- (-1.5,-1.5) -- cycle;

\draw[line width=0.8mm, red, dash pattern={on 7pt off 7pt}] (-1.5, 1) -- (1.5, 1);

\node[draw=none] at (0,1.3) {$e = e'$};

\node[draw=none] at (-1,1.2) {$a$};
\node[draw=none] at (1,1.2) {$b$};

\end{scope}

\end{tikzpicture}}
     \caption{parallel-edge contraction}
     \label{fig:parallelEdgeExample}
 \end{subfigure}
    \caption{Examples of the three contractions that will be used to stitch up the discs and construct $G_{1}$ and $G_{2}$.}
    \label{fig:threeContractionsExample}
\end{figure}
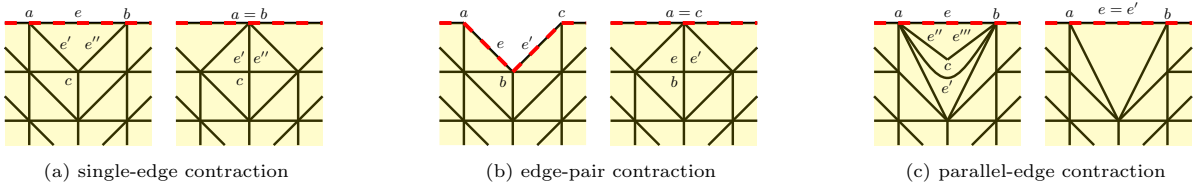

We start with the graph $G_{D_{1}}$, which is just the graph $G$ restricted to the disc $D_{1}$. 
In other words, it is the plane graph containing the vertices $V_{D_{1}}$, the edges $E_{D_{1}}$, as well as the faces $F_{D_{1}} \cup \{f^{*}\}$, where $f^{*}$ is the face that is bounded by the edges $E_{A}$, which can be thought of as the disc $D_2$ without internal edges in our spherical embedding of $G$. 
We will then repeatedly use the three contractions seen in \cref{fig:threeContractionsExample} to shrink the cycle $E_{A}$ away, until the face $f^{*}$ disappears (see \cref{fig:contractionExample} for an example of this stitching).

We use single-edge contractions (see \cref{fig:singleEdgeExample}) to shrink away edges of $E_{A}$ for which $\ell_{2}(e) = 0$, and edge-pair contractions (see \cref{fig:edgePairExample}) to merge together edges $e, e'$ that are such that $\ell_{2}(e) = e'$, and $\ell_{2}(e') = e$. We also add the merged edge to the ordered set of arcs $S_{1}$, so that we only count cycle that cross this edge in the correct orientation. Finally, parallel-edge contractions (see \cref{fig:parallelEdgeExample}) are used to ensure that the resulting graph $G_{1}$ has no loops.
As $|E_{A}|$ eventually shrinks and disappears, the face $f^{*}$ disappears, and we are left with the graph $G_{1}$.

We will leave the details of the exact construction of $G_{1}, S_{1}$ and $G_{2}, S_{2}$, including the exact definitions of the contractions, and the order in which they are to be applied, to \cref{sec:appendix_c}. The construction of these graphs allows us to prove the following theorem. Its proof can also be found in \cref{sec:appendix_c}.

\begin{restatable}{theorem}{contractionsReduce}\label{theorem:contractionsReduce}
    Given a graph $G = (V, E, F)$ with $|V| = \Omega(1)$ with a cycle $A$ dividing into two discs $D, D'$, a labelling $\ell \in \mathcal{L}_{A}$, and a set $I$ of ordered arcs such that $\varepsilon(I) \subseteq \tilde{E}_{D}$, we can 
    compute $\left\lvert \mathcal{C}_{D, I \cap \tilde{E}_{D}}(\ell) \right\rvert$ in polynomial time, or 
    construct a graph $G_{1} = (V_{1}, E_{1}, F_{1})$, and an ordered set of arcs $S_{1}$, such that $|V_{1}| \leq \frac{3|V|}{4}$, and $\left\lvert \mathcal{C}_{D, I \cap \tilde{E}_{D}}(\ell) \right\rvert$ can be computed
    from $\left\lvert \mathcal{W}_{G_{1}, S_{1}} \right\rvert$ in polynomial time.
\end{restatable}
\subsection{Computing $\left\lvert \mathcal{W}_{G} \right\rvert$}\label{sec:inductionSection}

For the purpose of complexity claims on our algorithm
to compute $\left\lvert \mathcal{W}_{G} \right\rvert$,
we may assume the number of vertices $m$ of $G$ is greater than a
fixed constant; smaller graphs can be handled
by brute-force  in constant time, which will
serve as our base case in an induction. 
As $\mathcal{W}_{G} = \mathcal{W}_{G, \emptyset}$, we have
\begin{multline*}
    \left\lvert \mathcal{W}_{G} \right\rvert = \sum_{\text{compatible } (\ell_{1}, \ell_{2})  \neq (0, 0)}\left(\sum_{\text{valid } I \in \mathcal{I}_{\emptyset, T(\ell_{1}, \ell_{2})}} \left\lvert \mathcal{C}_{D_{1}, I \cap \tilde{E}_{D_{1}}}(\ell_{1}) \right\rvert \times \left\lvert \mathcal{C}_{D_{2}, I \cap \tilde{E}_{D_{2}}}(\ell_{2}) \right\rvert  \right) \\
    + \Big(\left\lvert \mathcal{C}_{D_{1}, \emptyset}(0) \right\rvert + \left\lvert \mathcal{C}_{D_{2}, \emptyset}(0) \right\rvert - 1\Big).
\end{multline*}

As seen in \cref{theorem:contractionsReduce}, the problem of computing 
$\left\lvert \mathcal{C}_{D_{1}, I \cap \tilde{E}_{D_{1}}}(\ell_{1}) \right\rvert$ and $\left\lvert \mathcal{C}_{D_{2}, I \cap \tilde{E}_{D_{2}}}(\ell_{2}) \right\rvert$
can be reduced to the problems of the form of computing
$\left\lvert \mathcal{W}_{G_{1}, S_{1}} \right\rvert$. 
Now note that if $G_{1} = (V_{1}, E_{1}, F_{1})$, then \cref{lemma:contractionPlanarTriangulated} says that
if $|V_{1}| \geq 3$, $G_{1}$ 
is a planar triangulated graph with 
no vertices of degree less than 2, and the dual $H_{1}$ of $G_{1}$ is
a 3-regular planar multi-graph (with no loops).
We have therefore reduced the problem of computing $\left\lvert \mathcal{W}_{G} \right\rvert$ to that of computing $\left\lvert \mathcal{W}_{G_{1}, S_{1}} \right\rvert$ 
for a number of graphs with at most $\frac{3|V|}{4}$ vertices, or we have reduced it to a problem of $O(1)$ size.

So, when $G_{1}$ is large, one can recursively further reduce these smaller problems.
This allows us to construct a tree $\mathcal{T}$ with the problem $\left\lvert \mathcal{W}_{G} \right\rvert$ at the root, with the children of each node being the sub-problems 
that need to be solved to solve the parent problem. 
The leaves of this tree are problems of $O(1)$ size.

Starting at the root level $\left\lvert \mathcal{W}_{G_{0}, S_{0}} \right\rvert$ at level 0	
with $G_0=G$ and $S_0 = \emptyset$,
a problem $\left\lvert \mathcal{W}_{G_{k}, S_{k}} \right\rvert$ on level $k$ of the tree $\mathcal{T}$ has
$|V_{k}| = |V(G_k)| \leq m_k = \left(\nicefrac{3}{4}\right)^{k}|V|$.
So the height of the tree is $L = O(\log m)$.
Moreover, for a  problem on  level $k$ of
$\mathcal{T}$, the constructed ordered
set of arcs
for any compatible pair $(\ell_{1},\ell_{2})$ has size
$$|T_{k}| \leq |E_{A_{k}}| \leq \sqrt{8 m_k}.$$
This is the maximum number of edges that can be added to $S_{k}$.
Thus, 
$$ |S_{k + 1}| \leq |S_{k}| + |T_{k}| \leq |S_{k}| + t_{k}$$
where $t_{k} = \sqrt{8 m_k}$.
Since $|S_{0}| = 0$, this means that
$$|S_{k}| \leq \sum_{0 \leq i < k}t_{i}
< \sum_{i = 0}^{\infty} t_{i} = O(\sqrt{|V|}).$$

We now have the following lemma, whose proof can be found in \cref{sec:appendix_d}.

\begin{restatable}{lemma}{mainInduction}\label{hyp:mainInduction}
    There  is a constant $\alpha$, such that
    any problem $\left\lvert \mathcal{W}_{G_{l}, S_{l}} \right\rvert$ on level $l$ of the tree $\mathcal{T}$ can be computed in time
    $$\alpha^{\sqrt{m_{l}}}  \frac{(t_{0} + \dots + t_{L - 1})!}{(t_{0} + \dots + t_{l - 1})!(t_{l})! \cdots (t_{L - 1})!}.$$
\end{restatable}


Finally, we have our main theorem. Its proof can also be found in \cref{sec:appendix_d}.

\begin{restatable}{theorem}{mainThm}\label{theorem:mainThm}
	There exists a constant $\beta$ such that given any connected planar  3-regular multi-graph (with no loops) $H$ with $n$ vertices, the number of cycles on $H$ can be computed in $\beta^{\sqrt{n}}$ time.
\end{restatable}
\section{Extensions}\label{sec:extensions}


We will now extend the algorithm presented in \cref{sec:algorithm} to count the number of ways a planar triangulated graph can be partitioned into two regions such that the faces are simply connected (in other words, the dual graph is simply connected).

\begin{definition}\label{defn:mustCrossWalks}
	Let $X, Y \subseteq \tilde{E}$ be two disjoint subsets.
	Then,
	$$\mathcal{W}_{G}^{X, Y} = \{W \in \mathcal{W}_{G}: X \subseteq W, Y \cap W = \emptyset \}.$$
\end{definition}

In other words, 
$\mathcal{W}_{G}^{X, Y}$ 
is the set of cycles in $H$ that traverse every edge in $X$, 
and do not traverse any edge in $Y$.
We  define $\mathcal{P}_{D}^{X, Y}$, $\mathcal{W}_{G, S}^{X, Y}$, $\mathcal{P}_{D, S}^{X, Y}$,  $\mathcal{C}_{D, S}^{X, Y}(\ell)$ and
$\mathcal{C}_{S}^{X, Y}(\ell_{1}, \ell_{2})$ similarly.

The stipulation that cycles in $\mathcal{W}_{G}^{X, Y}$ must pass through
all edges in $X$ (and avoid $Y$) may appear to be of the same kind as
the requirement  in $\mathcal{W}_{G, S}$ that they must pass
through $S$. However, there is a  crucial difference, namely cycles 
 $W \in \mathcal{W}_{G, S}$ are 
required to traverse the edges $\varepsilon(S)$ in the directed order as arcs
and in the  order specified by $S$,
whereas there is no such requirement for $W \in \mathcal{W}_{G}^{X, Y}$
with respect to $X$. 
Our  $2^{O(\sqrt{n})}$ time algorithm is for $\lvert \mathcal{W}_{G, \emptyset}\rvert$.
With respect to an arbitrary  $S$, no  subexponential run time claim was made
for  $\lvert \mathcal{W}_{G, S} \rvert$  in its dependence on $S$.
However, this time bound  $2^{O(\sqrt{n})}$
does hold for counting $\lvert \mathcal{W}_{G}^{X, Y} \rvert$, which has the same complexity as counting $\lvert \mathcal{W}_{G, \emptyset}\rvert$,  
and we can extend our algorithm to this setting without much effort. A more detailed discussion of \cref{theorem:mustCrossMainThm} can be found in \cref{sec:appendix_e}.

\begin{theorem}\label{theorem:mustCrossMainThm}
	There exists a constant $\beta$ such that given any 3-regular planar multi-graph (with no loops) $H$ with $n$ vertices, 
	and any $X, Y \subseteq \tilde{E}$, 
	the number of cycles on $H$ that traverse each edge in $X$ and do not traverse any edge in $Y$ 
	can be computed in $\beta^{\sqrt{n}}$ time.
\end{theorem}

We can apply \cref{theorem:mustCrossMainThm} to count the number of ways of partitioning a planar triangulated graph into two simply connected graphs (i.e. the faces are connected).
Given a graph $H = (\tilde{F}, \tilde{E}, \tilde{V})$ with the border $B = (\tilde{F}_{B}, \tilde{E}_{B})$, 
we consider an arbitrary pair of vertices $a, b \in \tilde{F}_{B}$, such that $a \neq b$. 
Since $a, b \in \tilde{F}_{B}$ and $B$ is a cycle, by traversing this cycle in two different directions,
there exist two SAWs within $\tilde{E}_{B}$ between the vertices $a$ and $b$. 
Let $B_{a, b} \subseteq \tilde{E}_{B}$ be one of these SAWs with the endpoints $a, b$.

We can now let $X = B_{a, b}$, and $Y = \{\tilde{e} \in \tilde{E}: \tilde{e} \text{ is incident on } \tilde{F}_{B} \setminus \{a, b\} \} \setminus X$. 
Then, we can use \cref{theorem:mustCrossMainThm} to compute $\lvert \mathcal{W}_{G}^{X, Y} \rvert$. If we repeat this for all possible pairs of $a \neq b$, 
we will be done.

Using standard techniques, this exact counting algorithm can also be used to draw samples from the uniform distribution over all partitions into simply connected regions (or cycles in planar graphs).
Therefore, this algorithm yields a  $2^{O(\sqrt{n})}$ time algorithm to uniformly sample from the space of all partitions of a 3-regular planar graph into two simply connected regions (i.e. such that the faces of the regions are connected, or alternately, the vertices of the two regions in the dual graph are connected).

However, a vast number of boundaries of such partitions are likely to be space filling curves.
It is possible to extend the algorithm above so that it only counts partitions where the boundary is not space filling,
so we can sample from a space of more natural partitions.
A sketch of this extension is to be found in \cref{sec:appendix_e}.

One restriction of the above algorithm is that it is currently limited to 3-regular graphs.
However, the general problem can be reduced to a generalized version of the $3$-regular graph problem with only a little effort.
We can simply replace all vertices with degree $> 3$ with a path of vertices all with degree $3$.
The only issue this causes is that a cycle on this modified graph need not be a cycle on the original graph, 
as the path corresponding to a single vertex may be visited multiple times by a cycle.
However, this can be easily worked around, by making use of the parameters $X, Y$ on the paths that cut across the planar separator $A$.
It can be shown that if a graph has vertices with degrees bounded by some integer $d$, counting the number of cycles on the graph introduces only an additional factor of $d^{O(\sqrt{n})}$.

\bibliography{refs}

\newpage

\appendix 

\section{Appendix A}\label{sec:appendix_a}

\begin{remark}\label{remark:dualRemark}
    By the abstract definition of
    the dual graph, $\tilde{F},  \tilde{E}, \tilde{V}$ are just $F, E, V$ respectively (or in 1-1 correspondence), but we
    use the notation $\tilde{F},  \tilde{E}, \tilde{V}$ to indicate intuitively to which graph an object belongs.  E.g., by $v \in \tilde{F}$ we wish to
    indicate that it is a vertex $v$ in $H$ (but it can also be  considered as a triangular face in $G$), and by	
    $\{u, v\} \in \tilde{E}$ we think of it as an edge between two
    adjacent vertices  $u$ and $v$ in $H$ (and also two adjacent
    triangular faces  in $G$).
\end{remark}

\begin{remark}\label{remark:danglingEdgeRemark}
    In effect, we split $v_{\infty}$ into $|E_A|$ many
    dangling vertices, and think of them
    as lying in $D'$.
    When there is no confusion we will identify
    $\{u, v\} \in \tilde{E}_{A}$ with
    $\{u, v^*\}$ in (the modified) $H_{D}$, except if two adjacent edges of $E_{A}$ in $G$ are on the same triangular face in $D'$,
    in which case we will treat the dangling vertices as distinct and different, even though the two vertices are identified in the original graph.
\end{remark}

\labellingExists*
\begin{proof}			
	By definition, $P = W \cap \tilde{E}_{D}$ for some $W \in \mathcal{W}_{G}$.
	Clearly, $\tilde{e} \in W$.
	Starting from $u_{2} \in \tilde{F}_{D}$ in the direction away from $u_1$ we can follow the cycle $W$, identifying the vertices $u_{3}, \dots, u_{k-1} \in \tilde{F}_{D}$, until we reach a vertex $u_{k} \notin \tilde{F}_{D}$. Such a vertex $u_{k}$ exists, since $W$ is a  cycle and $u_{1} \notin \tilde{F}_{D}$ lies on it.
	
	Note that this choice of $u_{3}, \dots, u_{k}$ does not depend on our particular choice of $W$, but only depends on $P$ and the initial $\tilde{e}$.
	To see this, 
	suppose $P = W_{1} \cap \tilde{E}_{D} = W_{2} \cap \tilde{E}_{D}$. For either $W = W_1$
	or $W_2$, 	every vertex on the intersection
	$P = W \cap \tilde{E}_{D}$  within $D$ has degree 2, 
	and $\{u_{i}, u_{i + 1} \} \in P$ for all $1 \leq i < k$.
	So the same sequence $u_{3}, \dots, u_{k}$ will be selected.
\end{proof}

\begin{remark}\label{remark:labellingRemark}
    This is true even when $k=3$ and $u_3=u_1$ in which case
    $\tilde{e}'$ and $\tilde{e}$ are a pair of parallel edges between $u_1$ and $u_2$.
\end{remark}

\begin{lemma}\label{lemma:symmetryOfEll}
	Let $\ell_{P}$ be the labelling of $E_{A}$ defined by $P$ on $D$ for some $D \in \{D_{1}, D_{2}\}$. For any $e \in E_{A}$, if $\ell_{P}(e) = e'$ for some $e' \in E_{A}$, then $\ell_{P}(e') = e$.
	If $P = W \cap \tilde{E}_{D}$, then	the edges where $\ell_P$ is nonzero coincides with where $W$ crosses $A$; in particular, $\ell_P$ is identically 0 iff $P \cap \tilde{E}_{A} 
	= \emptyset$.
\end{lemma}
\begin{proof}
	If following a walk from  $\tilde{e}$ leads  to  $\tilde{e}'$, then  following the walk back from $\tilde{e}'$ leads to  $\tilde{e}$. Therefore, $\ell_{P}(e) = e'$ implies that $\ell_{P}(e') = e$.
	The last statement is obvious.
\end{proof}

\compatibilityPartitionSimple*
\begin{proof}
	We show that $\mathcal{W}_{G}$ 
	can be expressed as a disjoint union
	$$\mathcal{W}_{G} =  \bigcup_{\ell_{1} \in \mathcal{L}_{A}}\bigcup_{\ell_{2} \in \mathcal{L}_{A}}\mathcal{C}(\ell_{1}, \ell_{2}).$$
	
	By definition,  $\mathcal{C}(\ell_{1}, \ell_{2}) \subseteq 
	\mathcal{W}_{G}$, for every  $\ell_{1}, \ell_{2} \in \mathcal{L}_{A}$.
	Now, let $W \in \mathcal{W}_{G}$. Let $P_{1} = W \cap \tilde{E}_{D_{1}}$ and $P_{2} = W \cap \tilde{E}_{D_{2}}$. Then $\ell_{P_{1}}, \ell_{P_{2}} \in \mathcal{L}_{A}$, and
	$W \in \mathcal{C}(\ell_{P_{1}}, \ell_{P_{2}})$.
	This is a disjoint union, for if
	$W \in \mathcal{C}(\ell_{1}, \ell_{2})$ and also $W \in \mathcal{C}(\ell_{1}', \ell_{2}')$,
	then $\ell_1 = \ell_{W \cap \tilde{E}_{D_{1}}} =  \ell_{1}'$ and
	$\ell_2 = \ell_{W \cap \tilde{E}_{D_{1}}} =  \ell_{2}'$.
\end{proof}

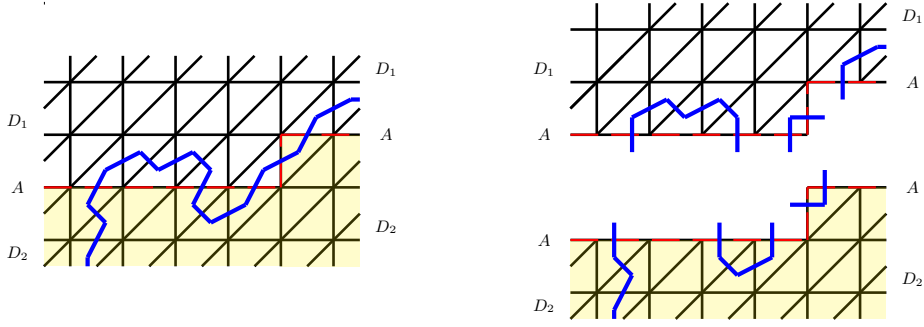
\begin{figure}
	\centering
	\scalebox{0.7}{\begin{tikzpicture}[line join=miter, draw opacity=1] 
\draw[line width=0.5mm, black] (-1, 1.5) -- (-1, -2.5);
\draw[line width=0.5mm, black] (0, 1.5) -- (0, -2.5);
\draw[line width=0.5mm, black] (1, 1.5) -- (1, -2.5);
\draw[line width=0.5mm, black] (2, 1.5) -- (2, -2.5);
\draw[line width=0.5mm, black] (3, 1.5) -- (3, -2.5);
\draw[line width=0.5mm, black] (4, 1.5) -- (4, -2.5);
\draw[line width=0.5mm, black] (-1.5, 1) -- (4.5, 1);
\draw[line width=0.5mm, black] (-1.5, -0) -- (4.5, -0);
\draw[line width=0.5mm, black] (-1.5, -1) -- (4.5, -1);
\draw[line width=0.5mm, black] (-1.5, -2) -- (4.5, -2);
\draw[line width=0.5mm, black] (-1.5, -0.5) -- (0.5, 1.5);
\draw[line width=0.5mm, black] (-1.5, -1.5) -- (1.5, 1.5);
\draw[line width=0.5mm, black] (-1.5, -2.5) -- (2.5, 1.5);
\draw[line width=0.5mm, black] (-0.5, -2.5) -- (3.5, 1.5);
\draw[line width=0.5mm, black] (0.5, -2.5) -- (4.5, 1.5);
\draw[line width=0.5mm, black] (1.5, -2.5) -- (4.5, 0.5);
\draw[line width=0.5mm, black] (2.5, -2.5) -- (4.5, -0.5);
\draw[line width=0.5mm, black] (3.5, -2.5) -- (4.5, -1.5);
\draw[line width=0.5mm, black] (-1.5, 2.5) -- (-1.5, 2.5);
\draw[line width=0.5mm, black] (-1.5, 0.5) -- (-0.5, 1.5);

\fill[opacity=0.2, yellow, even odd rule] (-1.5,-1) -- (3,-1) -- (3,-2.5) -- (-1.5,-2.5) -- cycle;
\fill[opacity=0.2, yellow, even odd rule] (3,0) -- (4.5,0) -- (4.5,-2.5) -- (3,-2.5) -- cycle;

\draw[line width=0.4mm, red, dash pattern={on 15pt off 7pt}] (-1.5, -1) -- (3, -1);
\draw[line width=0.4mm, red, dash pattern={on 15pt off 7pt}] (3, -1) -- (3, 0);
\draw[line width=0.4mm, red, dash pattern={on 15pt off 7pt}] (3, 0) -- (4.5, 0);

\node[draw=none] at (-2,-1) {$A$};
\node[draw=none] at (-2,-2.25) {$D_{2}$};
\node[draw=none] at (-2,0.25) {$D_{1}$};

\node[draw=none] at (5,0) {$A$};
\node[draw=none] at (5,-1.75) {$D_{2}$};
\node[draw=none] at (5,1.25) {$D_{1}$};

\draw[line width=0.8mm, blue] (-0.67, -2.5) -- (-0.67, -2.33);
\draw[line width=0.8mm, blue] (-0.67, -2.33) -- (-0.33, -1.67);
\draw[line width=0.8mm, blue] (-0.33, -1.67) -- (-0.67, -1.33);
\draw[line width=0.8mm, blue] (-0.67, -1.33) -- (-0.33, -0.67);
\draw[line width=0.8mm, blue] (-0.33, -0.67) -- (0.33, -0.33);
\draw[line width=0.8mm, blue] (0.33, -0.33) -- (0.67, -0.67);
\draw[line width=0.8mm, blue] (0.67, -0.67) -- (1.33, -0.33);
\draw[line width=0.8mm, blue] (1.33, -0.33) -- (1.67, -0.67);
\draw[line width=0.8mm, blue] (1.67, -0.67) -- (1.33, -1.33);
\draw[line width=0.8mm, blue] (1.33, -1.33) -- (1.67, -1.67);
\draw[line width=0.8mm, blue] (1.67, -1.67) -- (2.33, -1.33);
\draw[line width=0.8mm, blue] (2.33, -1.33) -- (2.67, -0.67);
\draw[line width=0.8mm, blue] (2.67, -0.67) -- (3.33, -0.33);
\draw[line width=0.8mm, blue] (3.33, -0.33) -- (3.67, 0.33);
\draw[line width=0.8mm, blue] (3.67, 0.33) -- (4.33, 0.67);
\draw[line width=0.8mm, blue] (4.33, 0.67) -- (4.5, 0.67);

\begin{scope}[xshift=10cm]
\draw[line width=0.5mm, black] (-1, 2.5) -- (-1, 0);
\draw[line width=0.5mm, black] (0, 2.5) -- (0, 0);
\draw[line width=0.5mm, black] (1, 2.5) -- (1, 0);
\draw[line width=0.5mm, black] (2, 2.5) -- (2, 0);
\draw[line width=0.5mm, black] (3, 2.5) -- (3, 0);
\draw[line width=0.5mm, black] (4, 2.5) -- (4, 1);
\draw[line width=0.5mm, black] (-1.5, 2) -- (4.5, 2);
\draw[line width=0.5mm, black] (-1.5, 1) -- (4.5, 1);
\draw[line width=0.5mm, black] (-1.5, 0) -- (3, 0);
\draw[line width=0.5mm, black] (-1.5, 0.5) -- (0.5, 2.5);
\draw[line width=0.5mm, black] (-1, 0) -- (1.5, 2.5);
\draw[line width=0.5mm, black] (0, 0) -- (2.5, 2.5);
\draw[line width=0.5mm, black] (1, 0) -- (3.5, 2.5);
\draw[line width=0.5mm, black] (2, 0) -- (4.5, 2.5);
\draw[line width=0.5mm, black] (4, 1) -- (4.5, 1.5);
\draw[line width=0.5mm, black] (-1.5, 0.5) -- (-0.5, 1.5);

\draw[line width=0.4mm, red, dash pattern={on 15pt off 7pt}] (-1.5, 0) -- (3, 0);
\draw[line width=0.4mm, red, dash pattern={on 15pt off 7pt}] (3, 0) -- (3, 1);
\draw[line width=0.4mm, red, dash pattern={on 15pt off 7pt}] (3, 1) -- (4.5, 1);

\node[draw=none] at (-2,0) {$A$};
\node[draw=none] at (-2,1.25) {$D_{1}$};

\node[draw=none] at (5,1) {$A$};
\node[draw=none] at (5,2.25) {$D_{1}$};

\draw[line width=0.8mm, blue] (-0.33, -0.33) -- (-0.33, 0.33);
\draw[line width=0.8mm, blue] (-0.33, 0.33) -- (0.33, 0.67);
\draw[line width=0.8mm, blue] (0.33, 0.67) -- (0.67, 0.33);
\draw[line width=0.8mm, blue] (0.67, 0.33) -- (1.33, 0.67);
\draw[line width=0.8mm, blue] (1.33, 0.67) -- (1.67, 0.33);
\draw[line width=0.8mm, blue] (1.67, 0.33) -- (1.67, -0.33);
\draw[line width=0.8mm, blue] (2.67, -0.33) -- (2.67, 0.33);
\draw[line width=0.8mm, blue] (2.67, 0.33) -- (3.33, 0.33);
\draw[line width=0.8mm, blue] (3.67, 0.67) -- (3.67, 1.33);
\draw[line width=0.8mm, blue] (3.67, 1.33) -- (4.33, 1.67);
\draw[line width=0.8mm, blue] (4.33, 1.67) -- (4.5, 1.67);

\end{scope}

\begin{scope}[xshift=10cm]

\draw[line width=0.5mm, black] (-1, -2) -- (-1, -3.5);
\draw[line width=0.5mm, black] (0, -2) -- (0, -3.5);
\draw[line width=0.5mm, black] (1, -2) -- (1, -3.5);
\draw[line width=0.5mm, black] (2, -2) -- (2, -3.5);
\draw[line width=0.5mm, black] (3, -1) -- (3, -3.5);
\draw[line width=0.5mm, black] (4, -1) -- (4, -3.5);
\draw[line width=0.5mm, black] (3, -1) -- (4.5, -1);
\draw[line width=0.5mm, black] (-1.5, -2) -- (4.5, -2);
\draw[line width=0.5mm, black] (-1.5, -3) -- (4.5, -3);
\draw[line width=0.5mm, black] (-1.5, -2.5) -- (-1, -2);
\draw[line width=0.5mm, black] (-1.5, -3.5) -- (0, -2);
\draw[line width=0.5mm, black] (-0.5, -3.5) -- (1, -2);
\draw[line width=0.5mm, black] (0.5, -3.5) -- (2, -2);
\draw[line width=0.5mm, black] (1.5, -3.5) -- (4, -1);
\draw[line width=0.5mm, black] (2.5, -3.5) -- (4.5, -1.5);
\draw[line width=0.5mm, black] (3.5, -3.5) -- (4.5, -2.5);

\fill[opacity=0.2, yellow, even odd rule] (-1.5,-2) -- (3,-2) -- (3,-3.5) -- (-1.5,-3.5) -- cycle;
\fill[opacity=0.2, yellow, even odd rule] (3,-1) -- (4.5,-1) -- (4.5,-3.5) -- (3,-3.5) -- cycle;

\draw[line width=0.4mm, red, dash pattern={on 15pt off 7pt}] (-1.5, -2) -- (3, -2);
\draw[line width=0.4mm, red, dash pattern={on 15pt off 7pt}] (3, -2) -- (3, -1);
\draw[line width=0.4mm, red, dash pattern={on 15pt off 7pt}] (3, -1) -- (4.5, -1);

\node[draw=none] at (-2,-2) {$A$};
\node[draw=none] at (-2,-3.25) {$D_{2}$};

\node[draw=none] at (5,-1) {$A$};
\node[draw=none] at (5,-2.75) {$D_{2}$};

\draw[line width=0.8mm, blue] (-0.67, -3.5) -- (-0.67, -3.33);
\draw[line width=0.8mm, blue] (-0.67, -3.33) -- (-0.33, -2.67);
\draw[line width=0.8mm, blue] (-0.33, -2.67) -- (-0.67, -2.33);
\draw[line width=0.8mm, blue] (-0.67, -2.33) -- (-0.67, -1.67);
\draw[line width=0.8mm, blue] (1.33, -1.67) -- (1.33, -2.33);
\draw[line width=0.8mm, blue] (1.33, -2.33) -- (1.67, -2.67);
\draw[line width=0.8mm, blue] (1.67, -2.67) -- (2.33, -2.33);
\draw[line width=0.8mm, blue] (2.33, -2.33) -- (2.33, -1.67);
\draw[line width=0.8mm, blue] (2.67, -1.33) -- (3.33, -1.33);
\draw[line width=0.8mm, blue] (3.33, -1.33) -- (3.33, -0.67);

\end{scope}
\end{tikzpicture}}
	\caption{Here, we have a segment of a cycle $W \in \mathcal{W}_{G}$, and the two induced PSAWs $W \cap \tilde{E}_{D_{2}}$ and $W \cap \tilde{E}_{D_{1}}$. Notice the dangling edges in the PSAWs that are incident on the other disc.}
	\label{fig:partialWalkExample}
\end{figure}

\begin{remark}\label{remark:PSAWRemark}
    Intuitively a PSAW is the intersection of a cycle on $H$ with $D$ (i.e. $W \cap \tilde{E}_{D}$ for $W \in \mathcal{W}_{G}$). However, in order to work effectively
    with this notion, we have to define it in a way that only refers to $H_{D}$, and does not refer to any unknown cycle in the graph $H$. See \cref{fig:partialWalkExample} for an example of a PSAW where this occurs. Even if two edges $e, e'$ from $W$ are incident on the same face of $D_{2}$, when we look at the same two dangling edges in $W \cap \tilde{E}_{D_{1}}$, we see that they are now no longer incident on the same vertex.
\end{remark}

\partialSAWwellDefined*
\begin{proof}
	Let $D' \in \{D_{1}, D_{2}\}$ be the disc other than $D$.
	Let $P = W \cap \tilde{E}_{D}$ for some  $W \in \mathcal{W}_{G}$.  If there exists an edge $\tilde{e} = \{u, v\} \in W$ such that $u \in \tilde{F}_{D}, v  \in \tilde{F}_{D'}$, we may equivalently consider this edge is  $\{u, \pi_{D}(e)\}$, which is incident on a dangling vertex instead of the vertex $v$. Now, let $Q = W \cap \tilde{E}_{D'}$, and consider $\ell_{Q}$. It follows from the construction of this function that $P \cup \mu_{D}(\ell_{Q})$ is a cycle. Therefore, $P \in \mathcal{P}_{D}$.
\end{proof}

\begin{remark}\label{remark:conformRemark}
    Even though the labelling $\ell_{2}$ does not uniquely determine the labelling $\ell_{1}$, the labelling $\ell_{2}$ together with a
    specific (cyclic) order to traverse  those edges from $\tilde{E}_{A}$
    indicated by $\ell_{2}$  
    does uniquely determine the labelling $\ell_{1}$.
\end{remark}

\section{Appendix B}\label{sec:appendix_b}

\incompatibilityBad*
\begin{proof}
Let $W \in \mathcal{C}_{S}(\ell_{1}, \ell_{2})$.
Then  $\ell_{W \cap \tilde{E}_{D_{1}}}= \ell_1$ and
$\ell_{W \cap \tilde{E}_{D_{2}}} = \ell_2$, and
since $W$ is a cycle on $H$, we have the same set
where they are nonzero on $\tilde{E}_A$ which is $W \cap \tilde{E}_A$.

Now suppose $\{e \in E_{A}: \ell_{1}(e) \not = 0\} = \{e \in E_{A}: \ell_{2}(e) \ne 0 \} \not = \emptyset$.
The set $M$ defined in \texttt{CompatibilityCheck} remains a subset of $E_{A}$
and strictly grows
during the {\bf while} loop.  As $E_{A}$ is finite,
the  {\bf while} loop must terminate.
Starting with some arbitrary $e_{1} \in E_{A}$, let $M(e_{1})$ be
the set $M$ that is obtained at the end of the 
{\bf while} loop if we start with the edge $e_{1}$. 
Let $e_{2} = \ell_{1}(e_{1})$. This means that  any $W \in \mathcal{C}_{S}(\ell_{1}, \ell_{2})$ when enters $\tilde{F}_{D_{1}}$ through the edge $\tilde{e}_{1}$ must exit through the edge $\tilde{e}_{2}$. 
As $\ell_1(e_2) \ne 0$, we also have $\ell_2 (e_2) \ne 0$.
Now, if we let $e_{3} = \ell_{2}(e_{2})$, we see that $W$ has to enter $\tilde{F}_{D_{2}}$ through $\tilde{e}_{2}$, and exit through $\tilde{e}_{3}$. We can repeat this process until we reach an edge $e_{k + 1}$ such that $\ell_{1}(e_{k + 1}) \in M$.  Since $W$ is a cycle, this must be  $e_{1}$, and  the cycle has been traversed. Therefore, we have found that $W \cap \tilde{E}_{A} = \{\tilde{e}_{1}, \dots, \tilde{e}_{k} \}$.
It follows that $M(e_1)$
has collected all the edges where $W$ crosses $A$,
$M(e_1) = \{e_1, e_2, \ldots, e_k\}$,
thus $M(e_1) = \{e \in E_{A}: \ell_{1}(e) \neq 0 \}$,
and \texttt{CompatibilityCheck}  returns {\bf true}.
\end{proof}

\begin{remark}\label{remark:TWellDefinedRemark}
	It should be noted that $T$ can also be viewed as an ordered set of arcs in $H_{D_{1}}$ and $H_{D_{2}}$, since $\varepsilon(T) \subseteq \tilde{E}_{A}$.
	If $T$ is viewed as an ordered set of arcs in $H_{D_{1}}$, then we can replace all the vertices in $\tilde{F}_{D_{2}}$ in the arcs with the corresponding dangling vertices in the graph, and the same applies for $H_{D_{2}}$.
	We will not distinguish between these cases any further, and use the same symbol $T$ in all three
	settings,  as the meaning will be clear from the context.
\end{remark}

\begin{lemma}\label{lemma:zeroSpecialCase}\leavevmode
	\begin{itemize}
		\item If $\varepsilon(S) \cap \tilde{E}_{D_{1}} \neq \emptyset$ and $\varepsilon(S) \cap \tilde{E}_{D_{2}} \neq \emptyset$, then $\left\lvert \mathcal{C}_{S}(0, 0) \right\rvert = 0$.
		\item If $\emptyset \neq \varepsilon(S) \subseteq \tilde{E}_{D}$ for $D \in \{D_{1}, D_{2} \}$, then $\left\lvert \mathcal{C}_{S}(0, 0) \right\rvert = \left\lvert \mathcal{C}_{D, S}(0) \right\rvert$.
		\item If $\varepsilon(S) = \emptyset$, then $\left\lvert \mathcal{C}_{S}(0, 0) \right\rvert = \left\lvert \mathcal{C}_{D_{1}, S}(0) \right\rvert + \left\lvert \mathcal{C}_{D_{2}, S}(0) \right\rvert - 1$.
	\end{itemize}
\end{lemma}
\begin{proof}
    If $W \in \mathcal{C}_{S}(0, 0)$, then  $W \cap \tilde{E}_{A} = \emptyset$, as otherwise there would be some edge $\tilde{e} \in W \cap \tilde{E}_{A}$, which would imply that $\ell_{W \cap \tilde{E}_{D_{1}}}(e) \neq 0$.
    Thus, $W$ cannot have edges in both 	$\tilde{E}_{D_{1}}$ and $\tilde{E}_{D_{2}}$.  In particular,
    if $\varepsilon(S) \cap \tilde{E}_{D_{1}} \neq \emptyset$, and $\varepsilon(S) \cap \tilde{E}_{D_{2}} \neq \emptyset$, then $\mathcal{C}_{S}(0, 0) = \emptyset$.
    	
    Now consider the case where $\emptyset \neq \varepsilon(S) \subseteq \tilde{E}_{D_1}$.
    For any $W \in \mathcal{W}_{G, S}$ since $\varepsilon(S) \subseteq W$,  it follows that $W \cap \tilde{E}_{D_{1}}  \neq \emptyset$.
    Suppose $W \in \mathcal{C}_{S}(0, 0)$. 
    Then, as we have already seen, $W \cap \tilde{E}_{A} = \emptyset$.
    Since $W$ is a cycle with $W \cap \tilde{E}_{D_{1}} \neq \emptyset$ and $W \cap \tilde{E}_{A} = \emptyset$, we have
     $W \subseteq \tilde{E}_{D_{1}}$,
     and as
    a consequence  $W = W \cap \tilde{E}_{D_{1}}$. 
    Applying \cref{def:PSAW}, 
    we see that $W \in \mathcal{P}_{D_1}$ with the empty $\mu_{D_1}(0)$ extension.
    And since $W$ conforms to $S$, 
    we have $W \in \mathcal{P}_{D_{1}, S}$.
    Finally as $\ell_W = \ell_{ W \cap \tilde{E}_{D_{1}}}= 0$,
    we get $W \in \mathcal{C}_{D_{1}, S}(0)$.
    
    Conversely, let $P \in \mathcal{C}_{D_{1}, S}(0)$. Then
    by definition $P \in \mathcal{P}_{D_{1}, S}$,
    $P \subseteq \tilde{E}_{D_{1}}$ and
    $\ell_{P} = \ell_{P \cap \tilde{E}_{D_{1}}} =0$, hence $P \cap \tilde{E}_{A} = \emptyset $. 
    By the definition of $\mathcal{P}_{D_{1}, S}$, there is  some
    $\ell \in \mathcal{L}_{A}$ such that
    $P \cup \mu_{D_1}(\ell)$
    is a cycle. We claim such $\ell$ must be 0. Indeed,  for any $e$ if $\ell(e) \not =0$ then this $e$ must belong to $P \cap \tilde{E}_{A}$, a contradiction to $P \cap \tilde{E}_{A} = \emptyset $. Hence this $\ell =0$, and $\mu_{D_{1}}(\ell) = \emptyset$.
    Therefore $P$ itself is a cycle. We have $P \in \mathcal{W}_{G, S}$, and since $P \subseteq \tilde{E}_{D_{1}}$, we have
    $\ell_{P \cap \tilde{E}_{D_{2}}}=0$ in addition to $\ell_{P \cap \tilde{E}_{D_{1}}}=0$.
    It follows that $P \in \mathcal{C}_{S}(0, 0)$. We have proved that, when
    $\emptyset \neq \varepsilon(S) \subseteq \tilde{E}_{D_1}$,
    $$\mathcal{C}_{S}(0, 0)
    = \mathcal{C}_{D_{1}, S}(0).$$
    
    Finally, consider the case  $\varepsilon(S) = \emptyset$.	
    The empty cycle consisting of no vertex and no edge is in  $\mathcal{C}_{\emptyset}(0, 0)$
    as well as in both $\mathcal{C}_{D_{1}, \emptyset}(0)$ and $\mathcal{C}_{D_{2}, \emptyset}(0)$.
    
    Now consider nonempty $W \in \mathcal{C}_{\emptyset}(0, 0)$. We know that $W \cap \tilde{E}_{A} = \emptyset$.
    Hence, $W \subseteq \tilde{E}_{D_{1}}$	or $W \subseteq \tilde{E}_{D_{2}}$, but not both.
    Say  $W \subseteq \tilde{E}_{D_{1}}$. 
    Then $W \in
    \mathcal{P}_{D_{1}} = \mathcal{P}_{D_{1}, \emptyset}$ with the empty extension
    $\mu_{D_{1}}(0) = \emptyset$.
    Hence, $W \in \mathcal{C}_{D_{1}, \emptyset}(0)$.
    Conversely, if $W \in \mathcal{C}_{D_{1}, \emptyset}(0)$, then $W \subseteq \tilde{E}_{D_{1}}$, and
    $\ell_W = \ell_{W \cap\tilde{E}_{D_{1}}} =0$.  Hence $W \cap \tilde{E}_{A}
    = \emptyset$.
    By the definition of $\mathcal{P}_{D_{1}}$,
    any labelling $\ell \in  \mathcal{L}_{A}$
    such that $W \cup \mu_{D_1}(\ell)$ is a cycle implies that $\ell =0$, and so
    $W$ is a cycle.
    It follows that $W \in  \mathcal{C}_{\emptyset}(0, 0)$.
    
    We have shown that a nonempty $W$ belongs to $\mathcal{C}_{\emptyset}(0, 0)$ iff it belongs to exactly one of $\mathcal{C}_{D_{1}, \emptyset}(0)$ or $\mathcal{C}_{D_{2}, \emptyset}(0)$.
    Taking into account the
    empty cycle,
    we have
    $\left\lvert \mathcal{C}_{\emptyset}(0, 0) \right\rvert = \left\lvert \mathcal{C}_{D_{1}, \emptyset}(0) \right\rvert + \left\lvert \mathcal{C}_{D_{2}, \emptyset}(0) \right\rvert - 1.$
\end{proof}

\PhiTDecomposition*
\begin{proof}
    \cref{lemma:WimpliesP1andP2} implies that if $W \in \mathcal{C}_{\emptyset}(\ell_{1}, \ell_{2})$, then $W \cap \tilde{E}_{D_{1}} \in \mathcal{C}_{D_{1}, T}(\ell_{1})$, and $W \cap \tilde{E}_{D_{2}} \in \mathcal{C}_{D_{2}, T}(\ell_{2})$.
    
    \cref{lemma:P1andP2implyW} implies that the converse is also true. Therefore, for  compatible $(\ell_{1}, \ell_{2}) \neq (0, 0)$, the function $\phi: \mathcal{C}_{D_{1}, T}(\ell_{1}) \times \mathcal{C}_{D_{2}, T}(\ell_{2}) \rightarrow \mathcal{C}_{\emptyset}(\ell_{1}, \ell_{2})$
    defined as
    $$\phi(P_{1}, P_{2}) = P_{1} \cup P_{2}$$
    is a well-defined surjective function.
    It is easy to see that the function is also injective, since $P_{1} \cup P_{2} = P'_{1} \cup P'_{2}$ implies that
    $$P_{1} = (P_{1} \cup P_{2}) \cap \tilde{E}_{D_{1}} = (P'_{1} \cup P'_{2}) \cap \tilde{E}_{D_{1}} = P'_{1},$$
    and similarly, $P_{2} = P'_{2}$. Therefore, $\phi$ is a bijection.  It follows that,
    for  compatible $(\ell_{1}, \ell_{2}) \neq (0, 0)$,
    $$ \left\lvert \mathcal{C}_{\emptyset}(\ell_{1}, \ell_{2}) \right\rvert = \left\lvert \mathcal{C}_{D_{1}, T}(\ell_{1}) \right\rvert \times \left\lvert \mathcal{C}_{D_{2}, T}(\ell_{2}) \right\rvert.$$
\end{proof}

\begin{lemma}\label{lemma:WimpliesP1andP2}
	If $W \in \mathcal{C}_{\emptyset}(\ell_{1}, \ell_{2})$ for $(\ell_{1}, \ell_{2}) \neq (0, 0)$, then $W \cap \tilde{E}_{D_{1}} \in \mathcal{C}_{D_{1}, T}(\ell_{1})$, and $W \cap \tilde{E}_{D_{2}} \in \mathcal{C}_{D_{2}, T}(\ell_{2})$.
\end{lemma}
\begin{proof}
	Let $P_{1} = W \cap \tilde{E}_{D_{1}}$.
	We know from \cref{lemma:partialSAWwellDefined} that $P_{1} \in \mathcal{P}_{D_{1}}$.
	It also follows that $P_{1} \in \mathcal{C}_{D_{1}}(\ell_{1})$, since $\ell_{P_{1}} = \ell_{1}$, by our choice of $W$.
	Similarly, $P_{2} \in \mathcal{C}_{D_{2}}(\ell_{2})$, where $P_{2} = W \cap \tilde{E}_{D_{2}}$.
	Moreover, it follows from the construction of $P_{1}$, that $P_{1} \cup  \mu_{D_{1}}(\ell_{2})$ is a cycle.
	Now, we show that $P_{1} \cup  \mu_{D_{1}}(\ell_{2})$ (and similarly, $P_{2} \cup \mu_{D_{2}}(\ell_{1})$) conforms to $T$.
	Assume that $T = ((u_{1} \rightarrow v_{1}), \dots, (u_{k}\rightarrow v_{k}))$. 
	
    For convenience, we will use $e_{i} \in E_{A}$ to refer to the edge for which $\tilde{e}_{i} = \{u_{i}, v_{i} \}$. 
   From the construction of $T$, we know that $\ell_{1}(e_{1}) = e_{2}$.
   Therefore, from the definition of the function $ \ell_{P_{1}} = \ell_{1}$, we can see that $P_{1}$ must enter $\tilde{F}_{D_{1}}$ through the edge $(u_{1}, v_{1})$, and exit through the edge $(u_{2}, v_{2})$, and visit the vertices $u_{1}, v_{1}, u_{2}, v_{2}$, in that order.
   
	Once again, from the construction of $T$, we know that $\ell_{2}(e_{2}) = e_{3}$.
	Therefore, we know that $\{v_{2}, u_{3}\} \in  \mu_{D_{1}}(\ell_{2})$.
	So, it picks up from $P_{2}$ at the dangling vertex $v_{2}$, and joins it again at $u_{3}$.
	Combining with our earlier observation, we see that $P_{1} \cup  \mu_{D_{1}}(\ell_{2})$ visits the vertices $u_{1}, v_{1}, u_{2}, v_{2}, u_{3}, v_{3}$ in that order.
	This pattern then repeats for $e_{3}, \dots, e_{k}$.
	So, we know that $\varepsilon(T) \subseteq W$, and since $P_{1} \cup  \mu_{D_{1}}(\ell_{2})$ is a cycle, we can see that it visits the vertices $u_{1}, v_{1}, \dots, u_{k}, v_{k}$ in that order.
	Therefore, $P_{1} \in \mathcal{C}_{D_{1}, T}(\ell_{1})$. By an analogous argument, we  also have  $P_{2} \in \mathcal{C}_{D_{2}, T}(\ell_{2})$.
\end{proof}

\begin{lemma}\label{lemma:P1andP2implyW}
	If $P_{1} \in \mathcal{C}_{D_{1}, T}(\ell_{1})$, and $P_{2} \in \mathcal{C}_{D_{2}, T}(\ell_{2})$ for compatible $(\ell_{1}, \ell_{2}) \neq (0, 0)$, then $P_{1} \cup P_{2} \in \mathcal{C}_{\emptyset}(\ell_{1}, \ell_{2})$.
\end{lemma}
\begin{proof}
	Note that since $P_{1} \subseteq \tilde{E}_{D_{1}}$ and $P_{2} \subseteq \tilde{E}_{D_{2}}$, it follows that $P_{1} \cup P_{2} \subseteq \tilde{E}$.
	Here  a dangling edge $\tilde{e} \in P_{1}$ (or $P_{2}$) is identified with the corresponding edge in $\tilde{E}$.
	
	First, we  show that $P_{1} \cup P_{2}$ contains a cycle in  $\mathcal{W}_{G, \emptyset}$.
	Note that $P_{1} \cap P_{2} \subseteq \tilde{E}_{A}$.
    The  compatibility of $\ell_{1}, \ell_{2} \in \mathcal{L}_{A}$ implies that 
    \[P_{1} \cap \tilde{E}_{A} = \{e \in E_A: \ell_1(e) \ne 0 \}
    =  \{e \in E_A: \ell_2(e) \ne 0 \} = P_{2} \cap \tilde{E}_{A},\] 
    and hence both are equal to $P_{1} \cap P_{2}$.
    	
    Just like before, we may assume that $T = ((u_{1} \rightarrow v_{1}), \dots, (u_{k} \rightarrow v_{k}))$, such that $\tilde{e}_{i} = \{u_{i}, v_{i}\}$ for $e_{i} \in E_{A}$.
    Since $\ell_{1}$ and $\ell_{2}$ are compatible and $T = T(\ell_{1}, \ell_{2})$, we have $\ell_{1}(e_{1}) = e_{2}, \ell_{2}(e_{2}) = e_{3}, \ell_{1}(e_{3}) = e_{4}, \dots,\ell_{2}(e_{k}) = e_{1}$.
    Importantly, we also have that $\{\tilde{e}_{1}, \dots, \tilde{e}_{k} \} = P_{1} \cap P_{2}$.
    
    So, starting with $e_{1}$, we know that since $\ell_{1}(e_{1}) = e_{2}$, $P_{1}$ enters $\tilde{F}_{D_{1}}$ through $\tilde{e}_{1}$, and exits through $\tilde{e}_{2}$.
    Then, since $\ell_{2}(e_{2}) = e_{3}$, we know that $P_{2}$ enters $\tilde{F}_{D_{2}}$ through $\tilde{e}_{2}$, and exits through $\tilde{e}_{3}$.
    Repeating this process until we reach $e_{k}$, we see that $P_{1} \cup P_{2}$ contains a cycle $W$ that passes through $\{\tilde{e}_{1}, \dots, \tilde{e}_{k}\}$.
    
    Now we show that  $P_{1} \cup P_{2}$ 
    is  itself the cycle $W$ we have identified, i.e., no edge of $P_{1} \cup P_{2}$ is left out from the cycle $W$.  Let $e \in P_1$ be an arbitrary edge of $P_1$. 
    Since $P_{1} \in \mathcal{C}_{D_{1}, T}(\ell_{1})$,  there  exists some $\ell \in \mathcal{L}_{A}$ such that $P_{1} \cup  \mu_{D_{1}}(\ell)$ is a cycle. This $\ell$ and $\ell_1$ must be compatible, so they share all the nonzero locations on $\tilde{E}_{A}$.
    We are given $\ell_{1} \neq 0$, and thus $e$ is traversed
    on the cycle $P_{1} \cup  \mu_{D_{1}}(\ell)$ on a simple path segment
    beginning and ending at some edges on $\tilde{E}_{A}$. As $P_{1} \cap \tilde{E}_{A} = P_{2} \cap \tilde{E}_{A}$, and this is the same as the underlying edge set $\varepsilon(T)$,
    when we traverse the cycle $P_{1} \cup  \mu_{D_{1}}(\ell)$ 
    exactly the same simple path segments of $P_1$
    as indicated by $\ell_1$ are
    traversed, as in  the cyclic traversal of $W$ indicated by $T$,
    possibly in some different order. 
    In particular,  $e \in P_1$ is traversed in the cycle $W$.
    Similarly all edges of $P_2$ are also contained in $W$.

	Therefore, $W= P_{1} \cup P_{2}$ is a cycle,
	and $W \in \mathcal{W}_{G, \emptyset}$.
	Finally, as $W \cap \tilde{E}_{D_{1}} = P_{1}$, by construction, we have $\ell_{P_{1}} = \ell_{1}$.
	Similarly, $\ell_{P_{2}} = \ell_{2}$. Therefore,  $P_{1} \cup P_{2} \in \mathcal{C}_{\emptyset}(\ell_{1}, \ell_{2})$.
\end{proof}

Another immediate corollary of \cref{lemma:WimpliesP1andP2} and \cref{lemma:P1andP2implyW} is the following:
\begin{corollary}\label{corollary: compatibilitySimple}
    Given compatible $(\ell_{1}, \ell_{2}) \neq (0, 0)$,
    $$\mathcal{C}_{D_{1}}(\ell_{1}) = \bigcup_{\substack{\ell_{2} \in \mathcal{L}_{A}:\\
    (\ell_{1}, \ell_{2}) \text{ compatible}}} \mathcal{C}_{D_{1}, T(\ell_{1}, \ell_{2})}(\ell_{1}).$$
\end{corollary}

\begin{lemma}\label{lemma:interleavingsBounded}
    Let $t\ge 1$.  
    If $|S| \leq s$ and $|T| \leq t$, then
    $$\left \rvert  \mathcal{I}_{S, T} \right\rvert \leq 2 t{\binom{s + t}{t}}.$$
\end{lemma}
\begin{proof}
	Without loss of generality $|S|= s$ and $|T| = t$,
	as the upper bound is monotonic in both $s$ and $t$.
	If $s=0$ then the upper bound is obvious. Let $s \ge 1$.
	Since $\mathcal{I}_{S, T}$ only contains one representative of each equivalence class, we may choose it so that if $I \in \mathcal{I}_{S, T}$, then $I \cap \varepsilon(S) = S$.
	Picking the first element of $S$ as the starting position,  after that
	in  the forward  direction  of the cyclic order of $S$ we can set  $s-1+t$ positions,
	of which we may pick any $t$ positions to be occupied by $\varepsilon(T)$.
	Then there are $t$ choices for the starting element of $T$ and two possible cyclic directions for $T$
	relative to the cyclic order of $S$. This gives us an upper bound 
	\[2t \binom{s+t-1}{t} \le 2 t {\binom{s + t}{t}}.\]
\end{proof}

\begin{lemma}\label{lemma:interleavingsPartition}
    Let $S$ be an ordered set of arcs in the graph $H$.
	Given compatible labellings $\ell_{1}, \ell_{2} \in \mathcal{L}_{A}$,
	let $T = T(\ell_{1}, \ell_{2})$ be the ordered set of arcs in $H$ defined above.
	Then,
	$$\left\lvert \mathcal{C}_{S}(\ell_{1}, \ell_{2}) \right\rvert = \sum_{I \in \mathcal{I}_{S, T}}\left\lvert \mathcal{C}_{I}(\ell_{1}, \ell_{2}) \right\rvert.$$
\end{lemma}
\begin{proof}
    We show that  $\mathcal{C}_{S}(\ell_{1}, \ell_{2})$ 	is a disjoint union,
    $$\mathcal{C}_{S}(\ell_{1}, \ell_{2}) =  \bigcup_{I \in \mathcal{I}_{S, T}}\mathcal{C}_{I}(\ell_{1}, \ell_{2}).$$

	It is easy to see that if $W \in \mathcal{C}_{I}(\ell_{1}, \ell_{2})$ for some $I \in \mathcal{I}_{S, T}$,
	then $W \in \mathcal{C}_{S}(\ell_{1}, \ell_{2})$ since $S$ is an ordered subset of $I$.
	Similarly, if $W \in \mathcal{C}_{S}(\ell_{1}, \ell_{2})$, note that $W$ conforms to $T$ by construction of $T = T(\ell_{1}, \ell_{2})$.
	Therefore, there is some $I \in \mathcal{I}_{S, T}$ such that $W \in \mathcal{C}_{I}(\ell_{1}, \ell_{2})$.
	
	To show that it is a disjoint union,
    let $W \in \mathcal{C}_{I_{1}}(\ell_{1}, \ell_{2}) \cap  \mathcal{C}_{I_{2}}(\ell_{1}, \ell_{2})$ 
	for $I_{1}, I_{2} \in \mathcal{I}_{S, T}$. 
	If either $S$ or $T = \emptyset$, then
    it is clear that $I_1 \sim I_2$.
	Suppose $S, T \not = \emptyset$.
	By a cyclic shift and reversal we may assume
	in both $I_{1}$ and $I_2$, the ordered set of arcs  $S$ appears
	as 
    $((u_{1} \rightarrow v_{1}), \dots, (u_{k} \rightarrow v_{k}))$, with $(u_{1} \rightarrow v_{1})$ appearing first. 
    
    We pick a traversal of $W$ to start at 
    $(u_{1} \rightarrow v_{1})$ in $S$, in the order 
    given by  $I_1$ above.
    Then we follow the appearances in $W$ of the arcs whose edges are
    from $\varepsilon(S)$ and $\varepsilon(T)$.
    The sequence of these appearances shows that
    $I_1 \sim I_2$.
    This shows that the union is disjoint, one  term per each equivalence class in $\mathcal{I}_{S, T}$.
\end{proof}

\invalidBad*
\begin{proof}
	Note that condition (1) is not dependent on the specific choice of $I$. 
	By the construction of $T = T(\ell_{1}, \ell_{2})$, we know that if $W \in \mathcal{C}_{I}(\ell_{1}, \ell_{2}) \subseteq \mathcal{C}_{\emptyset}(\ell_{1}, \ell_{2})$, 
	then $W \cap \tilde{E}_{A} = \varepsilon(T)$. 
	So, it is not possible to find an edge $\tilde{e}'$ in $W$ such that $\tilde{e}' \in \big(\varepsilon(S) \cap \tilde{E}_{A}\big) \setminus \varepsilon(T)$. 
	Moreover, if $W \in \mathcal{C}_{I}(\ell_{1}, \ell_{2})$, it has to conform to both $S$ and $T$. 
	So, if $S$ and $T$ require that the vertices be visited in different orders, no walk can conform to both $S$ and $T$. 
	Therefore, if condition (1) is not satisfied, $\mathcal{C}_{I}(\ell_{1}, \ell_{2}) = \emptyset$.
	
	Now, let us consider condition (2). 
	Since $\tilde{E}_{A}$ is a cut between $\tilde{F}_{D_{1}}$ and $\tilde{F}_{D_{2}}$, 
	if a walk $W$ goes from a vertex in $\tilde{F}_{D}$ to another vertex not in $\tilde{F}_{D}$, 
	it has to go through some edge in $\tilde{E}_{A}$. 
	Since $\varepsilon(T) = W \cap \tilde{E}_{A}$ for all $W \in \mathcal{C}_{I}(\ell_{1}, \ell_{2})$, 
	it can be seen immediately that if condition (2) is violated, then $\mathcal{C}_{I}(\ell_{1}, \ell_{2}) = \emptyset$.
	
	Therefore, from \cref{lemma:interleavingsPartition} we have that
    $$\left\lvert \mathcal{C}_{S}(\ell_{1}, \ell_{2}) \right\rvert = \sum_{\text{valid } I \in \mathcal{I}_{S, T}}\left\lvert \mathcal{C}_{I}(\ell_{1}, \ell_{2}) \right\rvert,$$
    where $T = T(\ell_{1}, \ell_{2})$
\end{proof}

\validIDecomposition*
\begin{proof}
    For a compatible $(\ell_{1}, \ell_{2}) \neq (0, 0)$, and a valid $I \in \mathcal{I}_{S, T}$ (where $T = T(\ell_{1}, \ell_{2})$), 
    consider the following function $\phi_{I}: \mathcal{C}_{D_{1}, I \cap \tilde{E}_{D_{1}}}(\ell_{1}) \times \mathcal{C}_{D_{2}, I \cap \tilde{E}_{D_{2}}}(\ell_{2}) \rightarrow \mathcal{C}_{I}(\ell_{1}, \ell_{2})$ defined as:
    $$\phi_{I}(P_{1}, P_{2}) = P_{1} \cup P_{2}$$
    Now, from \cref{lemma:phiIbijective}, we see that
    $$\left\lvert \mathcal{C}_{S}(\ell_{1}, \ell_{2}) \right\rvert = \sum_{\text{valid } I \in \mathcal{I}_{S, T}}\left\lvert \mathcal{C}_{D_{1}, I \cap \tilde{E}_{D_{1}}}(\ell_{1}) \right\rvert \times \left\lvert \mathcal{C}_{D_{2}, I \cap \tilde{E}_{D_{2}}}(\ell_{2}) \right\rvert.$$
\end{proof}

\begin{lemma}\label{lemma:phiIbijective}
	The function $\phi_{I}$ is bijective.
\end{lemma}
\begin{proof}
	First, we need to verify that the function is well-defined. 
	Note that \cref{lemma:P1andP2implyW} implies that since $\mathcal{C}_{D_{1}, I \cap \tilde{E}_{D_{1}}}(\ell_{1}) \subseteq \mathcal{C}_{D_{1}, T}(\ell_{1})$ 
	and $\mathcal{C}_{D_{2}, I \cap \tilde{E}_{D_{2}}}(\ell_{2}) \subseteq \mathcal{C}_{D_{2}, T}(\ell_{2})$, 
	it follows that $P_{1} \cup P_{2} \in \mathcal{C}_{\emptyset}(\ell_{1}, \ell_{2})$.
	
	Now, consider the cycle $W = P_{1} \cup P_{2}$ in $H$. 
	Since $I$ is a valid interleaving, we may assume that 
	$I = \big((u^{T,{1}} \rightarrow v^{T,{1}}), (u^{S,{1}}_{1} \rightarrow v^{S,{1}}_{1}), \dots, (u^{S,{1}}_{t_{1}} \rightarrow v^{S,{1}}_{t_{1}}), (u^{T,{2}} \rightarrow v^{T,{2}}), (u^{S,{2}}_{1} \rightarrow v^{S,{2}}_{1}), \dots, (u^{S,{2}}_{t_{2}} \rightarrow v^{S,{2}}_{t_{2}}), \dots, (u^{T,{k}} \rightarrow v^{T,{k}}), (u^{S,{k}}_{1} \rightarrow v^{S,{k}}_{1}), \dots, (u^{S,{k}}_{t_{k}} \rightarrow v^{S,{k}}_{t_{k}})\big)$, 
	where
	$$\big\{\{u^{T,{1}}, v^{T,{1}}\}, \dots, \{u^{T,{k}}, v^{T,{k}}\} \big\} = \varepsilon(I) \cap \tilde{E}_{A}$$
	and $k$ is even, being the number of times a cycle crosses $A$.
	We may further assume without loss of generality that $u^{T,{1}}, v^{T,{2}}, u^{T,{3}}, v^{T,{4}}, \dots, v^{T,{k}} \in \tilde{F}_{D_{2}}$, $v^{T,{1}}, u^{T,{2}}, \dots, u^{T,{k}} \in \tilde{F}_{D_{1}}$,
	$u^{S,{1}}_{i}, v^{S,{1}}_{i}, u^{S,{3}}_{i}, v^{S,{3}}_{i}, \dots, u^{S,{k - 1}}_{i}, v^{S,{k - 1}}_{i} \in \tilde{F}_{D_{1}}$, and that $u^{S,{2}}_{i}, v^{S,{2}}_{i}, u^{S,{4}}_{i}, v^{S,{4}}_{i}, \dots, u^{S,{k}}_{i}, v^{S,{k}}_{i} \in \tilde{F}_{D_{2}}$ for all $i$.
	
	In this case, as $W = P_{1} \cup P_{2}$, since $P_{1} \in \mathcal{C}_{D_{1}, I \cap \tilde{E}_{D_{1}}}(\ell_{1})$ and $P_{2} \in \mathcal{C}_{D_{2}, I \cap \tilde{E}_{D_{2}}}(\ell_{2})$, 
	it follows that $W \supseteq \varepsilon(I)$. 
	Moreover, $W$ visits the vertices $u^{T,{1}}, v^{T,{1}}, u^{S,{1}}_{1}, v^{S,{1}}_{1}, \dots, u^{S,{1}}_{t_{1}}, v^{S,{1}}_{t_{1}}, u^{T,{2}}, v^{T,{2}}$ in $\tilde{E}_{D_{1}}$ in that order.
	From $u^{T,{2}}, v^{T,{2}}$, it then visits the vertices $u^{S,{2}}_{1}, \dots, v^{S,{2}}_{t_{2}}, u^{T,{3}}, v^{T,{3}}$ in $\tilde{E}_{D_{2}}$ in that order, and so on.
	Therefore, when we put these segments of $W$ together, we see that $W$ conforms to $I$. 
	Therefore, $W \in \mathcal{C}_{I}(\ell_{1}, \ell_{2})$, and $\phi_{I}$ is well-defined.
	
	Now, we show that this function is injective. 
	If there exist $(P_{1}, P_{2})$ and $(P'_{1}, P'_{2})$ such that $P_{1} \cup P_{2} = P'_{1} \cup P'_{2}$, we know that
	$$P_{1} = (P_{1} \cup P_{2}) \cap \tilde{E}_{D_{1}} = (P'_{1} \cup P'_{2}) \cap \tilde{E}_{D_{1}} = P'_{1}$$
	Therefore, $P_{1} = P'_{1}$. 
	Similarly, $P_{2} = P'_{2}$. Therefore, the function is injective.
	
	Finally, we show that $\phi_{I}$ is surjective. 
	Given $W \in \mathcal{C}_{I}(\ell_{1}, \ell_{2}) \subseteq \mathcal{C}_{\emptyset}(\ell_{1}, \ell_{2})$, \cref{lemma:WimpliesP1andP2} tells us that 
	$W \cap \tilde{E}_{D_{1}} \in \mathcal{C}_{D_{1}, T}(\ell_{1})$, and $W \cap \tilde{E}_{D_{2}} \in \mathcal{C}_{D_{2}, T}(\ell_{2})$.
	Now, let $P_{1} = W \cap \tilde{E}_{D_{1}}$, and $P_{2} = W \cap \tilde{E}_{D_{2}}$. 
	Since $W$ conforms to $I$, and therefore $I \cap \tilde{E}_{D_{1}}$, it is clear that $P_{1} \cup  \mu_{D_{1}}(\ell_{2})$ conforms to $I \cap \tilde{E}_{D_{1}}$ as well. 
	Therefore, $P_{1} \in \mathcal{C}_{I \cap \tilde{E}_{D_{1}}}(\ell_{1})$, and similarly, $P_{2} \in \mathcal{C}_{I \cap \tilde{E}_{D_{2}}}(\ell_{2})$. 
	Since, $P_{1} \cup P_{2} = W$ by construction, it follows that $\phi_{I}$ is surjective.
\end{proof}
\section{Appendix C}\label{sec:appendix_c}

We will go through the construction of the graph $G_{1}$, and the ordered set of arcs $S_{1}$ in more detail here. The construction of $G_{1}$ occurs inductively.
Starting from $D^{0} = G_{D_{1}}$, we iteratively construct $D^{1}, \dots, D^{\eta} = G_{1}$. 
Assume that $D^{i} = (V^{i}, E^{i}, F^{i})$. 
We know that $D^{0}$ contains the cycle $A^{0} = A$ which bounds the face $f^{*}$. 
During each iteration, we construct $D^{i + 1}$ by modifying $D^{i}$ with boundary $A^i$, 
so that $D^{i + 1}$ is a graph on a disc with  boundary $A^{i + 1}$.
We will abuse the notation a bit, and refer to the face bounded by the edges $E_{A}^{i}$
of $A^i$ in each $D^{i}$ still as $f^{*}$.

First, we will formally describe the contractions we will use to transform $D^{i}$ into $D^{i + 1}$.

\begin{enumerate}		
	\item
	\textbf{Single-edge contraction}: We are given some edge $e = \{a, b\} \in E_{A}^{i}$. As $e$ is an edge on the boundary $A^{i}$ of
	$D^{i}$, there is a unique triangular face within $D_1$ containing this edge $e$. 
	Let $e' = \{a, c\}, e'' = \{b, c\}$ both in $E^{i}$, such that $\{e, e', e''\} \in F^{i}$ is that face. We assume $\deg(c) >2$, i.e.,  there is no parallel edge $e^* = \{a, b\} \in E^{i}$ such that $\{e^*, e', e''\} \in F^{i}$ (otherwise we can not perform  single-edge contraction on $e$). 
	To perform single-edge contraction on $e$, we identify the vertices $a$ and $b$ with each other, and the edges $e'$ and $e''$ with each other. We also delete the edge $e$ and the face $\{e, e', e''\}$. The degree of  $c$ is decreased by 1. In this case, $E_{A}^{i + 1} = E_{A}^{i} \setminus \{e\}$; however, if $E_{A}^{i} = \{e, e', e''\}$, then $E_{A}^{i + 1} = \emptyset$. See \cref{fig:singleEdgeExample}.

	\item 
	\textbf{Edge-pair contraction}: We have a pair of consecutive edges $e, e' \in E_{A}^{i}$  on $A^i$, $e = \{a, b\}$ and $e' = \{b, c\}$. Moreover, there is no $e'' = \{a, c\} \in E^{i}$ such that $\{e, e', e''\} \in F^{i}$, in other words, $\deg(b) > 2$. To perform edge-pair contraction on $\{e, e'\}$, we identify the vertices $a$ and $c$ with each other, and the edges $e$ and $e'$ with each other.  The degree of  $b$ is decreased by 1. In this case, $V_{A}^{i + 1}$ is  $V_{A}^{i} \setminus \{b\}$
	with $a$ and $c$ identified, and $E_{A}^{i + 1} = E_{A}^{i} \setminus \{e, e'\}$.  See \cref{fig:edgePairExample}.
	
	\item
	\textbf{Parallel-edge contraction}: We are given two parallel edges  $e = \{a, b\} \in E_{A}^{i}, e' = \{a, b\} \in E^{i}$, and two adjacent faces $\{ e, e'', e'''\}, \{e', e'', e'''\} \in F^{i}$, where  $e'' = \{a, c\}, e''' = \{c, b\} \in E^{i}$.
	To perform parallel-edge contraction on $\{e, e'\}$, 
	we delete the vertex $c$, and the edges $e'', e'''$, and identify the edges $e$ and $e'$ with each other.
	The two faces $\{ e, e'', e'''\}, \{e', e'', e'''\} \in F^{i}$ are also removed. In this case, $E_{A}^{i + 1} = E_{A}^{i}$ (with
	the identified $e$ and $e'$ taking the place of $e$), but $|V^{i + 1}| < |V^{i}|$. However, if $E_{A}^{i} = \{e, e'\}$, then $E_{A}^{i + 1} = \emptyset$. See \cref{fig:parallelEdgeExample}.
\end{enumerate}

Now, let us see the precise order in which these contractions are applied. 
For the graph
$D^{0}$, we define the function $\ell_{2}^{0} = \ell_{2}$ on $E_{A}^{0} = E_{A}$.
	Inductively assume that $\ell_{2}^{i} \in \mathcal{L}_{A^{i}}$, 
 we will
 define $\ell_{2}^{i + 1}$ as we construct $D^{i + 1}$ from $D^i$.

To construct $D^{i + 1}$, we first find an edge 
$e = \{a, b\} \in E_{A}^{i}$ such that $\ell_{2}^{i}(e) = 0$,
if such an edge exists. 
We check if this edge $e$ is eligible for a parallel-edge contraction 
(i.e., we check if there is a parallel edge $e' = \{a, b\} \in E^{i}$, 
and edges $e'' = \{a, c\}, e''' = \{c, b\} \in E^{i}$ such that $\{e, e'', e'''\}, \{e', e'', e'''\} \in F^{i}$.)
 See \cref{fig:parallelEdgeExample}.
If this is the case, we apply a parallel-edge contraction to obtain $D^{i + 1}$, and let $\ell_{2}^{i + 1} = \ell_{2}^{i}$. Clearly, $\ell_{2}^{i + 1} \in \mathcal{L}_{A^{i + 1}}$ in this case.
If a parallel-edge contraction is not possible for the edge $e$, then
let the unique triangular face in $F^i$ containing $e$ be $\{e, e', e''\}$,
and $e'=\{a, c\}, e''=\{b, c\}$, we must have $\deg(c) >2$, and single-edge contraction on  $e$ is applicable.
 See \cref{fig:singleEdgeExample}.
We apply a single-edge contraction on the edge $e$, and let $\ell_{2}^{i + 1} = \ell_{2}^{i}|_{E_{A}^{i + 1}}$. Since $\ell_{2}^{i}(e) = 0$, and $E_{A}^{i + 1} = E_{A}^{i} \setminus \{e\}$, it is clear that $\ell_{2}^{i + 1} \in \mathcal{L}_{A^{i + 1}}$.

Now we assume that  there are no edges $e \in E_{A}^{i}$ such that $\ell_{2}^{i}(e) = 0$,
but $E_{A}^{i} \neq \emptyset$. 
Since $\ell_{2}^{i} \in \mathcal{L}_{A^{i}}$,
the pairing of edges by $\ell_{2}^{i}$ corresponds to
a  Motzkin  path which denotes a set of non-intersecting chords.
We claim that	there are two  adjacent edges $e, e' \in E_{A}^{i}$ such that
 $\ell_{2}^{i}(e) = e'$ and  $\ell_{2}^{i}(e') = e$.
To see that, start with any paired  
edges $e_1, e_2 \in E_{A}^{i}$ 
with $\ell_{2}^{i}(e_1) = e_2$ and $\ell_{2}^{i}(e_2) = e_1$.
If $e_1, e_2$ are not adjacent edges in $E_{A}^{i}$, then they
separate $A^i$ into  two segments, and  the
pairings according to $\ell_{2}^{i}$ are confined within 
each segment, or else there would be intersecting chords.
Restricted within each segment,  $\ell_{2}^{i}$  still denotes
a Motzkin  path which assigns no 0 values, and by induction we can get  
two  adjacent paired edges $e, e' \in E_{A}^{i}$  as claimed.
 
Assume $e = \{a, b\}, e' = \{b, c\}$. 
If there exists some edge $e'' = \{a, c\} \in E^{i}$, 
such that $\{e, e', e''\} \in F^{i}$,
i.e., $\deg(b) =2$, then we do not construct $D^{i + 1}$ at all.
We can directly claim that $\lvert \mathcal{C}_{D_{1}, I \cap \tilde{E}_{D_{1}}}(\ell_{1}) \rvert = 1$, 
if $\varepsilon(I \cap \tilde{E}_{D_{1}}) = \{\tilde{e}, \tilde{e}'\}$;  we set $\lvert \mathcal{C}_{D_{1}, I \cap \tilde{E}_{D_{1}}}(\ell_{1}) \rvert = 0$, otherwise (see \cref{lemma:specialCaseContractions}).
On the other hand, if there is no such face in $F^{i}$, then edge-pair contraction on $e, e'$ is applicable;
 see \cref{fig:edgePairExample}.  We perform an edge-pair contraction on $e, e'$ to construct $D^{i + 1}$, 
and let $\ell_{2}^{i + 1} = \ell_{2}^{i}|_{E_{A}^{i + 1}}$. 
By our choice that $e, e'$ are paired adjacent edges, it follows that $\ell_{2}^{i + 1} \in \mathcal{L}_{A^{i + 1}}$.

This describes the construction of  $D^{i + 1}$ when $E_{A}^{i} \neq \emptyset$, with the
one  exception 
of that special case  as specified, which
immediately concludes the computation of $\lvert \mathcal{C}_{D_{1}, I \cap \tilde{E}_{D_{1}}}(\ell_{1}) \rvert$.  
Since during each contraction step the graph $D^i$ does not
increase but either $|E_{A}^{i}|$ or $|V^{i}|$ strictly decreases, 
this process will eventually terminate. 
Assume this process continues until $i = \eta$, at which point $E_{A}^{\eta} = \emptyset$, and  we let $G_{1} = D^{\eta}$. 
We note that our description of constructing
$D^{\eta}$ from $D^0$ depends on a particular order
of applicable contraction
steps  and does not assert that the process 
produces a canonical or unique graph
$G_{1}$. However, we will not need this uniqueness.

\begin{lemma}\label{lemma:contractionPlanarTriangulated}
	The graph $G_{1} = (V_{1}, E_{1}, F_{1})$ is a planar triangulated graph. Moreover, if $\lvert V_{1} \rvert \geq 3$, then $G_{1}$ has no vertices of degree less than $2$.
\end{lemma}
\begin{proof}
Clearly starting from  a plane graph the contraction   steps
	maintain planarity.
	Therefore $G_{1}$ is planar.
	
	Now we  show that $G_{1}$ is triangulated 
	and that when $|V_{1}| \geq 3$, $G_{1}$ has no vertex of degree $1$. 
    We will inductively show that for each $i$, 
	the faces $F^{i} \setminus \{f^{*}\}$ are triangles. Since $f^{*}$ does not exist in $D^{\eta} = G_{1}$,
	 this would imply that $G_{1}$ is triangulated. 
	 Similarly, we will inductively show that the degrees of all vertices stay greater than 1, as long as $|V^{i}| \geq 3$.
	 Since $|V^{i}|$ decreases as $i$ increases, this completes the proof.

    Consider the case where a parallel-edge contraction (\cref{fig:parallelEdgeExample}) is applied to $D^{i}$.
    Consider the edge $e'$. 
    If $e' \in E_{A}^{i}$, then $E_{A}^{i} = \{e, e'\}$, as they form a cycle, and $E^{i} = \{e, e', e'', e'''\}$ with $V^{i} = \{a, b, c\}$.
    Therefore, after the contraction, $E_{A}^{i + 1} = \emptyset$, 
    and $V^{i + 1} = V^{\eta} = V_{1} = \{a, b\}$.
    In this case, $G_{1}$ is vacuously triangulated.
    
    Now, assume $e' \notin E_{A}^{i}$.
    Then, $e'$ is shared by two triangular faces. One triangular face is
    $\{e', e'', e'''\}$. Let the other
    one be $\{e', e_1, e_2\}$. The effect of this parallel-edge contraction step is simply to make
    $\{e, e_1, e_2\}$ a new triangular face replacing three triangular faces, and nothing outside of
    $\{e, e_1, e_2\}$ is changed
    (see \cref{fig:parallelEdgeExample}). Since all the other faces in
    $F^{i} \setminus \{f^{*} \}$ were triangular,
    every $f \in F^{i + 1} \setminus \{f^{*} \}$ is also triangular.
    
    Moreover, the only vertices in $V^{i + 1}$ coming from $V^i$ whose degrees are changed by the parallel-edge contraction are the vertices $a$ and $ b$. 
    But we know that when $e' \notin E_{A}^{i}$,
    the edges $e, e_{1}$ and the edges $e, e_{2}$ are incident to $a, b$ respectively, in $D^{i + 1}$. 
    Therefore, all vertices in $V^{i + 1}$ have degree at least 2.
	
	Next consider the case where a single-edge contraction (\cref{fig:singleEdgeExample}) is applied to $D^{i}$. 
	Let $e = \{a, b\} \in E_{A}^{i}, e' = \{a, c\}, e'' = \{b, c\} \in E^{i}$ be the edges involved.
	If $e', e'' \in E_{A}^{i}$, as well, then, $E_{A}^{i} =\{e, e', e''\}$,
	as these edges form a cycle.
	Then, after the contraction, $E_{A}^{i + 1} = \emptyset$, and $V^{i + 1} = V^{\eta} = V_{1} = \{a, c\}$.
	So, $G_{1}$ is vacuously triangulated.
	
	Now, assume without loss of generality that $e' \notin E_{A}^{i}$.
	So, $e'$ is shared by two triangular faces in $D^{i}$.
	One of these faces is $\{e, e', e''\}$. 
	Let the other one be $\{e', e_{1}, e_{2}\}$.
	Note that we perform this single-edge contraction step only after we have
	performed all parallel-edge contraction steps.
	So in particular, $e'' \notin \{e', e_{1}, e_{2}\}$.
	Moreover, since $e \in E_{A}^{i}$, it belongs to only one face in $F^{i}$.
	Therefore, $e \notin \{e', e_{1}, e_{2}\}$.
	So, after the contraction, edges in $\{e', e_{1}, e_{2}\} $
	are not deleted or identified with each other,
	and the face remains a triangle in $F^{i + 1}$.
	Similarly, if $e'' \notin E_{A}^{i}$ is also true, the triangle $\{e'', e_{3}, e_{4}\}$
	(other than $\{e, e', e''\}$) that contains it remains a triangle in $F^{i + 1}$.
	Since these are the only faces in $F^{i + 1} \setminus \{f^{*}\}$
	that are affected by the contraction 
	(note that $\{e, e', e''\}$ is deleted),
	every face $f \in F^{i + 1} \setminus \{f^{*}\}$ is triangular.
	
	Moreover, the only vertices in $V^{i + 1}$ whose degrees are changed by the single-edge contraction are the vertices $a ( = b)$, and $c$.
	But we know that when $e' \notin E_{A}^{i}$,
	there exists a face $\{e', e_{1}, e_{2}\} \in F^{i + 1}$ that contains both vertices.
	So, their degree is at least $2$.
	Therefore, all vertices in $V^{i + 1}$ have degree at least 2.
	
	Finally consider the case where an edge-pair contraction (\cref{fig:edgePairExample}) is applied on $D^{i}$. 
	Let $e = \{a, b\}, e' = \{b, c\} \in E_{A}^{i}$ be the edges involved.
	Note that we perform the edge-pair contraction 
	only if there is no edge $e'' = \{a, c\}$ such that 
	$\{e, e', e''\} \in F^{i}$, i.e., $\deg(b) > 2$.
	Therefore, there exist $f_{1} = \{e, e_{1}, e_{2}\}, f_{2} = \{e', e_{3}, e_{4}\} \in F^{i}$ 
	such that $f_{1} \neq f_{2}$.
	Since $e' \notin f_{1}$, and no edges are deleted by the contraction, the triangle $f_{1}$ remains a triangle in $F^{i + 1}$.
	The same applies for $f_{2}$.
	Since these are the only faces in $F^{i + 1} \setminus \{f^{*}\}$
	which are affected by the contraction,
	every face $f \in F^{i + 1} \setminus \{f^{*}\}$ is triangular.
	
	Moreover, the only vertices in $V^{i + 1}$ whose degrees are changed by the edge-pair contraction are the vertices $a (= c)$ and $ b$. For $b$ since we had $\deg(b) >2$ in
	$F^i$, the degree remains no less than 2.
	For the combined vertex formed by the merging of $a$ and $c$ the two triangular faces $f_1$ and $f_2$ remain two distinct faces incident to it, its degree also  remains  at least 2.
\end{proof}

Before proceeding further, let us also justify why the construction of $G_{1}$ was unnecessary in the special case during the edge-pair contraction, where we found the face $\{e, e', e''\} \in F^{i}$.

\begin{lemma}\label{lemma:specialCaseContractions}
	While performing an edge-pair contraction on a graph $D^{i}$ on $(e, e')$, where $e = \{a, b\} \in E^i_A$ and $e' = \{b, c\} \in E^i_A$, if there exists some $e'' = \{a, c\}$ such that $\{e, e', e''\} \in F^{i}$,
	then $\left\lvert \mathcal{C}_{D_{1}, I \cap E_{D_{1}}}(\ell_{1}) \right\rvert = 1$ iff $\varepsilon(I \cap \tilde{E}_{D_{1}}) = \{\tilde{e}, \tilde{e}' \}$, and $\left\lvert \mathcal{C}_{D_{1}, I \cap E_{D_{1}}}(\ell_{1}) \right\rvert = 0$ otherwise.
\end{lemma}
\begin{proof}
    Let $f  = \{e, e', e''\} \in F^{i}$.
    Observe that we
    only  perform
    edge-pair contraction when
    there are no 0 labelled edges on
    $E^i_A$, which implies that $|E^i_A|$ is even as  nonzero labels are paired. In particular, 
    it is not the case that $E^i_A$ consists of precisely these three edges $\{e, e', e''\}$.
    Having excluded this
    particular possibility, 
    we observe that  none of the three types of operations (see \cref{fig:threeContractionsExample}) creates
    a triangular face containing  two consecutive edges of the boundary curve. It follows that $f$ is already a face in $F^0$.
    
    Now consider $D^{0} = G_{D_{1}}$.
    The condition that led us to perform this 
    edge-pair  contraction step on $(e, e')$
    is because $\ell_2(e)=e'$ and $\ell_2(e')=e$.
    This stipulates that a cycle enters and then immediately exits $D_1$ on the two edges $\{e, e'\}$ of the same face $f$, and the cycle must complete
    itself in the disc $D_2$ as stipulated by the labelling $\ell_2$ matching $e$ and $e'$. Thus the unique possibility for the cycle is to have a simple path
    in $D_2$ connecting $\tilde{e}$ and $\tilde{e}'$ and finish off within $D_1$
    by the trivial incursion in $D_1$ with $f$.  This happens if and only if $\varepsilon(I \cap \tilde{E}_{D_{1}}) = \{\tilde{e}, \tilde{e}' \}$ (and in fact as a consequence in this case,
    they are consecutive in $I$ and the arc orientations of $\tilde{e}, \tilde{e}'$ in $I$  match). 

 	Therefore, $\left\lvert \mathcal{C}_{D_{1}, I \cap E_{D_{1}}}(\ell_{1}) \right\rvert = 1$ 
	if $\varepsilon(I \cap \tilde{E}_{D_{1}}) = \{\tilde{e}, \tilde{e}'\}$, 
	and is 0 otherwise.
\end{proof}

With the graph $G_{1}$ constructed as above, we now need to construct an ordered set $S_{1}$ using $I \cap \tilde{E}_{D_{1}}$.
We will construct $S_{1}$ in conjunction with $G_{1}$. 
We will start with $S^{0} = I \cap \tilde{E}_{D_{1}}$, and as we construct $D^{i }$ using a contraction,
we will also iteratively construct $S^{i}$, until we reach $S^{\eta} = S_{1}$.

In addition to constructing $S^{i}$, 
we need to construct a function $\ell_{1}^{i}$ (the construction of $\ell_{2}^{i}$  was already described
 earlier  for $D^{i}$)  on the set $E_{A}^{i}$, 
such that $\ell_{1}^{i}$ and $\ell_{2}^{i}$ are compatible, 
and $T(\ell_{1}^{i}, \ell_{2}^{i}) = \varepsilon(S^{i} \cap \tilde{E}_{A}^{i})$. 
We start with $\ell_{1}^{0} = \ell_{1}$, 
which clearly satisfies all requirements. 
Then, if $D^{i + 1}$ is constructed by a parallel-edge contraction,
we let $\ell_{1}^{i + 1} = \ell_{1}^{i}$. If $D^{i + 1}$ is constructed by a single-edge contraction, 
we let $\ell_{1}^{i + 1} = \ell_{1}^{i}|_{E_{A}^{i + 1}}$. 
If $D^{i + 1}$ is constructed by an edge-pair contraction, 
we have two cases to deal with. Let $e, e' \in E_{A}^{i}$ be the two edges that are identified with each other. 
If $\ell_{1}^{i}(e) = e'$ (and vice versa), 
we simply let $\ell_{1}^{i + 1} = \ell_{1}^{i}|_{E_{A}^{i + 1}}$. 
Otherwise, if $\ell_{1}^{i}(e) = e''$ and $\ell_{1}^{i}(e') = e'''$ (and vice versa), 
we let $\ell_{1}^{i + 1}(e'') = e'''$ (and vice versa), 
and let $\ell_{1}^{i + 1}$ be the same as $\ell_{1}^{i}$ for all other edges. 
It is apparent from the construction, that $\ell_{1}^{i + 1}, \ell_{2}^{i + 1}$ are compatible, 
if $\ell_{1}^{i}, \ell_{2}^{i}$ are compatible. Note that after this edge-pair contraction on $(e, e')$, these two edges are removed from $E^i_A$, and $\ell_2^{i+1}$ is the restriction of $\ell_2^{i}$ on the remaining edges of $E^i_A$, namely $E_A^{i+1}$.
 
Now, getting back to the construction of $S^{i + 1}$, 
we will ensure that $S^{i + 1}$ is an ordered set of arcs on the dual $H^{i + 1}$ of the graph $D^{i + 1}$.
Moreover, we will ensure that during the construction of $S^{i + 1}$, 
we will never remove or change the order in which arcs appear in $S^{i}$. 
At most, we will identify arcs with each other, if the underlying edges are identified in $D^{i + 1}$, 
and we will remove duplicates of the same arc if the duplicates appear consecutively after identification.
We will then establish a recursive relation between 
$\lvert \mathcal{C}_{D^{i}, S^{i}}(\ell_{1}^{i}) \rvert$ and 
$\lvert \mathcal{C}_{D^{i + 1}, S^{i + 1}}(\ell_{1}^{i + 1}) \rvert$.

However, before we show that, let us consider a few  cases where the construction of $S^{i + 1}$, and with it the construction of $G_{1}$, can be easily terminated.

\begin{lemma}\label{lemma:Si+1SpecialCaseContractions}\leavevmode
	\begin{enumerate}
		\item 
		Let $D^{i + 1}$ be constructed by a single-edge contraction. Let $e$ be the edge that is deleted, and let $e', e''$ be the edges that are identified with each other  (see \cref{fig:singleEdgeExample}). Let $f$ be the face that is deleted by the identification of $e'$ with $e''$, and $\tilde{e}' = \{\tilde{f}_{1}, \tilde{f}\}$, and $\tilde{e}'' = \{\tilde{f}, \tilde{f}_{2} \}$.
		\begin{enumerate}
			\item If $\tilde{e} \in \varepsilon(S^{i})$, then $\lvert \mathcal{C}_{D^{i}, S^{i}}(\ell_{1}^{i}) \rvert = 0$.	
							
			\item 
			If $\{\tilde{e}', \tilde{e}'' \} \subseteq \varepsilon(S^{i})$, but $((\tilde{f}_{1} \rightarrow \tilde{f}), (\tilde{f} \rightarrow \tilde{f}_{2}))$ does not 
			appear as consecutive arcs in the cyclic order of $S^i$, 
			then $\lvert \mathcal{C}_{D^{i}, S^{i}}(\ell_{1}^{i}) \rvert = 0$.				
		\end{enumerate}
	
		\item 
		Let $D^{i + 1}$ be constructed by a parallel-edge contraction. Let $e, e'$ be the parallel edges, and let $e'', e'''$ be the edges that are deleted (see \cref{fig:parallelEdgeExample}).
		\begin{enumerate}
			\item 
			If $(\ell^{i}_{1}, \ell^{i}_{2}) \neq (0, 0)$, and $\{\tilde{e}'', \tilde{e}''' \} \cap \varepsilon(S^{i}) \neq \emptyset$, then $\lvert \mathcal{C}_{D^{i}, S^{i}}(\ell_{1}^{i}) \rvert = 0$.
			
			\item 
			If $(\ell^{i}_{1}, \ell^{i}_{2}) = (0, 0)$, and $\{\tilde{e}'', \tilde{e}''' \} \cap \varepsilon(S^{i}) \neq \emptyset$, but $\varepsilon(S^{i}) \not\subseteq \{\tilde{e}'', \tilde{e}''' \}$, then $\lvert \mathcal{C}_{D^{i}, S^{i}}(\ell_{1}^{i}) \rvert = 0$.
			
			\item 
			If $(\ell^{i}_{1}, \ell^{i}_{2}) = (0, 0)$, and $\emptyset \neq \varepsilon(S^{i}) \subseteq \{\tilde{e}'', \tilde{e}''' \}$, then $\lvert \mathcal{C}_{D^{i}, S^{i}}(\ell_{1}^{i}) \rvert = 1$.
			
			\item 
		If 
		$\{\tilde{e}, \tilde{e}' \} \cap \varepsilon(S^{i}) \neq \emptyset$, then $\lvert \mathcal{C}_{D^{i}, S^{i}}(\ell_{1}^{i}) \rvert = 0$.
		\end{enumerate}		
	\end{enumerate}		
\end{lemma}
\begin{proof}
    Recall that we perform  single-edge contraction or parallel-edge contraction  only when $\ell^i_2(e)=0$.	In case (1a), this contradicts the requirement that it must pass $\tilde{e}$.
    Similarly, 		in case (1b), given that
    no passage is allowed at $\tilde{e}$, 
    (by $\ell_2^i(e)=0$),
    any passage using one of  $\{\tilde{e}',  \tilde{e}'' \}$
    must also use the other, and they
    must be for consecutive steps in the cyclic order of the interleaving, and in the  correct orientation of the arcs.
	
	In case (2) part (a)(b)(c), consider a PSAW $P \in \mathcal{C}_{D^{i}, S^i}(\ell_{1}^{i})$. It has either 
	$\tilde{e}''$ or $\tilde{e}'''$. 
    Since a parallel-edge contraction was applied,
	$\ell_{1}^{i}(e) = \ell_{2}^{i}(e) = 0$. 
	Given any vertex (other than the vertex $\tilde{f}^{*}$) in $\tilde{F}^{i}$, $P$ must have either 0 or 2 edges incident on it. Thus $P$ has both 	$\tilde{e}''$ or $\tilde{e}'''$.
	But since $\tilde{e}''$ and $\tilde{e}'''$ form a cycle in $H^{i}$, 
	it must be the case that $P = \{\tilde{e}'', \tilde{e}''' \}$.
	
	This immediately proves  (2a), (2b), and (2c).
	For (2d), given 	$\{\tilde{e}, \tilde{e}' \} \cap \varepsilon(S^{i}) \neq \emptyset$,
	any passage of a cycle using one of
	$\{\tilde{e}, \tilde{e}'\}$
	must use both. But this  contradicts $\ell_2(e)=0$,
	which is	the reason we performed this 
	parallel-edge contraction on $e$.
\end{proof}

\begin{remark*} 
    Item 2 part (a)(b)(c) of \cref{lemma:Si+1SpecialCaseContractions} combined takes care of the case 
    $\{\tilde{e}'', \tilde{e}''' \} \cap \varepsilon(S^{i}) \neq \emptyset$ in a
    parallel-edge contraction step.
    After all cases in item  2 part (a)(b)(c)(d) are taken care of,
    we may assume 	$\{\tilde{e}, \tilde{e}', \tilde{e}'', \tilde{e}''' \} \cap \varepsilon(S^{i}) = \emptyset$ in a
    parallel-edge contraction step.
\end{remark*}

We now see how to construct $S^{i + 1}$, assuming none of these special cases occur.	
First, let us consider the case where $D^{i + 1}$ is constructed  from $D^{i}$ by performing an edge-pair contraction on $e$ and $e' \in E_{A}^{i}$. This step identifies 
$e$ and $e'$ with each other, 
but all the other edges stay as before.
With a little abuse of notation we may assume that $e, e' \in E_{A}^{0} = E_{A}$  (as no contraction step creates
any edge on the boundary), such that $\ell_{2}(e) = e'$ (and vice versa).
This means that $\tilde{e}, \tilde{e}' \in \varepsilon(T)$ by construction of $T$. Let $\tilde{e} = \{u_{1}, v_{1}\}$, 
and $\tilde{e} = \{u_{2}, v_{2}\}$ with $u_{1}, v_{2} \in \tilde{F}_{D_{1}}$, and $v_{1}, u_{2}\notin \tilde{F}_{D_{1}}$ are dangling vertices.
Since $\ell_{2}(e) = e'$, by the construction of $T$, we can see that 
$((u_{1} \rightarrow v_{1}), (u_{2} \rightarrow v_{2}))$ are consecutive arcs in $T$. 
Moreover, since $I \in \mathcal{I}_{S, T(\ell_{1}, \ell_{2})}$ is a valid interleaving, we know that $((u_{1} \rightarrow v_{1}), (u_{2} \rightarrow v_{2}))$ are consecutive arcs in $I \cap \tilde{E}_{D_{1}}$
as any arcs between these two in $I$ must belong to $\tilde{E}_{D_{2}} \setminus \tilde{E}_{A}$.

By the induction hypothesis, we assumed that during the construction of $S^{1}, \dots, S^{i}$, these arcs were not removed, or had their order changed.
Therefore, $((u_{1} \rightarrow v_{1}), (u_{2} \rightarrow v_{2}))$ must also be consecutive arcs in $S^{i}$. 
Now, if we assume that $f_{1}$ is the unique face in $F^{i}$ that contains the edge $e$, and $f_{2}$ is the unique face containing $e'$, 
we know by our choice of $u_{1}, v_{2}$, that $\tilde{f}_{1} = u_{1}$ and $\tilde{f}_{2} = v_{2}$. 
Moreover, we have already taken care of the special case where $f_{1} = f_{2}$ in \cref{lemma:specialCaseContractions}, so here, we may assume that $f_{1} \neq f_{2}$.
Now, when we identify the edges $e$ and $e'$ with each other in $D^{i + 1}$, 
we see that $f_{1}$ and $f_{2}$ are now adjacent to each other. 
In other words, $(u_{1}, v_{2}) \in \tilde{E}^{i + 1}$. 
Therefore, in $S^{i + 1}$, we can replace the consecutive arcs $((u_{1} \rightarrow v_{1}), (u_{2} \rightarrow v_{2}))$ 
with the arc $((u_{1} \rightarrow v_{2}))$. 
Note that $S^{i + 1}$ now satisfies the induction hypothesis.

Now, let us consider the case where $D^{i + 1}$ is constructed by performing a parallel-edge construction on $D^{i}$.
Note that items 2 (a)(b)(c) in \cref{lemma:Si+1SpecialCaseContractions} dealt with the case when $\{\tilde{e}'', \tilde{e}''' \} \cap \varepsilon(S^{i}) \not = \emptyset$, and then after items 2 (d), 
we only need to consider case $\{\tilde{e}, \tilde{e}', \tilde{e}'', \tilde{e}''' \} \cap \varepsilon(S^{i}) = \emptyset$. 
But in this case no edge in $S^{i}$ is affected by the construction of $D^{i + 1}$.
So, we may let $S^{i + 1} = S^{i}$.

Finally, let us consider the case where $D^{i + 1}$ is constructed by performing a single-edge contraction on $D^{i}$. 
We identify the edges $e' = \{a, c\}, e'' = \{b, c\} \in E^{i}$ with each other, and the edge $e = \{a, b\} \in E^{i}$ is deleted. 
But, all other edges stay as before.
Since the special cases from \cref{lemma:Si+1SpecialCaseContractions} have already been dealt with, we may assume that $\tilde{e} \notin \varepsilon(S^{i})$.	
As for $\tilde{e}'$ and $\tilde{e}''$, unlike in the case of the edge-pair contractions, these edges are not guaranteed to belong to $\varepsilon(S^{i})$. 
So, there are various possibilities in which these edges can appear (or not) in $\varepsilon(S^{i})$. 
If neither of the edges appear, then we have nothing more to do, and the induction is already complete. 
Let us consider the case, where exactly one of the edges (say $\tilde{e}'$ appears in $\varepsilon(S^{i})$). 
Let $\tilde{e}' = \{u_{1}, v_{1}\}$, and $\tilde{e}'' = \{u_{2}, v_{2}\}$. 
Since $e'$ and $e''$ both belong to the face $f = \{e, e', e''\}$, we may further assume that $\tilde{f} = v_{1} = u_{2}$, and that $\tilde{f}_{1} = u_{1}, \tilde{f}_{2} = v_{2}$. 
In this case, note that because of the contraction, the faces $f_{1}$ and $f_{2}$ are now adjacent to each other. 
So, $\{u_{1}, v_{2}\} \in \tilde{E}^{i + 1}$. 
Therefore, in $S^{i + 1}$, we can replace any instance of $(u_{1} \rightarrow v_{1})$ with $(u_{1} \rightarrow v_{2})$, 
and any instance of $(v_{1} \rightarrow u_{1})$ with $(v_{2} \rightarrow u_{1})$. 
As we can see, in this case too, $S^{i + 1}$ satisfies the induction hypothesis.

Finally, let us consider the case where $\tilde{e}$ and $\tilde{e}'$ both belong to $\varepsilon(S^{i})$.
Once again, since the special cases from \cref{lemma:Si+1SpecialCaseContractions} are already taken care of, 
we know that $((u_{1} \rightarrow v_{1}), (u_2 \rightarrow v_{2}))$ (here $v_1=u_2$) are consecutive arcs in $S^{i}$. 
So, we can replace these arcs with $(u_{1} \rightarrow v_{2})$.
We have already seen how this satisfies our induction hypothesis.

Now that we have constructed $G_{1}$ and $S_{1}$, we need to prove that they work as we intended.

\begin{lemma}\label{lemma:contractionsCorrect}
	Let the graph $D^{i}$ and the ordered set $S^{i}$ be such that they do not fall into the special cases from \cref{lemma:Si+1SpecialCaseContractions}. 
	Let $D^{i + 1}$, and $S^{i + 1}$ be constructed as above. 
	If $D^{i + 1}$ is constructed by an edge-pair contraction, or single-edge contraction, then
	$$\left\lvert \mathcal{C}_{D^{i}, S^{i}}(\ell_{1}^{i}) \right\rvert = \left\lvert \mathcal{C}_{D^{i + 1}, S^{i + 1}}(\ell_{1}^{i + 1}) \right\rvert.$$
	If $D^{i + 1}$ is constructed by a parallel-edge contraction,  then
	$$\left\lvert \mathcal{C}_{D^{i}, S^{i}}(\ell_{1}^{i}) \right\rvert = \left\lvert \mathcal{C}_{D^{i + 1}, S^{i + 1}}(\ell_{1}^{i + 1}) \right\rvert + \mathbbm{1}_{\{(\ell_{1}^{i}, \ell_{2}^{i}) = (0, 0);\ \varepsilon(S^{i}) = \emptyset\}}.$$
\end{lemma}
\begin{proof}
	Consider $P^{i} \in \mathcal{C}_{D^{i}, S^{i}}(\ell_{1}^{i})$.
	Since $\ell_{1}^{i}$ is compatible with $\ell_{2}^{i}$ by construction, it follows that $P^{i} \cup \mu_{D^{i}}(\ell_{2}^{i})$ is a cycle in $H^{i} = H_{D^{i}}$. 
	We will now use $P^{i}$ to construct a $P^{i + 1} \in \mathcal{C}_{D^{i + 1}, S^{i + 1}}(\ell_{1}^{i + 1})$.
	
	First, let us consider the case where $D^{i + 1}$ was constructed using an edge-pair contraction. 
	As we have seen during the construction of $S^{i + 1}$, this means that the edges $\tilde{e} = \{u_{1}, v_{1}\} \in \tilde{E}_{A}^{i}$ and 
	$\tilde{e}' = \{u_{2}, v_{2}\} \in \tilde{E}_{A}^{i}$ are replaced with the edge $\{u_{1}, v_{2}\}$. 
	Now, since $\ell_{2}^{i}(e) \neq 0$ and $\ell_{2}^{i}(e') \neq 0$ (this is because we  perform edge-pair contraction only for the paired edge by $\ell_2$), we know that $\big\{\{u_{1}, v_{1}\}, \{u_{2}, v_{2}\} \big\} \subseteq P^{i}$. 
	We construct $P^{i + 1}$ by just replacing these two edges in $P^{i}$ with $\{u_{1}, v_{2}\}$.
	
	Clearly, $P^{i + 1} \subseteq \tilde{E}_{D^{i + 1}}$. 
	Consider $P^{i + 1} \cup \mu_{D^{i + 1}}(\ell_{2}^{i + 1})$. 
	We already know that $P^{i} \cup \mu_{D^{i}}(\ell_{2}^{i})$ is a cycle in $H^{i}$. 
	In this cycle, we just need to replace the segment $\big\{\{u_{1}, v_{1}\}, \{v_{1}, u_{2}\}, \{u_{2}, v_{2} \} \big\}$ 
	(where $\{u_{1}, v_{1} \}$ and $\{u_{2}, v_{2} \}$ belong to $P^{i}$ and $\{v_{1}, u_{2} \}$ belongs to $\mu_{D^{i}}(\ell_{2}^{i})$) 
	with the edge $\{u_{1}, v_{2}\}$ to get $P^{i + 1} \cup \mu_{D^{i + 1}}(\ell_{2}^{i + 1})$. So, $P^{i + 1} \in \mathcal{P}_{D^{i + 1}}$.
	
	Now, we need to show that $P^{i + 1} \cup \mu_{D^{i + 1}}(\ell_{2}^{i + 1})$ conforms to $S^{i + 1}$. 
	By our choice of $S^{i + 1}$, this is apparent, 
	since we replaced the consecutive arcs $\big((u_{1} \rightarrow v_{1}), (u_{2} \rightarrow v_{2})\big)$ in $S^{i}$ with $(u_{1} \rightarrow v_{2})$. 
	So, if $P^{i}$ conforms to $S^{i}$, $P^{i + 1}$ has to conform to $S^{i + 1}$. 
	Moreover, by definition of $\ell_{1}^{i + 1}$, it is easy to see that $\ell_{P^{i + 1}} = \ell_{1}^{i + 1}$. 
	Therefore, $P^{i + 1} \in \mathcal{C}_{D^{i + 1}, S^{i + 1}}(\ell_{1}^{i + 1})$.
	
	Clearly, this is an invertible transformation. 
	So, when $D^{i + 1}$ is constructed using an edge-pair contraction,
	$$\left\lvert \mathcal{C}_{D^{i}, S^{i}}(\ell_{1}^{i}) \right\rvert = \left\lvert \mathcal{C}_{D^{i + 1}, S^{i + 1}}(\ell_{1}^{i + 1}) \right\rvert.$$
	
	Now, let us consider the case where $D^{i + 1}$ was constructed using a single-edge contraction. 
	Once again, as we saw during the construction of $S^{i + 1}$, the edges $\{u_{1}, v_{1}\}, \{v_{1}, v_{2}\} \in \tilde{E}^{i}$ are replaced by the edge $\{u_{1}, v_{2}\} \in \tilde{E}^{i + 1}$. 
	If $P^{i}$ does not contain either of the deleted edges, we have nothing to do. 
	It is immediate that $P^{i + 1} \in \mathcal{C}_{D^{i}, S^{i}}(\ell_{1}^{i})$. 
	On the other hand, assume that $P^{i}$ contains one of these edges (say $\{u_{1}, v_{1}\}$). 
	We know that $P^{i} \cup \mu_{D^{i}}(\ell_{2}^{i})$ is a cycle in $H^{i}$. 
	So, we know that there has to be another edge in $P^{i}$ that is incident on $v_{1}$. 
	But this edge cannot be $\tilde{e}$, since $\ell_{2}^{i}(e) = 0$ by our choice of $e$ (as the edge where we perform a single-edge contraction). 
	Therefore, $P^{i}$ must contain the only other edge $\{v_{1}, v_{2}\}$ that is incident on $v_{1}$. 
	Therefore, in $P^{i + 1}$, we may replace these two edges with $\{u_{1}, v_{2}\}$.
	
	We need to show that $P^{i + 1} \in \mathcal{P}_{D^{i + 1}}$. 
	To do that, consider $P^{i + 1} \cup \mu_{D^{i + 1}}(\ell_{2}^{i + 1})$. 
	Note that in the cycle $P^{i} \cup \mu_{D^{i}}(\ell_{2}^{i})$, we are replacing the 
	segment $\big\{ \{u_{1}, v_{1}\}, \{v_{1}, v_{2}\} \big\}$ with the edge $\{u_{1}, v_{2}\}$ 
	to obtain $P^{i + 1} \cup \mu_{D^{i + 1}}(\ell_{2}^{i + 1})$. So, $P^{i + 1} \in \mathcal{}P_{D^{i + 1}}$. 
	Moreover, since the special case from \cref{lemma:Si+1SpecialCaseContractions} has not occured, 
	we know that even if $\big\{\{u_{1}, v_{1}\}, \{v_{1}, v_{2}\} \big\} \subseteq \varepsilon(S^{i})$, from the construction of $S^{i + 1}$ and $\ell_{1}^{i + 1}$, we see that
	$P^{i + 1} \in \mathcal{C}_{D^{i + 1}, S^{i + 1}}(\ell_{1}^{i + 1})$.
	
	Clearly, this is also an invertible transformation. 
	So, when $D^{i + 1}$ is constructed using a single-edge contraction,
	$$\left\lvert \mathcal{C}_{D^{i}, S^{i}}(\ell_{1}^{i}) \right\rvert = \left\lvert \mathcal{C}_{D^{i + 1}, S^{i + 1}}(\ell_{1}^{i + 1}) \right\rvert.$$
	
	Finally, we consider the case where $D^{i + 1}$ was constructed using a parallel-edge contraction.
	Since the special cases from \cref{lemma:Si+1SpecialCaseContractions} have not come to pass, 
	we know that $\{\tilde{e}, \tilde{e}', \tilde{e}'', \tilde{e}''' \} \cap \varepsilon(S^{i}) = \emptyset$. 
	In this case, as we saw in the proof of \cref{lemma:Si+1SpecialCaseContractions}, 
	if $\tilde{e}''$ or $\tilde{e}'''$ belongs to $P^{i}$, then in fact, $P^{i} = \{\tilde{e}'', \tilde{e}''' \}$. 
	So, if $(\ell_{1}^{i}, \ell_{2}^{i}) \neq (0, 0)$, 
	or if $\varepsilon(S^{i}) \neq \emptyset$, 
	it follows that $\{\tilde{e}'', \tilde{e}''' \} \cap P^{i} = \emptyset$. 
	Since each vertex in $H^{i}$ can have either 0 or 2 edges from $P^{i}$ incident on it, 
	we also find that $\tilde{e}, \tilde{e}' \notin P^{i}$. 
	Therefore, we can let $P^{i + 1} = P^{i}$.
	
	Clearly, $P^{i + 1} \cup \mu_{D^{i + 1}}(\ell_{2}^{i + 1})$ is a cycle, 
	and it conforms to $S^{i + 1}$, since $S^{i + 1} = S^{i}$, 
	and $\{\tilde{e}, \tilde{e}', \tilde{e}'', \tilde{e}''' \} \cap \varepsilon(S^{i}) = \emptyset$. 
	Therefore, $P^{i + 1} \in \mathcal{C}_{D^{i + 1}, S^{i + 1}}(\ell_{1}^{i + 1})$.
	
	Clearly, this is an invertible transformation. 
	So, when $D^{i + 1}$ is constructed using a parallel-edge contraction, when $(\ell_{1}^{i}, \ell_{2}^{i}) \neq (0, 0)$ or $\varepsilon(S^{i}) \neq \emptyset$,
	$$\left\lvert \mathcal{C}_{D^{i}, S^{i}}(\ell_{1}^{i}) \right\rvert = \left\lvert \mathcal{C}_{D^{i + 1}, S^{i + 1}}(\ell_{1}^{i + 1}) \right\rvert.$$
	
	The last case to deal with is when $(\ell_{1}^{i}, \ell_{2}^{i}) = (0, 0)$, and $\varepsilon(S^{i}) = \emptyset$.
	Assume $P^{i} \neq \{ \tilde{e}'', \tilde{e}''' \}$.
	Once again, we let $P^{i + 1} = P^{i}$.
	Clearly, $P^{i + 1} \in \mathcal{C}_{D^{i + 1}, S^{i + 1}}(\ell_{1}^{i + 1})$, just as we saw in the previous case. 
	So, this transformation from $\mathcal{C}_{D^{i}, S^{i}}(\ell_{1}^{i}) \setminus \big\{\{\tilde{e}'', \tilde{e}'''\} \big\}$ to $\mathcal{C}_{D^{i}, S^{i}}(\ell_{1}^{i})$ is invertible. 
	However, in this case, $P^{i} = \{\tilde{e}'', \tilde{e}'''\}$ is also possible.
	Therefore, when $(\ell_{1}^{i}, \ell_{2}^{i}) = (0, 0)$, and $\varepsilon(S^{i}) = \emptyset$,
	$$\left\lvert \mathcal{C}_{D^{i}, S^{i}}(\ell_{1}^{i}) \right\rvert = \left\lvert \mathcal{C}_{D^{i + 1}, S^{i + 1}}(\ell_{1}^{i + 1}) \right\rvert + 1$$
\end{proof}

\contractionsReduce*
\begin{proof}
	In the case where $G_{1}$ is constructed, this follows directly from \cref{lemma:contractionsCorrect}, 
	since the construction takes polynomial time, and we know that $\left\lvert \mathcal{C}_{D^{0}, S^{0}}(\ell_{1}^{0}) \right\rvert = \left\lvert \mathcal{C}_{D_{1}, I \cap \tilde{E}_{D_{1}}}(\ell_{1}) \right\rvert$, 
	and that $\left\lvert \mathcal{C}_{D^{\eta}, S^{\eta}}(\ell_{1}^{\eta}) \right\rvert = \left\lvert \mathcal{W}_{G_{1}, S_{1}} \right\rvert$.
	
	There are two cases in which $G_{1}$ would not be constructed. 
	The first is the case when applying an edge-pair contraction on edges $e, e' \in \tilde{E}_{A}^{i}$, 
	we found another edge $e''$, such that $\{e, e', e''\} \in F^{i}$. 
	The correctness in this case was established in \cref{lemma:specialCaseContractions}.
	
	The other case is when we came across the special cases from \cref{lemma:Si+1SpecialCaseContractions} during the construction of $S^{i + 1}$. 
	Since $\varepsilon(S^{i}) \neq \emptyset$ in this case anyway, 
	from \cref{lemma:contractionsCorrect}, we see that $\left\lvert \mathcal{C}_{D_{1}, I \cap \tilde{E}_{D_{1}}}(\ell_{1}) \right\rvert = \left\lvert \mathcal{C}_{D^{0}, S^{0}}(\ell_{1}^{0}) \right\rvert = \left\lvert \mathcal{C}_{D^{i}, S^{i}}(\ell_{1}^{i}) \right\rvert$, 
	and the rest follows from \cref{lemma:Si+1SpecialCaseContractions}.
\end{proof}
\section{Appendix D}\label{sec:appendix_d}

\mainInduction*
\begin{proof}	
    As leaf nodes have constant size,
    the base case is clearly true.
    Assume 
    the lemma statement is true for all problems at level $>k $. 
    Consider 	 a problem $\left\lvert \mathcal{W}_{G_{k}, S_{k}} \right\rvert$ at level $k$ of $\mathcal{T}$.
    		
    We know from \cref{fact:MotzkinBound} that $\left\lvert \mathcal{L}_{(A_{k})} \right\rvert$ is the number of Motzkin paths on $E_{D_{k}}$,
    which is at most $\kappa^{t_{k}}$ for a fixed constant $\kappa$.
    Now, we know that $|S_{k}| \leq t_{0} + \dots + t_{k - 1}$, and  for any compatible $(\ell_{1}, \ell_{2})$, we have $|T(\ell_{1}, \ell_{2})| \leq t_{k}$. Therefore, from \cref{lemma:interleavingsBounded}, we get that for each compatible $(\ell_{1}, \ell_{2})$,
    
    $$\left\lvert \mathcal{I}_{S_{k}, T(\ell_{1}, \ell_{2})} \right\rvert \leq 2  t_{k}
    \binom{t_{0} + \dots + t_{k}}{t_k}. $$
    
    Since each compatible $(\ell_{1}, \ell_{2})$ and a valid $I$ result in at most two subproblems, 
    the number of children of this problem at level $k$ is at most
    $$4(\kappa^{t_{k}})^{2} t_{k}
    \binom{t_{0} + \dots + t_{k}}{t_k}.$$
    
    By inductive hypothesis we know that solving each of these subproblems  takes time at most
    
    $$\alpha^{\sqrt{m_{k + 1}}}  \frac{(t_{0} + \dots + t_{L - 1})!}{(t_{0} + \dots + t_{k})!(t_{k + 1})! \cdots (t_{L - 1})!}.$$
    
    Now, note that given the problem $\left\lvert \mathcal{W}_{G_{k}, S_{k}} \right\rvert$, 
    checking if any given $\ell_{1}, \ell_{2} \in \mathcal{L}_{(A_{k})}$ are compatible, 
    and checking if $I \in \mathcal{I}_{S_{k}, T(\ell_{1}, \ell_{2})}$ is valid takes polynomial time (as a function of the number of vertices in the graph). 
    So does the reduction from $\left\lvert \mathcal{C}_{(D_{k}), S_{k}}(\ell) \right\rvert$ to $\left\lvert \mathcal{W}_{G_{k + 1}, S_{k + 1}} \right\rvert$. 
    In total, we can assume that determining and finally combining the solutions to each of the children of the problem 
    $\left\lvert \mathcal{W}_{G_{k}, S_{k}} \right\rvert$ takes $p(m_{k})$ time, where $p(\cdot)$ is a polynomial function.
    
    Combining everything, the total time taken to solve the problem at level $k$ is at most:
    \begin{align*}
    &4t_{k}p(m_{k}) \left(\kappa^{2t_{k}} \frac{(t_{0} + \dots + t_{k})!}{(t_{0} + \dots + t_{k - 1})!(t_{k})!}\right) 
    \left(\alpha^{\sqrt{m_{k + 1}}} \frac{(t_{0} + \dots + t_{L - 1})!}{(t_{0} + \dots + t_{k})!(t_{k + 1})! \cdots (t_{L - 1})!} \right) \\
    &= 4t_{k}p(m_{k})  \left(\kappa^{2t_{k}} \alpha^{\sqrt{m_{k + 1}}}\right)  \left( \frac{(t_{0} + \dots + t_{L - 1})!}{(t_{0} + \dots + t_{k - 1})!(t_{k})! \cdots (t_{L - 1})!} \right)
    \end{align*}
    
    Note that  $4t_{k}p(m_{k}) = 4(\sqrt{8  m_{k}})p(m_{k})$ can be bounded by a polynomial of $m_{k}$.
    So there is a constant $\gamma$,
    such that the time taken to solve the problem at level $k$ is at most
    \[\gamma^{\sqrt{m_{k}}}
    \alpha^{\sqrt{\frac{3}{4}m_{k}}}
    \frac{(t_{0} + \dots + t_{L - 1})!}{(t_{0} + \dots + t_{k - 1})!(t_{k})! \cdot (t_{L - 1})!},\]
    and $\gamma^{\sqrt{m_{k}}}
    \alpha^{\sqrt{\frac{3}{4}m_{k}}}
    < \alpha^{\sqrt{m_{k}}}$
    for a large constant $\alpha$.
    This completes the induction.
\end{proof}

\mainThm*
\begin{proof}
    Let $G = (V, E, F)$ be the dual of the graph $H$ with $|V| = m$.
    By the 3-regularity of $H$,  and 
    by Euler's formula,  $m = n/2 + 2$.
    By \cref{hyp:mainInduction},
    computing $\left\lvert \mathcal{W}_{G} \right\rvert$ takes time
    $$\alpha^{\sqrt{m}} \frac{(t_{0} + \dots + t_{L - 1})!}{(t_{0})! \cdots (t_{L - 1})!}.$$
    Applying \cref{lemma:factorialsBound} with 
    $r = \sqrt{\nicefrac{3}{4}}$, and $a = \sqrt{8m}$ competes the proof.
\end{proof}

\begin{lemma}\label{lemma:factorialsBound}
	For any $0 \leq r < 1$, there exists a constant $\delta$ such that for any $a \geq 0$ and $k \geq 0$,
	$$\frac{(a +  ar +\dots + ar^{k})!}{(a)!  (ar)! \cdots (ar^{k})!}  \leq \delta^{a + ar + \dots + ar^{k}}.$$
\end{lemma}
\begin{proof}
	Take  $\delta = 2^{\frac{1}{1 - r}}$.
	We  prove this by induction on $k$. When $k = 0$, the statement is clearly true as $\delta \geq 1$. Assume that the statement is true for $k - 1$. Then,
	\begin{align*}
	\frac{(a +  ar +\dots + ar^{k})!}{(a)!  (ar)! \cdots (ar^{k})!} 
		&= \frac{(a +  ar +\dots + ar^{k})!}{a! (ar +\dots + ar^{k})!}   \frac{(ar +\dots + ar^{k})!}{(ar)! 
		\cdots (ar^{k})!}\\
		&\leq 2^{a + \dots + ar^{k}} \delta^{ar + \dots + ar^{k}}\\
		&\le \left(2^{\frac{1}{1 - r}} \right)^{a}\delta^{ar + \dots + ar^{k}}\\
		&= \delta^{a + ar + \dots + ar^{k}}.
	\end{align*}
\end{proof}
\section{Appendix E}\label{sec:appendix_e}


In order to prove \cref{theorem:mustCrossMainThm}, we will need the following analogue of \cref{lemma:compatibilityPartition}:

\begin{lemma}\label{lemma:mustCrossCompatibilityPartition}
	$$\left\lvert \mathcal{W}_{G, S}^{X, Y} \right\rvert = \sum_{\text{\rm compatible }\ell_{1}, \ell_{2} \in \mathcal{L}_{A}} \left\lvert \mathcal{C}_{S}^{X, Y}(\ell_{1}, \ell_{2}) \right\rvert.$$
	Let $X_{D} = X \cap \tilde{E}_{D}, Y_{D} = Y \cap \tilde{E}_{D}$ for $D \in \{D_{1}, D_{2}\}$. If $(\ell_{1}, \ell_{2}) \neq (0, 0)$ is a compatible pair of labellings then
	$$\left\lvert \mathcal{C}_{S}^{X, Y}(\ell_{1}, \ell_{2}) \right\rvert = \sum_{\text{\rm valid } I \in \mathcal{I}(S, T(\ell_{1}, \ell_{2}))}\left\lvert \mathcal{C}_{D_{1}, I \cap \tilde{E}_{D_{1}}}^{X_{D_{1}}, Y_{D_{1}}}(\ell_{1}) \right\rvert \times \left\lvert \mathcal{C}_{D_{2}, I \cap \tilde{E}_{D_{2}}}^{X_{D_{2}}, Y_{D_{2}}}(\ell_{2}) \right\rvert$$
	If $(\ell_{1}, \ell_{2}) = (0, 0)$, then
	\begin{itemize}
		\item If $\emptyset \neq \varepsilon(S) \subseteq \tilde{E}_{D}$ and $X \subseteq \tilde{E}_{D}$ 
		for $D \in \{D_{1}, D_{2} \}$, then $\left\lvert \mathcal{C}_{S}^{X, Y}(0, 0) \right\rvert = \left\lvert \mathcal{C}_{D, S}^{X, Y_{D}}(0) \right\rvert$.
		\item If $\varepsilon(S) = \emptyset$ and $\emptyset \neq X \subseteq \tilde{E}_{D}$ for $D \in \{D_{1}, D_{2}\}$, 
		then $\left\lvert \mathcal{C}_{S}^{X, Y}(0, 0) \right\rvert = \left\lvert \mathcal{C}_{D, S}^{X, Y_{D}}(0) \right\rvert$.
		\item If $\varepsilon(S) = \emptyset$ and $X = \emptyset$, 
		then $\left\lvert \mathcal{C}_{S}^{X, Y}(0, 0) \right\rvert = \left\lvert \mathcal{C}^{X, Y_{D_{1}}}_{D_{1}, S}(0) \right\rvert + \left\lvert \mathcal{C}^{X, Y_{D_{2}}}_{D_{2}, S}(0) \right\rvert - 1$.
		\item In all other cases, $\left\lvert \mathcal{C}_{S}^{X, Y}(0, 0) \right\rvert = 0$.
	\end{itemize}
\end{lemma}

The problem of computing $\left\lvert \mathcal{C}_{D_{1}, I \cap \tilde{E}_{D_{1}}}^{X_{D_{1}}, Y_{D_{1}}}(\ell_{1}) \right\rvert$ 
and $\left\lvert \mathcal{C}_{D_{2}, I \cap \tilde{E}_{D_{2}}}^{X_{D_{2}}, Y_{D_{2}}}(\ell_{2}) \right\rvert$ for any $I \in \mathcal{I}(S, T(\ell_{1}, \ell_{2}))$ is also similar to the previous setting. 
The stitching technique used in \cref{sec:stitchingSection} can be used to compute these quantities. 
There are a few things we need to be careful about. 
During the construction of $G_{1}$ (and similarly, $G_{2}$), while performing a single-edge contraction or parallel-edge contraction, 
if we delete an edge $e$ from the graph, we need to verify that $\tilde{e} \notin X$.
We also need to ensure that while performing any contraction, we never identify two edges $e$ and $e'$ with each other such that $\tilde{e} \in X$ and $\tilde{e}' \in Y$. 
If these two events occur, then $\left\lvert \mathcal{C}_{D_{1}, I \cap \tilde{E}_{D_{1}}}^{X_{D_{1}}, Y_{D_{1}}}(\ell_{1}) \right\rvert = 0$.
Otherwise, the reduction works similar to \cref{lemma:contractionsCorrect}.
Overall, if we account for $X, Y$ while choosing the values for the base cases, 
the proof of \cref{theorem:mustCrossMainThm} is nearly identical to that of \cref{theorem:mainThm}.

We will now extend the algorithm to count cycles of a fixed length. We will need to define a few terms to do this.

\begin{definition}\label{defn:fixedLengthWalks}
	Let $X, Y \subseteq \tilde{E}$, and let $l \geq 0$. Then,
	$$\mathcal{W}_{G}^{X, Y, l} = \{W \in \mathcal{W}_{G}^{X, Y}: \lvert W \rvert = l \}.$$
	Let $X, Y \subseteq \tilde{E}_{D}$ for some $D \in \{D_{1}, D_{2}\}$, and let $l \geq 0$. Then,
	$$\mathcal{P}_{D}^{X, Y, l} = \{P \in \mathcal{P}_{D}^{X, Y}: \lvert P \rvert = l \}.$$
\end{definition}

Similarly, we  define $\mathcal{W}_{G, S}^{X, Y, l}$,
${\mathcal{P}_{D, S}^{X, Y, l}}$,
${\mathcal{C}_{D, S}^{X, Y, l}(\ell)}$, and
${\mathcal{C}_{S}^{X, Y, l}(\ell_{1}, \ell_{2})}$.

\begin{lemma}\label{lemma:fixedLengthCompatibilityPartition}
	$$\left\lvert {\mathcal{W}_{G, S}^{X, Y, l}} \right\rvert = \sum_{\text{\rm compatible }\ell_{1}, \ell_{2} \in \mathcal{L}_{A}} \left\lvert {\mathcal{C}_{S}^{X, Y, l}(\ell_{1}, \ell_{2})} \right\rvert.$$
	Let $X_{D} = X \cap \tilde{E}_{D}, Y_{D} = Y \cap \tilde{E}_{D}$ for $D \in \{D_{1}, D_{2}\}$. 
	If $(\ell_{1}, \ell_{2}) \neq (0, 0)$ is a compatible pair of labellings then
	$$\left\lvert {\mathcal{C}_{S}^{X, Y, l}(\ell_{1}, \ell_{2})} \right\rvert = \sum_{\text{\rm valid } I \in \mathcal{I}(S, T(\ell_{1}, \ell_{2}))}\sum_{i = 0}^{l - l'} \left\lvert {\mathcal{C}_{D_{1}, I \cap \tilde{E}_{D_{1}}}^{X_{D_{1}}, Y_{D_{1}}, l' + i}(\ell_{1})} \right\rvert \times \left\lvert {\mathcal{C}_{D_{2}, I \cap \tilde{E}_{D_{2}}}^{X_{D_{2}}, Y_{D_{2}}, l' + (l - i)}(\ell_{2})} \right\rvert$$
	where $l' = \lvert \{e \in E_{A}: \ell_{1}(e) \neq 0 \} \rvert$.
	
	If $(\ell_{1}, \ell_{2}) = (0, 0)$ then
	\begin{itemize}
		\item If $\emptyset \neq \varepsilon(S) \subseteq \tilde{E}_{D}$ and $X \subseteq \tilde{E}_{D}$ for $D \in \{D_{1}, D_{2} \}$, then $\left\lvert {\mathcal{C}_{S}^{X, Y, l}(0, 0)} \right\rvert = \left\lvert {\mathcal{C}_{D, S}^{X, Y_{D}, l}(0)} \right\rvert$.
		\item If $\varepsilon(S)= \emptyset$ and $\emptyset \neq X \subseteq \tilde{E}_{D}$ for $D \in \{D_{1}, D_{2}\}$, then $\left\lvert {\mathcal{C}_{S}^{X, Y, l}(0, 0)} \right\rvert = \left\lvert {\mathcal{C}_{D, S}^{X, Y_{D}, l}(0)} \right\rvert$.
		\item If $\varepsilon(S) = \emptyset$ and $X = \emptyset$, then $\left\lvert {\mathcal{C}_{S}^{X, Y, l}(0, 0)} \right\rvert = \left\lvert {\mathcal{C}^{X, Y_{D_{1}}, l}_{D_{1}, S}(0)} \right\rvert + \left\lvert {\mathcal{C}^{X, Y_{D_{2}}, l}_{D_{2}, S}(0)} \right\rvert - \mathbbm{1}_{\{l = 0\}}$,
		where $\mathbbm{1}_{\{l = 0\}} = 1$ if $l=0$, and 0 otherwise.
		\item In all other cases, $\left\lvert {\mathcal{C}_{S}^{X, Y, l}(0, 0)} \right\rvert = 0$.
	\end{itemize}
\end{lemma}

Now, consider the problem of computing 
$\left\lvert {\mathcal{C}_{D_{1}, I \cap \tilde{E}_{D_{1}}}^{X_{D_{1}}, Y_{D_{1}}, j}}(\ell_{1}) \right\rvert$, where $T = T(\ell_{1}, \ell_{2})$, 
and $I \in \mathcal{I}_{S, T}$. We use the same stitching technique detailed in \cref{sec:stitchingSection} 
to construct the graph $G_{1}$. 
However, (referring to notations in \cref{sec:stitchingSection}) because of the way edges in $D$ may get identified with each other, 
the relations between the lengths of the cycles in $G_{1}$, 
and the lengths of PSAWs in $D$ can get complicated. 
Let
$$\psi: \mathcal{C}_{D_{1}, I \cap \tilde{E}_{D_{1}}}^{X_{D_{1}}, Y_{D_{1}}}(\ell_{1}) \rightarrow \mathcal{W}_{G_{1}, S_{1}}^{X_{D_{1}}, Y_{D_{1}}}$$
be the invertible function, whose existence is guaranteed by \cref{theorem:contractionsReduce}.

Let $R$ be the set of all edges in $G_{D_{1}}$ that are either merged with another edge, or deleted, 
by a single-edge contraction or edge-pair contraction, 
during the construction of $G_{1}$.
Note that
$$R \subseteq \bigcup \Big\{ \tilde{e}, \tilde{e}', \tilde{e}'' : \{e, e', e'' \} \in F_{D_{1}}, e \in E_{A}  \Big\}$$
i.e., $R$ is a subset of the set of all edges belonging to the faces in $D_{1}$ that contain edges from $A$. 
Therefore, $|R| \leq 3|E_{A}|$.
We have a disjoint union	$${\mathcal{C}_{D_{1}, I \cap \tilde{E}_{D_{1}}}^{X_{D_{1}}, Y_{D_{1}}, l}}(\ell_{1}) = \bigsqcup_{R_{X} \subseteq R}{\mathcal{C}_{D_{1}, I \cap \tilde{E}_{D_{1}}}^{X_{D_{1}} \cup R_{X}, Y_{D_{1}} \cup (R \setminus R_{X}), l}}(\ell_{1})$$

Consider some $R_{X} \subseteq R$.
For simplicity, let $R_{Y} = R \setminus R_{X}$.
We now want to compute $\left\lvert {\mathcal{C}_{D_{1}, I \cap \tilde{E}_{D_{1}}}^{X_{D_{1}} \cup R_{X}, Y_{D_{1}} \cup R_{Y}, l}}(\ell_{1}) \right\rvert$.
The reduction to $G_{1}$ is the same as before, but this time, 
we have a much better handle on the size of the cycles and PSAWs.
Just like with $S^{i + 1}$ and $X^{i + 1}, Y^{i + 1}$, 
we also iteratively construct $R_{X}^{i + 1} \subseteq R_{X}^{i}$
such that $R_{X}^{i + 1}$ is a set of all the distinct edges in $R_{X}^{i}$ from $E^{i + 1}$.
We construct $R_{Y}^{i + 1}$ similarly.
Following the same ideas as in  \cref{lemma:Si+1SpecialCaseContractions} and \cref{lemma:contractionsCorrect}, we can compute $\left\lvert {\mathcal{C}_{D_{1}, I \cap \tilde{E}_{D_{1}}}^{X_{D_{1}} \cup R_{X}, Y_{D_{1}} \cup R_{Y}, l}}(\ell_{1}) \right\rvert$.

The only thing that would have to be changed in \cref{lemma:Si+1SpecialCaseContractions} is in case (2c). 
When we apply a parallel-edge contraction, and delete the edges $e'', e'''$. 
Let $d'', d'''$ be the  number of distinct edges in $E_{D_{1}}$ that are identified with $e'', e'''$ respectively, in the graph $D^{i}$. 
We modify case (2c) to say $\left\lvert \mathcal{C}_{D^{i}, S^{i}}^{X^{i} \cup R_{X}^{i}, Y^{i} \cup R_{Y}^{i}, l} \right\rvert = 1$ 
if and only if $l = d'' + d'''$, and otherwise, 
it is $0$.
It goes without saying that we should also account for $X^{i}, Y^{i}$ here, 
and claim that it is $1$ only if $X^{i} \subseteq \{\tilde{e}'', \tilde{e}'''\} $.

Then, we use the following lemma to compute $\left\lvert {\mathcal{C}_{D_{1}, I \cap \tilde{E}_{D_{1}}}^{X_{D_{1}} \cup R_{X}, Y_{D_{1}} \cup (R \setminus R_{X}), l}}(\ell_{1}) \right\rvert$.

\begin{lemma}\label{lemma:fixedLengthContractionsCorrect}
	Let the graph $D^{i + 1}$ be such that it does not fall into the special cases from \cref{lemma:Si+1SpecialCaseContractions}. 
	Let $S^{i + 1}$, $X^{i + 1}, Y^{i+ 1}$, $R_{X}^{i + 1}, R_{Y}^{i + 1}$ be constructed as explained. Let $r = |R_{X}^{i}|$, and $r' = |R_{X}^{i + 1}|$.
	If $D^{i + 1}$ is constructed by an edge-pair contraction, or single-edge contraction, then
	$$\left\lvert \mathcal{C}_{D^{i}, S^{i}}^{X^{i} \cup R_{X}^{i}, Y^{i} \cup R_{Y}^{i}, l}(\ell_{1}^{i}) \right\rvert = \left\lvert \mathcal{C}_{D^{i + 1}, S^{i + 1}}^{X^{i} \cup R_{X}^{i}, Y^{i} \cup R_{Y}^{i}, l + r' - r}(\ell_{1}^{i + 1}) \right\rvert.$$
	If $D^{i + 1}$ is constructed by a parallel-edge contraction, with edges $e'', e'''$ deleted from $D^{i}$, let $d'', d'''$ be the number of edges in $E_{D_{1}}$ identified with $e'', e'''$ in $D^{i}$, then
	$$\left\lvert \mathcal{C}_{D^{i}, S^{i}}^{X^{i} \cup R_{X}^{i}, Y^{i} \cup R_{Y}^{i}, l}(\ell_{1}^{i}) \right\rvert = \left\lvert \mathcal{C}_{D^{i + 1}, S^{i + 1}}^{X^{i} \cup R_{X}^{i}, Y^{i} \cup R_{Y}^{i}, l + r' - r}(\ell_{1}^{i + 1}) \right\rvert + \mathbbm{1}_{\{(\ell_{1}^{i}, \ell_{2}^{i}) = (0, 0);\  \varepsilon(S^{i}) = \emptyset;\ l = d'' + d'''\}}.$$
\end{lemma}

Therefore, the algorithm in \cref{sec:inductionSection} can be easily adapted to compute 
$\lvert {\mathcal{W}_{G}^{X, Y, l}} \rvert$. 
During each level of the tree we constructed in \cref{sec:inductionSection}, 
given a problem $\lvert \mathcal{W}^{X_{k}, Y_{k}}_{G_{k}, S_{k}} \rvert$, 
we compute $\lvert {\mathcal{W}^{X_{k}, Y_{k}, l}_{G_{k}, S_{k}}} \rvert$ for all $l \leq \lvert \tilde{E}_{k} \rvert$. 
The correctness of this algorithm can be easily verified. 
Let us now consider the time complexity of this algorithm.

In \cref{sec:inductionSection}, we saw that a problem at level $k$ of the tree has at most $4(\kappa^{t_{k}})^{2} t_{k}\binom{t_{0} + \dots + t_{k}}{t_{k}} $ subproblems as children. 
In this setting, the number of children can be bounded by
$$4(\kappa^{t_{k}})^{2} t_{k} \binom{t_{0} + \dots + t_{k}}{t_{k}} O(m_{k}2^{3t_{k}})$$
The $2^{3t_{k}}$ factor is because that is the maximum possible number of choices for $R_{X} \subseteq R$, 
and the factor $O(m_{k})$ is a bound on all possible lengths of cycles we have to consider. 
It is easy to see from the proof of \cref{theorem:mainThm} that this extra factor can be absorbed into the $\beta^{\sqrt{n}}$ term.

\begin{theorem}\label{theorem:fixedLengthMainThm}
	There exists a constant $\beta$ such that given any connected 3-regular planar multi-graph (with no loops) $H$ with $n$ vertices, and any $X, Y \subseteq \tilde{E}$, the number of cycles on $H$ of a fixed length $l$, that traverse each edge in $X$, and do not traverse any edge in $Y$, can be computed in $\beta^{\sqrt{n}}$ time.
\end{theorem}

Just as we are able to count the number of cycles on $H$ of a fixed length,
we can count the number of cycles on $H$ that divide $H$ into two subgraphs with an $p$ faces on one subgraph, 
and $|\tilde{V}| - p$ faces on the other. 
The extra factor this produces can be absorbed into the $\beta^{\sqrt{n}}$ term.

\end{document}